\documentclass[preprint,12pt]{elsarticle}
\usepackage{amssymb}
\usepackage{amsmath}
\usepackage{booktabs}
\usepackage{tabularx}
\usepackage{array}
\usepackage{longtable}
\usepackage{placeins}
\usepackage{caption}
\usepackage[monochrome]{xcolor}
\usepackage[a4paper,left=2cm,right=2cm,top=2.5cm,bottom=2.5cm]{geometry}

\newif\ifrevisionmarked
\ifdefined\CleanVersion\revisionmarkedfalse\else\revisionmarkedtrue\fi
\newcommand{\rev}[1]{{\ifrevisionmarked\color{red}\else\color{black}\fi #1}}

\usepackage{tocloft}

\definecolor{PRblue}{RGB}{0,102,153}

\usepackage[
    colorlinks=true,
    linkcolor=PRblue,
    citecolor=PRblue,
    urlcolor=PRblue,
    linktoc=all,
    bookmarksnumbered=true
]{hyperref}

\journal{Physics Reports}

\begin{document}

\begin{frontmatter}
\biboptions{sort&compress}

\title{Source Detection in Complex Networks: Observability, Identifiability, and Inference Frameworks}

\author[aff1]{Xiu-Xiu Zhan\fnref{fn1}}
\author[aff1]{Qiao Ke\fnref{fn1}}
\author[aff1]{Chengjun Zhang}
\author[aff1]{Miao He}
\author[aff2]{Chenwei Xie}
\author[aff2]{Seba Contreras}
\author[aff2,aff4]{Zhongliang Zhao\corref{cor1}}
\author[aff2]{Yixiu Kong\corref{cor1}}
\author[aff3]{Cong Li\corref{cor1}}
\author[aff1]{Chuang Liu\corref{cor1}}

\affiliation[aff1]{
    organization={Center for Complexity Sciences, Hangzhou Normal University},
    city={Hangzhou},
    postcode={311121},
    country={China}
}

\affiliation[aff2]{
    organization={International Research Center for Complexity Sciences, Hangzhou International Innovation Institute, Beihang University},
    city={Hangzhou},
    postcode={311115},
    country={China}
}

\affiliation[aff3]{
    organization={Adaptive Networks and Control Lab, College of Future Information Technology, Fudan University},
    city={Shanghai},
    postcode={200433},
    country={China}
}

\affiliation[aff4]{
    organization={School of Electronic and Information Engineering, Beihang University},
    city={Beijing},
    postcode={100191},
    country={China}
}
\fntext[fn1]{These authors contributed equally to this work.}

\cortext[cor1]{Corresponding authors.}

\begin{abstract}
Tracing diffusion to its origin is a central inverse problem in complex networks, with applications ranging from epidemic investigation and information spreading to cyber incidents and disruptions in interconnected systems. Although source-detection methods have advanced along diverse methodological lines, their capabilities cannot be assessed independently of the evidence on which they operate. Network structure and diffusion dynamics determine which propagation histories are possible, while the observation mechanism determines which traces of those histories are retained. This review develops an information-centric and identifiability-aware perspective that connects these elements to source inference. We organize existing research around three overlapping observation settings: complete state snapshots, incomplete or noisy observations, and sensor-based observation and design. Within this organization, we examine structural, probabilistic, message-passing, simulation-based, and learning-based methods, comparing their required inputs, underlying assumptions, inference mechanisms, and computational demands. System-theoretic observability, statistical identifiability, and information-theoretic constraints provide complementary ways to examine what the available observations can reveal, while separating limitations of the evidence from those of estimation. Drawing on applications across epidemiological, information, biological, neural, and socioeconomic systems, we identify challenges in uncertainty assessment, model robustness, transfer across networks, and the acquisition of informative observations. 
This synthesis clarifies the conditions under which diffusion sources can be identified and how inference methods should be selected according to the available observations.
\end{abstract}


\end{frontmatter}

\begingroup
\small
\setlength{\parskip}{0pt}
\tableofcontents
\endgroup

\vspace{1em}

\section{Introduction}

Diffusion processes shape the evolution of complex networked systems, from epidemic outbreaks and information cascades to cyber incidents and disruptions in production and trade~\cite{keeling2005networks,pastor2015epidemic,shah2011rumors,shi2026temporal}. Much of network diffusion research starts from a specified initial condition and asks how a process will unfold. In many applications, however, the direction of inquiry is reversed: investigators observe part of a spreading process and seek to determine where it began. This is the problem of source detection~\cite{shah2011rumors,jiang2016identifying}. Its target is the true initiating node or source set, which need not coincide with the first reported case, the most central node, or the most influential participant in the observed cascade~\cite{fan2026influential}.

The inverse nature of this task introduces difficulties that cannot be attributed to an inference algorithm alone. Stochastic transmission, alternative propagation paths, recovery, and changes in network connections may produce different histories consistent with the same observed pattern. At the same time, the measurements available to an investigator rarely provide a complete account of those histories. A final snapshot can reveal the states of all nodes while concealing the order of transmission; a limited set of monitors may record informative arrival times or directions without observing most nodes. Measurement errors, missing records, and uncertain topology introduce further ambiguities~\cite{altarelli2014bayesian,antulov2015identification}. Whether these ambiguities prevent identification of the source depends on the diffusion dynamics, the candidate sources under consideration, and what is actually recorded. In particular, a source that cannot be identified from one observation may be distinguishable under a different measurement protocol. The central issue is therefore not simply how to search for a source, but which candidate origins the available evidence can distinguish.

\begin{figure}[!t]
\centering
\includegraphics[width=\linewidth]{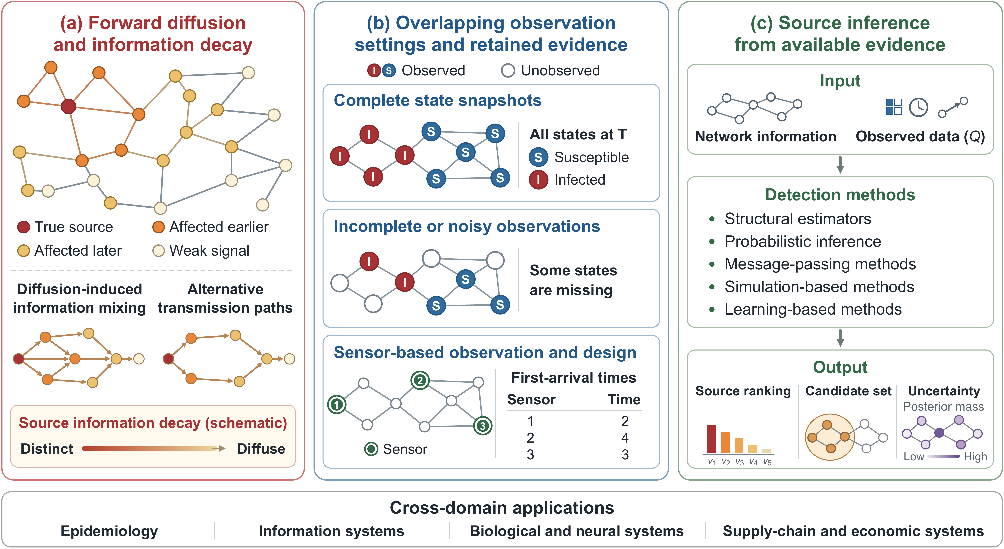}
\caption{ 
Information-centric overview of source detection in complex networks. (a) Forward diffusion may follow multiple transmission paths, potentially weakening the evidence that distinguishes the initiating source. The illustrated decline in source information is schematic and depends on the diffusion dynamics and observation mechanism. (b) Three overlapping observation settings: complete state snapshots, incomplete or noisy observations, and sensor-based observation and design. In the state diagrams, red nodes labeled (I) and blue nodes labeled (S) indicate recorded infected and susceptible states, respectively, while unfilled nodes represent unobserved states. Complete snapshots do not necessarily reveal transmission histories, whereas sensor observations may provide local temporal information, such as first-arrival times. (c) Source inference uses network information and observed data ($Q$) through structural, probabilistic, message-passing, simulation-based, and learning-based methods, producing source rankings, candidate sets, or uncertainty estimates. The bottom panel presents the application domains covered in this review.
}
\label{fig:roadmap}
\end{figure}

This relationship between diffusion, observation, and inference provides the organizing perspective of the review. As illustrated in Fig.~\ref{fig:roadmap}, network propagation generates the histories that connect an initiating source to later system states, while the observation mechanism retains only selected features of the process. Inference methods then use those records to evaluate possible origins. The amount of data collected, however, does not by itself determine its value for source detection. Two observation schemes may cover the same nodes yet record different variables, and measurements that reveal a change in state may be more informative than repeated records of an unchanged state. The distinction also applies to sensor deployment: locating monitors to maximize spatial coverage or obtain an early warning is not necessarily equivalent to placing them for accurate source detection.

Source-detection research has developed a broad range of methods. Structural approaches exploit the geometry of observed spreading patterns, while probabilistic methods evaluate candidates under specified diffusion models~\cite{shah2011rumors,altarelli2014bayesian}. Message-passing and simulation-based procedures handle uncertain propagation histories through different approximations or evaluations, while learning-based methods infer origins from representations obtained through training data. These approaches have been studied on static, temporal, multilayer, and higher-order networks, often under substantially different assumptions about the available observations~\cite{antulov2015identification,lokhov2014inferring,battiston2021physics,ke2026source,lambiotte2019higherorder}. Earlier reviews, including Jiang et al.~\cite{jiang2016identifying}, have surveyed the major methodological developments. Yet comparisons organized primarily around algorithmic families can obscure important differences between the inference problems being addressed. A method supplied with infection times or recorded transmission directions does not face the same task as one using only a state snapshot, even when both are evaluated on the same network. Likewise, improved performance may arise from additional information, more accurate diffusion assumptions, or a more effective inference procedure. Distinguishing these factors is essential for meaningful comparisons across methods.

We develop an information-centric and identifiability-aware perspective on source detection, emphasizing the relationship between available observations and the conclusions that can be drawn from them. Existing studies are organized into three overlapping settings: complete state snapshots, incomplete or noisy observations, and sensor-based observation and design. These settings do not form a hierarchy of information completeness. Sensor records are often incomplete, whereas complete spatial snapshots may leave the transmission history unknown. To characterize the observations more precisely, we consider six dimensions: spatial coverage, temporal sampling, recorded variables, topology knowledge, measurement errors, and acquisition design. This organization allows methods using different forms of evidence to be examined together without overlooking differences in their observation assumptions.

The theoretical discussion examines the conditions under which distinct sources can be distinguished and the uncertainty that remains when the observations are inconclusive. System-theoretic observability concerns whether the initial state of a specified dynamical system can be determined from its measured outputs~\cite{astrom2008feedback}. Statistical identifiability concerns whether different candidate sources induce distinct observation distributions, whereas posterior uncertainty describes how probability is distributed across candidates after a particular observation~\cite{zhu2015information}. Information-theoretic bounds further characterize the limits of inference under specified source priors and observation models~\cite{cover2006elements}. These distinctions separate limitations of the available evidence from numerical instability, model mismatch, and computational difficulties. They also preserve the objective of source detection: when the observations cannot identify a unique origin, a ranking, candidate set, or probability distribution expresses the remaining uncertainty rather than changing the target of inference.

We compare representative methods in terms of detection accuracy, required observations, diffusion assumptions, inference outputs, computational demands, and conditions of applicability. Particular attention is given to the extension from single-source to multiple-source detection, where overlapping spreading regions and joint source-set inference introduce additional difficulties~\cite{prakash2014efficiently}. Applications in epidemiology, information systems, biological and neural networks, and supply-chain and economic systems illustrate how both the definition of a source and the evidence available for identifying it vary across domains. These comparisons also reveal unresolved challenges in uncertainty assessment, robustness to incomplete or misspecified models, generalization across networks, and the design of informative observations.

The remainder of the review is organized as follows. Section~\ref{sec:foundations} introduces the source-detection problem, observation models, evaluation principles, and theoretical limits. Sections~\ref{sec:full-observability}--\ref{sec:sensor-based} examine methods using complete state snapshots, incomplete or noisy observations, and sensor-based observations and design, respectively. Section~\ref{sec:data-driven} discusses data-driven developments and cross-domain applications, while Section~\ref{sec:unifying-future} synthesizes the main challenges and future research directions.

\section{Foundations of Source Detection}
\label{sec:foundations}
\subsection{Diffusion Processes in Complex Systems}

A network describes possible interactions between nodes, while a diffusion model specifies how states evolve through those interactions~\cite{masuda2017random,rosvall2014memory,lambiotte2015memory,zhang2016dynamics}. In epidemic models, for example, a contact network identifies potential transmission routes but does not reveal which contacts actually led to infection~\cite{keeling2005networks}. The realized transmission network may therefore differ from the contact network, and the observed network may differ from both because links are missing or incorrectly recorded~\cite{zhan2019backbones}.

The choice of diffusion model affects what can be inferred about past infections from observed node states. Under Susceptible--Infected (SI) dynamics, infected nodes remain infected; under Susceptible--Infected--Recovered (SIR) dynamics, infected nodes can recover; and under Susceptible--Infected--Susceptible (SIS) dynamics, recovery returns nodes to the susceptible state, allowing reinfection~\cite{pastor2015epidemic}. Independent-cascade models instead describe diffusion through probabilistic activation attempts. These differences matter when identifying candidate sources. Restricting the search to currently infected nodes may be appropriate under single-source SI dynamics, but can exclude the true source under SIR or SIS~\cite{shah2011rumors,lokhov2014inferring,altarelli2014bayesian}. The transition parameters, recovery rules, and observation time must also be specified when evaluating the likelihood of a candidate source.

Diffusion dynamics can be described at different levels of approximation. For Markovian SIS dynamics with specified transition rates, the full joint-state process admits an exact finite-state description, whereas the N-intertwined model uses a mean-field approximation to describe the evolution of node infection probabilities~\cite{vanmieghem2009virus,wang2013interconnected}. Sahneh et al.~\cite{sahneh2013generalized} develop a generalized epidemic mean-field formulation for multistate spreading on multilayer networks. These approximations affect the predicted state probabilities used to evaluate candidate sources.

Temporal network models capture changes in transmission opportunities as network connections evolve, whereas static models assume fixed connections~\cite{delvenne2015diffusion,holme2015modern}. These changes may occur independently of spreading or arise from responses to it, as when awareness of infection leads individuals to reduce contacts~\cite{pastor2015epidemic,wang2017awareness}. Such adaptive behavior alters the possible transmission paths and must be considered when inferring sources from observations collected over time.

The available observations rarely provide a complete account of how diffusion unfolded. Epidemic surveillance may record diagnoses rather than infection times, while online platforms capture posts and reposts without necessarily recording every exposure. Security logs likewise document selected events rather than complete attack histories. These differences between the underlying process and the recorded evidence must be considered when specifying an observation model. Application-specific examples are discussed in Section~\ref{sec:data-driven}.

\subsection{Source Detection as an Inverse Problem of Diffusion}

The distinction between the diffusion process and its observations leads to an inverse problem. In the forward setting, a source and a propagation model determine how an outbreak evolves and what may be observed. Source detection works in the opposite direction: given the available records \(Q\), the aim is to identify the node \(s^*\) at which propagation began. The true source is the target of recovery, regardless of whether the observations are sufficient to identify it. Statistical estimators provide possible ways to infer this source, but no particular estimation rule defines the task~\cite{jiang2016identifying,zhu2015information}.

In a single-source epidemic without additional introductions, the source is the first individual infected. This need not be the first person diagnosed or reported, since infection may precede detection by an unknown interval and some cases may never be recorded. The source is also distinct from nodes selected for their centrality or expected spreading influence. Such measures describe network position or propagation potential rather than the historical origin of a particular outbreak.

Whether the true source can be identified depends on the diffusion dynamics and the observation mechanism. Different sources may induce identical observation distributions, making them statistically indistinguishable. Even when these distributions differ, a single observed outbreak may remain compatible with several candidate sources. This ambiguity is distinct from difficulties arising from model misspecification, numerical instability, or the computational cost of searching the candidate space. Exact source identification is possible under suitable observation conditions, for example when the labeled initial state is recorded. Source identifiability must therefore be considered under explicit assumptions about the diffusion process and the available observations.

With multiple initiating sources, the inference target becomes a source set \(\mathcal S^*\). The problem depends on whether the number of sources and their initiation times are known, as well as whether the spreading processes evolve independently or interact. When their spreading regions overlap, it may be difficult to determine which source accounts for a particular observation. Propagation from one source may also obscure evidence associated with another. These difficulties are compounded by the number of candidate source sets, which can grow combinatorially with network size. When the observations remain compatible with several candidates, a method may report alternative source sets or their estimated probabilities. Such outputs describe the uncertainty in the available evidence, while the true initiating source set remains the target of inference.

\subsection{Mathematical Formulation and Problem Definition}
\label{sec:problem_definition}

Consider a network \(G=(V,E)\) with \(N=|V|\) nodes and diffusion parameters \(\Theta\). We begin with the single-source setting, where the true source \(s^*\) belongs to an admissible candidate set \(\mathcal C_1\subseteq V\). This set may exclude nodes known not to be sources or incorporate constraints established independently of the unknown source identity. Knowledge of the network and diffusion parameters does not, however, imply knowledge of the realized node states, transmission history, or initiating source.

The diffusion dynamics on \(G\) are modeled as a stochastic process that governs the temporal evolution of node states. Let \(X_i(t)\) denote the state of node \(v_i\) at time \(t\), where the state space depends on the underlying diffusion mechanism. For example, epidemic models distinguish susceptible, infected, and recovered states, whereas information diffusion and cybersecurity applications adopt different state definitions according to their respective propagation dynamics. The global network state is represented by
$\mathbf{X}(t)=\bigl(X_1(t),X_2(t),\ldots,X_N(t)\bigr),$
which characterizes the system configuration at time \(t\).


The network constrains possible transmission routes, while the realized diffusion trajectory depends on the transmission events that occur along those routes. A breadth-first-search (BFS) tree retains one minimum-hop path to each reachable node; it is a structural representation of the graph and need not reproduce the transmission history. Minimum-hop paths coincide with earliest-arrival routes when transmission along every edge has the same fixed positive delay. Random delays, failed transmission attempts, recovery, or changing connections require the corresponding diffusion model to determine which histories are possible~\cite{pastor2015epidemic}.

For multiple initiating sources, let \(\mathcal S^*\subseteq V\) denote the true source set and \(\mathcal C\subseteq 2^V\) the family of admissible candidate source sets. If the number of sources \(k\) is known, \(\mathcal C\) is restricted to sets of size \(k\). Otherwise, the source count must also be inferred, for example through model selection, a prior, or a suitable penalty. The single-source case corresponds to singleton candidates \(\{v\}\), where \(v\in\mathcal C_1\). We use source-set notation when a formulation applies to both cases, while retaining node-valued estimates for methods designed to identify a single source.

To distinguish the underlying diffusion process from the data used for source inference, let \(\mathbf X_{[t_0,T]}\) denote the latent labeled state trajectory. The diffusion process and the resulting observations can be described by
\begin{equation}
\mathbf X_{[t_0,T]}\sim P_\Theta(\,\cdot\mid\mathcal S^*,G),
\qquad
Q\sim P_{\mathcal O}(\,\cdot\mid\mathbf X_{[t_0,T]},G),
\label{eq:unified_observation}
\end{equation}
where \(P_\Theta\) denotes the conditional distribution of the latent diffusion trajectory given the true source set \(\mathcal S^*\) and network \(G\), under the diffusion model with parameters \(\Theta\). The variable \(Q\) denotes the collection of records available for source inference, and \(P_{\mathcal O}\) denotes its conditional distribution given the latent trajectory and network \(G\) under the observation mechanism \(\mathcal O\). The observation mechanism \(\mathcal O\) determines which variables are recorded, at which nodes and times, and how missing or erroneous measurements arise. First-arrival times depend on the trajectory rather than the final state alone. Missing observations, categorical errors, censoring, and timestamp errors must be accounted for separately, as they cannot generally be represented by additive noise. Unknown initiation times or diffusion parameters must also be treated as nuisance quantities and estimated, conditioned on, or integrated out.

The distinction between latent states and recorded observations matters even when measurements consist only of node states. For instance, a binary infectious/noninfectious record of an SIR process groups susceptible and recovered nodes into the same observed category, although their possible infection histories differ. Source inference must therefore account for the relationship between the underlying states and the recorded labels. In particular, methods developed for infected subtrees under SI dynamics cannot be applied directly to models with recovery without adjusting the observation model and the corresponding inference assumptions.

An individual measurement records a particular variable, while \(Q\) denotes the collection of records available for source inference. For node-level observations, we write \(Q=\{(v_r,t_r,m_r,y_r)\}_{r=1}^{R}\), where \(v_r\), \(t_r\), \(m_r\), and \(y_r\) denote the sampled node, sampling time, measured variable, and recorded value, respectively. Measurements may include node states, transmission directions, or local events. For arrival-time records, the event used to define arrival must be specified. This representation accommodates repeated measurements, asynchronous sampling, and sensors that change locations over time. The sampling schedule is part of \(\mathcal O\), and any dependence on earlier observations must be included when measurements are selected adaptively.

The observation model allows source-detection methods to be compared under different measurement conditions. We organize the literature into three overlapping settings according to the available records and how they are collected. These settings are not mutually exclusive and do not represent successive levels of information completeness:
\begin{itemize}
\item \textbf{Complete state snapshots.} The observations \(Q=\{\mathbf X(T_j)\}_{j=1}^{J}\) contain the labeled states of all nodes at one or more specified times. Transmission events and state changes between snapshots generally remain unobserved. Complete spatial coverage therefore does not imply knowledge of the full diffusion history or system-theoretic observability.

\item \textbf{Incomplete or noisy observations.} Node states, event times, or network structure may be missing or recorded with errors. An observed susceptible or inactive state must be distinguished from a missing measurement. Records collected at different times also need not correspond to a common network state.

\item \textbf{Sensor-based observation and design.} Fixed, mobile, or adaptively selected sensors provide local measurements, including node states, first-arrival times, transmission directions, and other recorded events. These observations are generally incomplete. Sensor locations may be predetermined, optimized before diffusion begins, or adjusted as new measurements become available.
\end{itemize}

The three observation settings can be further characterized by six dimensions: spatial coverage, temporal sampling, recorded variables, topology knowledge, measurement errors, and acquisition design. These dimensions need not vary together. Measurements from the same nodes may differ in their ability to distinguish sources depending on whether they record final states, state transitions, or arrival times. Mobile sensors also require location histories to associate observations with network nodes, while asynchronous and adaptive measurements must retain their sampling times and acquisition rules. When sampling depends on earlier observations, the absence of a measurement may itself carry information. Thus, neither the number of measurements nor their spatial coverage alone determines the information available for source inference.

With the diffusion and observation models specified, Maximum A Posteriori (MAP) estimation and Maximum Likelihood Estimation (MLE) provide statistical approaches to identifying the true source. Given a specified prior and observation likelihood, the MAP estimator is defined as~\cite{antulov2015identification,altarelli2014bayesian}
\begin{equation}
    \hat{\mathcal{S}}
    =
    \operatorname*{argmax}_{\mathcal{S}\in\mathcal{C}}
    P(\mathcal{S}\mid Q,G,\Theta)
    =
    \operatorname*{argmax}_{\mathcal{S}\in\mathcal{C}}
    \frac{
    P(Q\mid \mathcal{S},G,\Theta)P(\mathcal{S})
    }{
    P(Q\mid G,\Theta)
    },
    \label{eq:map_source}
\end{equation}
where \(P(\mathcal S)\) denotes the prior over candidate sources conditional on the specified network and diffusion parameters. For an observation with positive evidence, \(P(Q\mid G,\Theta)\) is shared by all candidate sources. The MAP estimate can therefore be obtained by maximizing the unnormalized posterior:

\begin{equation}
    \hat{\mathcal{S}}
    =
    \operatorname*{argmax}_{\mathcal{S}\in\mathcal{C}}
    P(Q\mid \mathcal{S},G,\Theta)P(\mathcal{S}).
    \label{eq:posterior_source}
\end{equation}
When the prior distribution over the candidate source space is assumed to be uniform, the MAP estimator reduces to the MLE~\cite{snijders2007maximum,byshkin2018fast}:

\begin{equation}
\hat{\mathcal S}_{\mathrm{ML}}\in
\operatorname*{argmax}_{\mathcal S\in\mathcal C}
P(Q\mid\mathcal S,G,\Theta),
\label{eq:mle_source}
\end{equation}
where the likelihood describes the probability or probability density of the observed records under the assumed diffusion and measurement models. If several candidates attain the same maximum, a tie-breaking rule is needed to select a single estimate; alternatively, all maximizers can be retained. Under zero--one loss, MAP is the Bayes-optimal decision rule for the specified model and prior, while other loss functions may lead to different estimates. A high likelihood or a concentrated posterior does not guarantee identification of the true source, particularly when the model is misspecified. Computing the likelihood may require marginalizing over unobserved transmission histories, while multiple-source detection introduces an additional combinatorial search problem. Structural scores, simulation-based criteria, and learned rankings can also be used without assigning them a posterior probability interpretation.

Unknown initiation times and diffusion parameters introduce additional uncertainty into source inference. When \(t_0\) is unknown, it can be estimated jointly with the source or integrated out under a specified prior. These approaches may yield different source rankings: joint maximization selects the most favorable initiation time for each candidate, whereas marginalization accounts for the likelihood across possible initiation times. The same distinction applies to unknown transmission rates. Source and initiation time may also be dependent, for example when certain nodes can initiate diffusion only within specified time windows. Such restrictions can be incorporated into the joint prior or imposed as constraints in likelihood-based estimation. In either case, inference guarantees established for known parameters do not automatically carry over. Source estimates must therefore be interpreted in light of how these unknown quantities are treated.

Source-detection methods do not necessarily rely on explicit MAP or ML estimation. Structural methods rank candidate sources using the geometry of observed infections or feasible transmission paths. Message-passing methods propagate probabilistic constraints, whereas simulation-based methods compare observed data with diffusion outcomes generated from candidate sources. Minimum description length (MDL) approaches balance the coding costs of source configurations and the resulting infection patterns~\cite{prakash2014efficiently}. Learning-based methods construct source estimators from labeled or simulated training data. 
These methods may return individual nodes, source rankings, candidate sets, or uncertainty estimates, and their scores should not necessarily be interpreted as posterior probabilities.

\subsection{Performance Evaluation}
\label{performance evaluation}

The evaluation of source-detection methods depends on the detection task, estimator output, and observation setting. Meaningful comparisons require consistent experimental conditions, including network structure, diffusion dynamics, recorded variables, sampling and missingness mechanisms, measurement errors, and acquisition design. For sensor-based methods, the monitoring budget must also be specified. Detection accuracy, robustness, computational efficiency, scalability, and generalization should therefore be assessed under clearly defined network, diffusion, and observation conditions. We first introduce evaluation metrics for single-source and multiple-source detection, followed by the observation-dependent experimental controls summarized in Table~\ref{tab:evaluation_observation_paradigms}.

\subsubsection{Basic Evaluation Metrics}

Across observation settings, source inference targets the true source node or source set. Detection Accuracy (ACC), Average Error Distance (AED), candidate ranking, and source-set similarity assess different properties of an estimator's output~\cite{jiang2016identifying}. They should be interpreted under the same candidate domain, source-count convention, and observation protocol when methods are compared.

The following metrics are first introduced for the single-source detection problem. Assume that \(J\) independent experimental trials are conducted, where \(v^{*(j)}\) and \(\hat{v}^{(j)}\) denote the true and estimated source nodes, respectively, in the \(j\)-th trial. The ACC is defined as
\begin{equation}
    \mathrm{ACC}
    =\frac{1}{J}\sum_{j=1}^{J}
    \mathbb{I}\!\left(\hat{v}^{(j)}=v^{*(j)}\right),
    \label{eq:unified_source_acc}
\end{equation}
where \(\mathbb{I}(\cdot)\) denotes the indicator function, equal to \(1\) if the corresponding condition holds and \(0\) otherwise~\cite{shah2011rumors}. To provide a finer-grained evaluation, the diameter-normalized AED provides a quantitative measure of the topological distance between the estimated and true source nodes, i.e.,
\begin{equation}
    \varepsilon_d
    =\frac{1}{J}\sum_{j=1}^{J}
    \frac{d\!\left(\hat{v}^{(j)},v^{*(j)}\right)}
    {\operatorname{diam}(G)},
    \label{eq:unified_distance_error}
\end{equation}
where \(d(\cdot,\cdot)\) denotes the shortest-path distance between two nodes in \(G\), and \(\operatorname{diam}(G)\) denotes the diameter of \(G\). When a method outputs a ranked list of candidate sources, ranking-based
measures may also be reported~\cite{lokhov2014inferring}. The Top-\(k\) hit rate is the proportion of
independent trials in which the true source appears among the \(k\)
highest-ranked candidates. The mean reciprocal
rank (MRR) is the average
reciprocal rank assigned to the true source across all trials, with a rank of
\(1\) indicating the highest-ranked candidate. Higher values of either metric
indicate better ranking performance.

The diameter normalization assumes a connected network with positive finite diameter. For disconnected candidate domains, unreachable pairs need an explicit convention, and reporting raw hop error alongside normalized error avoids hiding that choice. Weighted or directed distances similarly require a specified distance definition.

For methods that estimate a set of source nodes, let \(\mathcal S^{*(j)}\) and \(\widehat{\mathcal S}^{(j)}\) denote the true and estimated source sets in independent trial \(j\), respectively, for \(j=1,\ldots,J\)~\cite{zhu2017catch}. Define \(P_j=|\mathcal S^{*(j)}\cap\widehat{\mathcal S}^{(j)}|/|\widehat{\mathcal S}^{(j)}|\) and \(R_j=|\mathcal S^{*(j)}\cap\widehat{\mathcal S}^{(j)}|/|\mathcal S^{*(j)}|\). With nonempty true source sets, an empty prediction is assigned zero precision, recall, and \(F_1\). We use trial averages:
\begin{equation}
\overline P=\frac1J\sum_{j=1}^{J}P_j,\quad
\overline R=\frac1J\sum_{j=1}^{J}R_j,\quad
\overline F_1=\frac1J\sum_{j=1}^{J}\frac{2P_jR_j}{P_j+R_j},
\label{eq:unified_multisource_metrics}
\end{equation}
where a zero denominator in the last term gives zero. The harmonic mean of averaged precision and recall is a different aggregation convention and should not be substituted without notice.

\subsubsection{Performance Evaluation under Different Observation Paradigms}

Table~\ref{tab:evaluation_observation_paradigms} summarizes the measurement conditions that accompany the common metrics. An accuracy value is interpretable only together with the candidate population, source prior, dynamics, outbreak stage, and recording mechanism. These conditions should be held fixed in comparisons or varied explicitly in sensitivity experiments.

\begin{table}[!htbp]
\ifrevisionmarked\color{red}\else\color{black}\fi
\centering
\caption{Observation settings and evaluation controls for source detection. The settings are not mutually exclusive.
}
\label{tab:evaluation_observation_paradigms}
\footnotesize
\setlength{\tabcolsep}{3.5pt}
\renewcommand{\arraystretch}{1.3}
\begin{tabularx}{\textwidth}{@{}p{0.20\textwidth}XX@{}}
\toprule
\textbf{Observation setting} & \textbf{Input and experimental controls} & \textbf{Additional outcomes or design criteria}\\
\midrule
Complete snapshots & All labeled states at one or several times; dynamics, source prior, time window, and network knowledge & Accuracy of source estimates and rankings, and sensitivity to outbreak stage and model mismatch\\
\addlinespace
Incomplete or noisy records & Masks, state errors, missing/spurious topology, asynchronous times, and censoring & Sensitivity of inference accuracy to each error mechanism, and size and coverage of any reported candidate set\\
\addlinespace
Sensor records and design & Budget, locations or mobility history, recorded variables, clock errors, fixed or adaptive acquisition & Source error and decision time, separately from alarm delay, coverage, or state-reconstruction objectives\\
\bottomrule
\end{tabularx}
\end{table}
A ranked output can be evaluated by the Top-\(k\) hit rate or MRR, while a source-set estimate can be evaluated by set-based measures. Probability calibration and uncertainty-set coverage require their own checks~\cite{gneiting2007proper,angelopoulos2023conformal}. These measures assess different outputs of the same source inference task.

For incomplete observations, vary node coverage and state error rates separately. Report whether missingness is random, degree dependent, or correlated with infection, and whether observed negative states are reliable. For uncertain networks, specify missing and spurious edges and missing nodes separately; each changes candidate paths differently. Temporal experiments should distinguish snapshot count, observation window, asynchronous sampling, censoring, and timestamp noise. These are controls on the inference input, not alternative accuracy metrics.

For sensor-based methods, report the monitor budget, acquisition variables, placement rule, and whether locations are fixed or adaptively changed~\cite{paluch2020optimizing}. The observation interval and treatment of sensors never reached by the process are also necessary. Evaluate acquisition objectives separately from source-estimation error: detecting any propagation event early does not by itself mean identifying its origin early.
 

For early-warning deployment, the delay between diffusion onset and the first sensor record is commonly used to evaluate monitoring performance~\cite{leskovec2007outbreak}:
\begin{equation}
    \bar{t}_{\mathrm{det}}
    =\frac{1}{J}\sum_{j=1}^{J}
    \left(
    \min_{o\in V_{\mathrm{sen}}}\tau_o^{(j)}-t_0^{(j)}
    \right),
    \label{eq:detection_time}
\end{equation}
where \(V_{\mathrm{sen}}\) denotes the set of deployed sensors, and \(t_0^{(j)}\) and \(\tau_o^{(j)}\) denote the diffusion onset time and the first recorded arrival at sensor \(o\) in trial \(j\), respectively. Eq.~(\ref{eq:detection_time}) gives the mean alarm delay when every trial produces at least one sensor record. If some outbreaks remain undetected within the observation window, the detection fraction and the treatment of these trials must be reported. Averaging over detected trials alone gives a conditional mean delay. The time required to identify the source should be evaluated separately, since an estimator may need additional observations after the first alarm.

Different deployment strategies optimize different observation objectives and are therefore evaluated using corresponding deployment-oriented criteria~\cite{spinelli2016observer,gajewski2022comparison}. For coverage-oriented deployment, a commonly adopted measure is the sensor coverage rate. Let \(\mathcal{N}(o)\) denote the neighborhood of sensor node \(o\). Then
\begin{equation}
    C_s(V_{\mathrm{sen}})
    =\frac{1}{N}
    \left|
    \bigcup_{o\in V_{\mathrm{sen}}}\mathcal{N}(o)
    \right|,
    \label{eq:sensor_coverage_rate}
\end{equation}
which quantifies the fraction of network nodes covered by the neighborhoods of the deployed sensors~\cite{zhang2016identification,gajewski2022comparison}.

Distance-based placement provides another benchmark for sensor deployment. The \(k\)-median strategy minimizes the total distance from network nodes to their nearest sensors~\cite{spinelli2016observer,paluch2020optimizing}:

\begin{equation}
    V_{\mathrm{sen}}^{k\text{-}\mathrm{Med}}
    =
    \operatorname*{argmin}_{|V_{\mathrm{sen}}|=K}
    \sum_{v\in V}
    \min_{o\in V_{\mathrm{sen}}}d(v,o),
    \label{eq:sensor_k_median}
\end{equation}
where \(d(v,o)\) denotes the shortest-path distance between node \(v\) and sensor \(o\), and \(K\) is the sensor budget. This objective evaluates the spatial distribution of sensors rather than source-detection accuracy, which must be assessed separately~\cite{spinelli2016observer,gajewski2022comparison}.

Computational efficiency should be assessed alongside network size, candidate space, observation length, and the costs of simulation and training, where applicable~\cite{ling2022source,paluch2018fast}. Reliable comparisons also require a distinction between independent diffusion realizations and repeated executions of an algorithm on the same realization. For learning-based methods, data partitioning is particularly important: training and test sets must be separated to avoid information leakage, and any overlap in network structures or diffusion settings should be made explicit~\cite{hou2025fgssi}. Evaluations on synthetic and real-world networks, where feasible, help establish whether performance extends beyond particular experimental conditions. Results should report variability across independent trials, with consistent source initialization, observation budgets, and parameter-tuning procedures across competing methods~\cite{jiang2016identifying,gajewski2022comparison}.

\subsection{A Unified View from Information to Identifiability}
\label{sec:information_identifiability}

Performance evaluation measures how accurately a method identifies diffusion sources under given conditions. A separate question is whether the observations themselves contain sufficient information to distinguish the true source from other candidates. This depends on the diffusion dynamics and the observation mechanism, independently of the estimator used. We examine this question through system-theoretic observability, information-theoretic constraints, and statistical identifiability, while distinguishing these properties from numerical stability and computational feasibility. These distinctions also help clarify how observation design affects source inference.

\subsubsection{A System-Theoretic Baseline}
\label{sec:linear_baseline}

System-theoretic observability offers a useful point of comparison. For a specified dynamical system, it concerns whether the initial state can be uniquely determined from measurements collected over a finite interval. This question is related to, but distinct from, identifying an initiating source within a restricted candidate set.

Consider the discrete-time linear system
\begin{equation}
x_{t+1}=Ax_t+Bu_t,\qquad y_t=Cx_t,
\end{equation}
where \(x_t\in\mathbb R^N\) is the state vector at time \(t\), \(u_t\in\mathbb R^p\) denotes the external input, and \(y_t\in\mathbb R^m\) is the measured output. The matrices \(A\in\mathbb R^{N\times N}\), \(B\in\mathbb R^{N\times p}\), and \(C\in\mathbb R^{m\times N}\) specify the state transitions, input effects, and measurements, respectively. When these matrices and the input sequence are known, the contribution of the inputs can be subtracted from the outputs. Stacking the remaining measurements over \(t=0,\ldots,L-1\) gives
\begin{equation}
\widetilde y_{0:L-1}=\mathsf O_Lx_0,\qquad
\mathsf O_L=
\begin{bmatrix}
C\\
CA\\
\vdots\\
CA^{L-1}
\end{bmatrix},
\label{eq:observability_matrix}
\end{equation}
where \(\widetilde y_{0:L-1}\) denotes the stacked measurements after removing the known input contribution, and \(\mathsf O_L\in\mathbb R^{mL\times N}\) is the observability matrix for a window of length \(L\).

For this noiseless linear system, the initial state \(x_0\) is uniquely determined by the measurements if and only if \(\operatorname{rank}\mathsf O_L=N\)~\cite{astrom2008feedback}. Source identification may impose a less restrictive requirement. Suppose a single source at node \(s\) is represented by \(x_0=e_s\), where \(e_s\) is the \(s\)-th standard basis vector. The resulting measurement vector is then the \(s\)-th column of \(\mathsf O_L\). Distinguishing candidate sources requires these columns to be pairwise distinct, even if \(\mathsf O_L\) does not have full column rank. However, distinct measurement vectors may be difficult to separate in the presence of noise, particularly when their differences are small. Similarly, a full-rank observability matrix can be poorly conditioned, making the estimation of arbitrary initial states sensitive to measurement errors.

Source identification within a restricted candidate set therefore does not require full-state observability. The rank conditions above apply to the specified linear system and do not directly establish source identifiability under stochastic SI, SIR, or SIS dynamics. In those settings, distinguishability must be assessed using the observation distributions induced by the candidate sources.

The distinction between mathematical invertibility and numerical stability also arises in stochastic epidemic models. Persoons and Van Mieghem~\cite{persoons2024patientzero} investigate backward evolution in SIS dynamics using the exact Markovian model and its N-intertwined approximation. For the exact model, the evolution of the full probability vector is mathematically invertible over a finite time interval when the generator is known, although backward computation can be numerically unstable. This result concerns the probability distribution over network states, rather than the identification of a source from a single observed epidemic realization. It therefore illustrates why invertibility of the underlying dynamics does not necessarily imply that the true source can be identified from the available observations.

\subsubsection{Information Constraints and Uncertainty}

Let \(S\) denote the unknown source, represented as a random node or source set with a specified prior over the candidate space \(\mathcal C\). Given an observation \(Q=q\), posterior uncertainty can be measured by
\begin{equation}
H_{\mathrm{post}}(q)
=-\sum_{s\in\mathcal C}P(S=s\mid q)\log P(S=s\mid q),
\qquad
N_{\mathrm{eff}}(q)=\exp H_{\mathrm{post}}(q).
\label{eq:posterior_entropy}
\end{equation}
Here and throughout this section, logarithms are natural, and conditioning on the known network and diffusion parameters is omitted for brevity. The effective candidate count \(N_{\mathrm{eff}}(q)\) summarizes how widely the posterior probability is distributed across candidate sources. It does not give the number of admissible candidates or indicate whether the true source has been identified. In particular, a concentrated prior can produce low posterior entropy even when the observation contributes little information about the source.

To distinguish posterior concentration from the information supplied by observations, consider the prior entropy and mutual information
\begin{equation}
H(S)=-\sum_sP(s)\log P(s),\qquad
I(S;Q)=H(S)-\mathbb E_Q[H_{\mathrm{post}}(Q)].
\label{eq:source_information}
\end{equation}
The posterior entropy \(H_{\mathrm{post}}(q)\) measures the uncertainty remaining after a particular observation, while \(H(S\mid Q)\) averages this uncertainty over all possible records. Their relation to the prior entropy gives \(I(S;Q)=H(S)-H(S\mid Q)\), the expected reduction in uncertainty about the source. This definition accommodates discrete, continuous, and mixed observations. For discrete \(Q\), mutual information can also be expressed as \(I(S;Q)=H(Q)-H(Q\mid S)\). When \(H(Q\mid S)\) is unchanged across observation designs, maximizing observation entropy also maximizes mutual information. Otherwise, the two criteria need not agree~\cite{cover2006elements}.

Information-theoretic bounds depend on the assumed probability model. If \(S\to X(t)\to X(t+h)\) forms a Markov chain, the data-processing inequality implies
\(I(S;X(t+h))\leq I(S;X(t))\).
This relation concerns successive system states under the stated Markov assumption. It does not describe the accumulation of observations, for which \(I(S;Q_1,Q_2)\geq I(S;Q_1)\). For a uniform prior over \(M\geq2\) candidate sources, Fano's inequality gives
\(P_e\geq1-[I(S;Q)+\log2]/\log M\),
where \(P_e\) is the probability of incorrectly identifying the source, averaged over the joint probability model. The bound applies to any estimator based on \(Q\), although it may be loose~\cite{cover2006elements}. Neither inequality determines the detection accuracy associated with a particular observation time.

Figure~\ref{fig:path_degeneracy_identifiability}(a) shows how source-detection accuracy varies with observation time in the experiments of Ke et al.~\cite{ke2026source}. As the proportion of infected nodes increases, the accuracy of the evaluated methods generally decreases. This behavior is specific to the diffusion models and inference methods considered. The observed decline in accuracy does not, by itself, establish a general loss of mutual information or statistical source identifiability.

The observation mechanism may discard information contained in the underlying diffusion trajectory. Even when source-related information remains in the recorded data, an estimator may fail to exploit it. Processing a fixed observation cannot increase its mutual information with the source under the same probability model. Additional training data may improve how an estimator uses the recorded observations, while external evidence or prior knowledge can provide information not contained in those records. Neither effect reverses information lost through the observation mechanism.

\subsubsection{Path Degeneracy and Identifiability Limits}

Under fixed diffusion dynamics, known parameters, and a specified observation mechanism, source identifiability requires every pair of distinct candidate sources to induce different distributions of the observed data. Equivalently, source identifiability can be expressed as
\begin{equation}
s\ne s'\ \Longrightarrow\
P(Q\in A\mid s)\ne P(Q\in A\mid s')
\quad\text{for some measurable }A.
\label{eq:statistical_identifiability}
\end{equation}
When nuisance parameters are unknown, the identifiability conditions depend on whether these parameters are treated as additional unknowns, fixed at specified values, or integrated out under a prior. If two admissible source--parameter combinations with different source labels induce the same observation distribution, the source is not identifiable over the full parameter space. Even when the distributions differ, they may assign positive likelihood to the same observation. Statistical identifiability therefore allows ambiguity in a record from a single outbreak.

Identifiability concerns the observation distributions specified by a model, rather than the outcome of a particular realization. If an observed record has positive likelihood under two candidate sources, neither can be ruled out from that record alone. When candidate sources induce different observation distributions, repeated independent realizations from the same unknown source can provide additional evidence for distinguishing them. Measurements taken at several times during one outbreak, however, do not constitute independent realizations. Node labels also matter: discarding them may make source hypotheses indistinguishable even when labeled observations can separate them. Claims of exact source identification or consistency of a source estimator must therefore specify the candidate space, observation mechanism, and assumptions about repeated observations.

For a particular observation \(Q\) with positive likelihood under both candidate source sets, consider the absolute log-likelihood difference
\begin{equation}
\Delta_Q(\mathcal S_1,\mathcal S_2)=
\left|
\log P(Q\mid\mathcal S_1,G,\Theta)
-
\log P(Q\mid\mathcal S_2,G,\Theta)
\right|.
\label{eq:likelihood_gap}
\end{equation}
This quantity compares the likelihoods of two candidates for a given observation, rather than their complete observation distributions. A small difference, including a value of zero, does not establish statistical non-identifiability. Similarly, equal likelihoods for two candidates do not necessarily imply substantial uncertainty across the entire candidate space. When the observations do not distinguish among plausible sources, an inference method may return a candidate set, ranking, or probability distribution. These outputs describe the uncertainty about the true initiating source or source set, which remains the target of inference.

Different transmission histories may lead to the same recorded state, but this does not necessarily make their initiating sources indistinguishable. Network symmetry may also contribute to source ambiguity, depending on the diffusion dynamics and the observation mechanism. For example, a graph automorphism may exchange two source nodes while their labeled observations remain distinct. Neither overlapping transmission paths nor network symmetry alone is sufficient to establish statistical non-identifiability. Such a conclusion requires comparing the observation distributions induced by the candidate sources under a specified measurement protocol.

\begin{figure}[htp]
\centering
\includegraphics[width=0.9\linewidth]{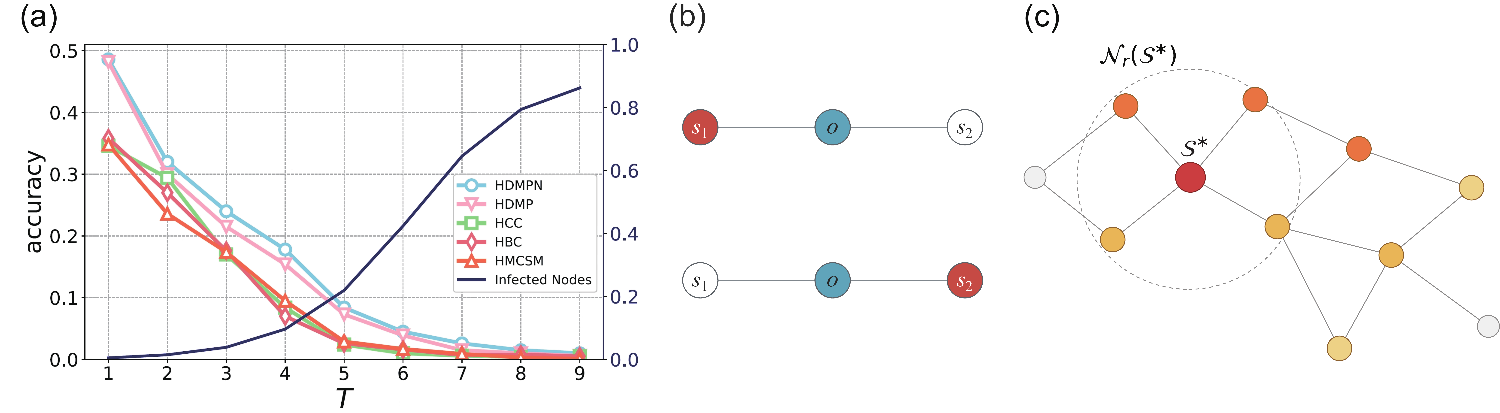}
\caption{ 
Source information and observational ambiguity.
(a) Source-detection accuracy (left axis) and the proportion of infected nodes (right axis) as functions of observation time \(T\), based on the experiments of Ke et al.~\cite{ke2026source}.
(b) Two single-source hypotheses on the labeled path \(s_1-o-s_2\), with \(s_1\) and \(s_2\) as the sources in the upper and lower rows, respectively (red nodes). The central node \(o\) (blue) is the only sensor. With a known onset time \(t_0=0\) and a deterministic transmission delay of one time unit per edge, both hypotheses yield the same first-arrival time \(\tau_o=1\) at the sensor and are indistinguishable from this measurement alone.
(c) Schematic probabilities assigned to candidate sources around the true source \(\mathcal{S}^{*}\) (red). The dashed circle marks the distance-tolerance neighborhood \(\mathcal{N}_r(\mathcal{S}^{*})\).
}
\label{fig:path_degeneracy_identifiability}
\end{figure}

Figure~\ref{fig:path_degeneracy_identifiability}(b) illustrates two single-source hypotheses on a labeled three-node path \(s_1-o-s_2\), with \(s^*=s_1\) in the upper row and \(s^*=s_2\) in the lower row. In both cases, infection begins at \(t_0=0\) from a single node, and transmission along each edge takes exactly one time unit. Only the first-arrival time \(\tau_o\) at the central sensor \(o\) is recorded. Since \(\tau_o=1\) under either hypothesis, the two sources induce identical observation distributions and cannot be distinguished from this measurement alone. Their labeled initial states, however, are different and would allow the true source to be identified if observed.

When exact identification is difficult, it is also useful to examine how much posterior probability is assigned to sources near the true origin. For a specified node distance or source-set matching distance \(d\), define
\begin{equation}
\mathcal N_r(\mathcal S^*)=\{\mathcal S\in\mathcal C:d(\mathcal S,\mathcal S^*)\leq r\}.
\label{eq:source_neighborhood}
\end{equation}

For a distance tolerance \(r\geq 0\) and \(0<\alpha<1\), posterior concentration near the true source can be assessed through the condition

\begin{equation}
P(\mathcal S\in\mathcal N_r(\mathcal S^*)\mid Q,G,\Theta)\geq1-\alpha.
\label{eq:posterior_concentration}
\end{equation}
When this condition holds, at least \(1-\alpha\) of the posterior probability is assigned to candidates within distance \(r\) of the true source. Since this neighborhood depends on the unknown source, the condition can be examined in simulations or retrospective analyses but cannot serve as an operational identifiability criterion. It also differs from posterior credible sets and frequentist confidence sets, which have distinct interpretations and coverage requirements
~\cite{khim2017confidence}.

\subsubsection{Observation Paradigms and Information Availability}

The theoretical distinctions developed above have direct implications for source inference under different observation conditions. A complete state snapshot records the states of all nodes but generally leaves the preceding transmission history unresolved. Additional snapshots can constrain possible propagation histories, although their value for identifying the source depends on the diffusion dynamics and observation times. Partial measurements may also distinguish candidate sources when they capture informative changes in the spreading process. The amount of recorded data alone is therefore insufficient to determine how well different origins can be distinguished.

The nature of missing information is equally important. Unobserved node states leave gaps in the recorded diffusion pattern, whereas uncertain network connections introduce ambiguity about the transmission paths available to the process. These uncertainties affect source inference differently and need to be reflected in the assumed diffusion and observation models. In particular, additional measurements are useful when they help separate candidate sources that remain difficult to distinguish from the existing records.

Sensor placement offers a way to select observations with this objective in mind. For a fixed sensor budget \(K\), an information-theoretic formulation seeks the sensor set that maximizes the mutual information between the source and the resulting measurements~\cite{krause2008sensor}:
\begin{equation}
V_{\mathrm{sen}}^*
\in
\operatorname*{argmax}_{V_{\mathrm{sen}}\subseteq V,\ |V_{\mathrm{sen}}|=K}
I(S;Q_{V_{\mathrm{sen}}}),
\label{eq:sensor_information_gain}
\end{equation}
where \(Q_{V_{\mathrm{sen}}}\) denotes the observations obtained from the selected sensors. This objective depends on the source prior and the assumed diffusion and measurement models. It measures the expected information provided by a sensor configuration, without guaranteeing that every observed realization will identify the source. Criteria based on network coverage, early warning, or distance separation address related but different objectives and need not favor the same sensor locations. Fixed placement determines the observation locations in advance, while adaptive strategies can use earlier measurements to guide subsequent acquisition.

These considerations connect the limits of source inference to the methods examined in Sections~\ref{sec:full-observability}--\ref{sec:sensor-based}. The observation settings overlap, but the information they retain and the assumptions required to interpret it can differ substantially. Comparing source-detection methods therefore requires attention to the evidence available to each method, rather than the observation setting alone.

\section{Source Detection from Complete State Snapshots}
\label{sec:full-observability}
 
A complete state snapshot provides the labeled states of all nodes at a given observation time, but does not reveal the transmission history that produced them. Consequently, several candidate sources may be compatible with a complete snapshot. This section reviews structural, probabilistic, simulation-based, and learning-based methods that infer diffusion origins from such observations. We focus primarily on methods using a single snapshot, while noting where multiple snapshots provide additional temporal evidence and require corresponding changes to the inference model.

\subsection{Detection Principles for Complete State Snapshots}

A complete labeled snapshot on a known network is represented by \(Q_T^{\mathrm{full}}=X(T)=(X_1(T),\ldots,X_N(T))\), where \(X_i(T)\) denotes the recorded state of node \(v_i\) at time \(T\). Complete spatial coverage does not imply knowledge of the transmission history, diffusion parameters, or the time elapsed since initiation. These quantities must be specified separately or accounted for during source inference.

For binary-state diffusion models, such as SI and binary formulations of rumor spreading, let \(V_I(T)=\{v_i\in V\mid X_i(T)=I\}\) denote the infected or activated nodes, and let \(G_I(T)=G[V_I(T)]\) be the corresponding induced subgraph. In the single-source setting, a candidate is represented by \(\mathcal{S}=\{v\}\). The infected subgraph captures the spatial arrangement and connectivity of observed infections, which are central to many structure-based detection methods. For multistate models such as SIR, a complete snapshot also distinguishes susceptible and recovered nodes, providing information that is lost when only the infected subgraph is retained.

Even with every node state recorded, the propagation events that produced the snapshot generally remain unknown. Different sources may lead to similar or identical observations through alternative transmission histories, while uncertainty in model parameters can further complicate their identification. Estimation errors may also arise from the limitations of the inference procedure itself. Complete snapshots therefore remove uncertainty about node states at the observation time, but do not necessarily resolve ambiguity about the initiating source.

The inference target remains \(s^*\) for a single source and \(\mathcal{S}^*\) for multiple sources. When an observation likelihood is available, the statistical estimators in Eqs.~\eqref{eq:map_source}--\eqref{eq:mle_source} can be applied. Other methods use structural criteria or learned prediction rules without explicitly evaluating a posterior distribution. Their treatment of snapshot information can be examined from three complementary perspectives.

\textbf{Geometry.} The spatial organization of infected nodes provides clues to possible source locations. Distances, infection boundaries, and the distribution of affected branches can be used to assess the compatibility of a candidate with the observed pattern. Jordan-type estimators minimize infection eccentricity, while Rumor Centrality counts compatible infection orders under specified spreading assumptions~\cite{shah2011rumors,zhu2014information,luo2014identify}. Susceptible nodes may provide additional evidence by constraining the regions that diffusion has reached. However, the relationship between these structural features and the true source depends on the network topology and diffusion model; a geometrically central node is not necessarily the origin of the observed outbreak~\cite{ali2025generalized}.

\textbf{Forward consistency.} A candidate source can also be evaluated by examining the spreading patterns it could generate under an assumed diffusion model. Dynamic message passing (DMP) estimates the evolution of marginal state probabilities, while belief propagation (BP) incorporates observations into probabilistic inference over candidate sources and latent variables. Simulation-based methods compare the observed snapshot with outbreaks generated from candidate origins~\cite{lokhov2014inferring,altarelli2014bayesian,antulov2015identification}. These approaches differ in how they account for unobserved transmission histories and approximate the relationship between a proposed source and the recorded states. Consequently, they may assign different scores to the same candidate even when supplied with identical observations.

\textbf{Learned prediction.} Data-driven methods infer source locations from patterns learned during training, often using simulated or observed source--snapshot pairs. Rather than evaluating every candidate through an explicit diffusion likelihood, these methods learn representations or prediction rules that associate observed states with possible origins~\cite{ling2022source,wang2022invertible}. Their predictions depend on the network structures, diffusion mechanisms, and source distributions represented in the training data. Performance on previously unseen outbreaks should therefore be distinguished from generalization to different networks, spreading dynamics, or observation mechanisms.

\subsection{Methodological Foundations}

\subsubsection{Structure-based Detection}

Structure-based methods rank diffusion sources using spatial or combinatorial features of the observed spreading pattern, including centrality, the infection-induced subgraph, propagation trees, boundaries, and local infection patterns. Their scores can be derived or justified under specified diffusion models even when candidate evaluation uses geometry or infection-order counts rather than explicitly summing over hidden spreading histories.

Classical network centrality measures, including degree, betweenness, closeness, and eigenvector centrality, have inspired a wide range of structure-based source detection methods~\cite{albert2000error,freeman1977set,freeman1978centrality,shah2010detecting,bonacich1987power,newman2018mathematics}. Representative examples include Rumor Centrality, the Jordan center, Propagation Centrality, and methods based on label propagation or structural entropy~\cite{shah2010detecting,zhu2014information,luo2014identify,yang2016locating,he2026hybrid}. These scores offer interpretable candidate comparisons whose computational burden depends on the network operations required. For example, distance calculations and tree recursions can be efficient, whereas exact infection-order counting on general networks can be prohibitive. The following methods differ both in which aspects of the snapshot they retain and in the assumptions supporting their source interpretation.

Jordan-type estimators provide a distinct geometric route. For observed infected nodes \(V_I\), the infection eccentricity of a candidate is the maximum shortest-path distance from that candidate to \(V_I\). Minimizing it selects a center covering the observed infections with the smallest radius. This is a minimax criterion, unlike a distance-sum center, which is dominated by the aggregate of all distances, or Rumor Centrality, which counts compatible orderings. If a deterministic equal-delay process has run for a specified time, the radius has a direct arrival-time interpretation. With stochastic delays and unknown duration, a center is instead evaluated through the model-specific arguments supporting that estimator~\cite{zhu2014information,luo2014identify}. The candidate domain matters: restricting the search to observed infected nodes is appropriate in some SI settings but can be wrong with hidden or recovered sources. On a network with cycles, shortest distances can underestimate the length of the realized infection route, although they still summarize geometric reachability. Multiple centers or tied eccentricities should be retained or resolved by a specified rule. These differences explain why a Jordan estimator and a likelihood estimator can agree under particular assumptions yet produce different rankings on the same general-network snapshot.

Independent-cascade inference illustrates why the state definition also matters. Berenbrink et al.~\cite{berenbrink2023inference} infer a single source from the nodes active at the observation step, with the start time unknown. On trees, feasible candidates are equidistant from all currently active nodes, and their estimator selects the closest such candidate. The analysis identifies regimes of weak and strong source detection on infinite regular and Poisson Galton--Watson trees, with general-network heuristics evaluated empirically. These results use a current activation frontier rather than the cumulative infected set of SI centralities.

\begin{figure}[htp]
    \centering
    \includegraphics[width=0.9\linewidth]{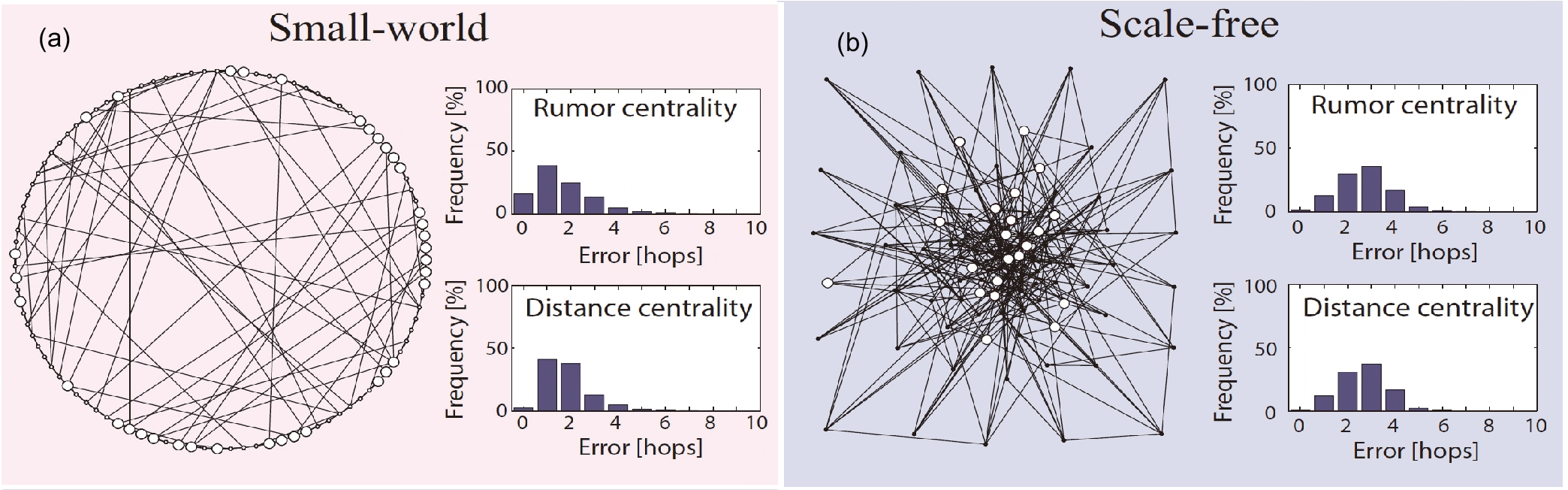}
    \caption{\rev{Comparison of RC and distance centrality on synthetic networks~\cite{shah2010detecting}. (a) Small-world network. (b) Scale-free network. Each panel pairs a network visualization with separate histograms of hop-distance errors between the estimated and true sources for the two estimators.}}

    \label{RC}
\end{figure}

Rumor Centrality (RC) is one of the most representative structure-based source detection methods~\cite{shah2010detecting,shah2016randomtrees}. It was originally developed for diffusion processes on tree networks, where the observed infected subgraph uniquely determines the underlying propagation tree. 
RC ranks candidate sources by counting the infection sequences compatible with the observed infected tree. The count reflects how many admissible spreading orders can originate from each candidate, while its interpretation as a likelihood depends on the assumptions of the diffusion model.
Specifically, if node \(v\) is the true source, the infected subgraph \(G_I(T)\) may be generated through multiple infection orders that satisfy the diffusion constraints.

\par\noindent\begin{minipage}{\linewidth}
\rev{Let \(n_I=|V_I(T)|\) be the size of the observed infected tree, and let \(n_u^{(v)}\) count the nodes in the subtree rooted at \(u\) when \(G_I(T)\) is rooted at \(v\), including \(u\) itself. In particular, \(n_v^{(v)}=n_I\). The RC count and corresponding source estimate are~\cite{shah2011rumors}}

\begin{equation}
    R(v;G_I(T))
    = \frac{n_I!}{\displaystyle\prod_{u\in V_I(T)} n_u^{(v)}},
    \qquad
    \hat v
    = \operatorname*{argmax}_{v\in V_I(T)} R(v;G_I(T)).
    \label{eq:rumor_centrality}
\end{equation}
 
\end{minipage}\par

The subtree product in RC has a combinatorial interpretation. After choosing a root, infection must enter each descendant subtree before visiting its deeper vertices. Within that constraint, orders from different branches can be interleaved. The factorial factor counts orderings before the subtree restrictions are imposed, and the subtree-size factors remove incompatible permutations. Re-rooting changes those sizes even though the observed infected set is unchanged. Thus the statistic contains more than the candidate's degree or mean distance to infected nodes. Its likelihood interpretation additionally requires compatible orders to receive the appropriate model weights. Heterogeneous edge rates can give two feasible orders different probabilities; finite network boundaries can also change how likely a given infection sequence is~\cite{yu2017boundary}. Counting alone then loses part of the likelihood. This separates the valid combinatorial statistic from the stronger statistical equivalence asserted only under its established assumptions.

RC equals an ML estimator in the canonical single-source SI model on infinite regular trees with independent identically distributed exponential edge transmission times and an observed infected subtree~\cite{shah2011rumors}. The result is not a theorem for every tree, heterogeneous rate model, recovery process, or sampling rule. On a general network, counting compatible orders and assigning them equal weight no longer reproduces the full likelihood. Figure~\ref{RC} contrasts feasible-order counting with distance-based ranking: the distinction is the information used, not a universal performance ordering.

Extending RC beyond tree networks introduces both modeling and computational difficulties. When the infected subgraph contains cycles, the transmission tree is no longer determined by the observed topology, and evaluating compatible spreading histories becomes considerably more involved. To overcome this limitation, several extensions and alternative structural methods have been proposed. One line of research generalizes RC from single-source to multiple-source settings~\cite{luo2012identifying,luo2012identifying2,luo2013identifying}. Another replaces explicit sequence counting with more tractable structural criteria, including the improved betweenness-based method of Comin et al.~\cite{comin2011identifying}, the Jordan center and reverse infection algorithm proposed by Zhu et al.~\cite{zhu2014information}, and spectral approaches that exploit network spectral properties together with node-age information~\cite{fioriti2014predicting}. A related direction approximates the propagation structure associated with each candidate source instead of enumerating all admissible infection sequences, leading to propagation centrality, which is introduced next.

\begin{figure}[htp]
    \centering
    \includegraphics[width=0.6\linewidth]{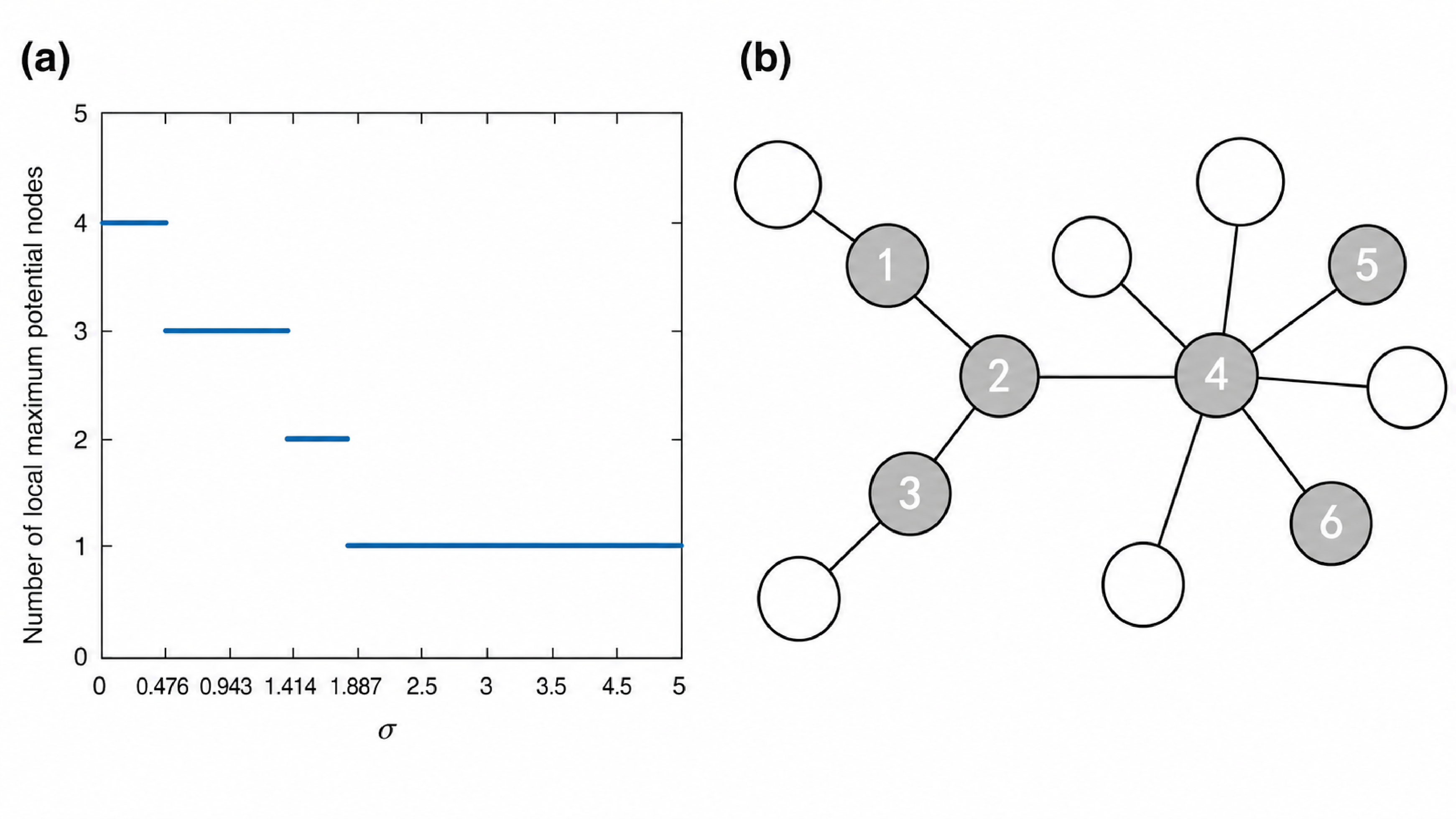}
    \caption{
Source-number estimation and candidate discrimination based on topological potential~\cite{wang2021localization}. (a) Variation in the number of local potential maxima with the impact factor \(\sigma\), illustrating how the estimated number of diffusion sources changes across different spatial scales. (b) Illustration of neighborhood-based candidate discrimination. Node 2 has a higher proportion of infected neighbors than node 4, providing additional information for source detection.}
    \label{PC}
\end{figure}

Propagation Centrality (PC) evaluates candidate sources through the local propagation structure associated with each node~\cite{yang2016locating}. Designed for single-source detection from a complete state snapshot, PC constructs a BFS tree \(\mathcal{T}^{\mathrm{BFS}}_v\) rooted at each infected candidate node \(v\). Let \(\ell_v(u)\) denote the depth of node \(u\) in \(\mathcal{T}^{\mathrm{BFS}}_v\). The method assesses whether the observed infected and uninfected nodes are distributed in a manner consistent with the layered propagation pattern implied by the BFS tree.

To quantify this propagation structure, PC defines three distance-based descriptors:
\begin{equation}
\begin{aligned}
\mathrm{FID}_v
=
\max_{u\in V_I(T)} \ell_v(u),
\mathrm{NUD}_v
=
\min_{u\in V\setminus V_I(T)} \ell_v(u),
\mathrm{FD}_v
=
\max_{u\in V(\mathcal{T}^{\mathrm{BFS}}_v)} \ell_v(u),
\end{aligned}
\label{eq:pc_fid_nud_fd}
\end{equation}
where \(\mathrm{FID}_v\) is the farthest infected distance from \(v\), representing the minimum BFS depth required to reach all observed infected nodes. \(\mathrm{NUD}_v\) is the nearest uninfected distance from \(v\), indicating the first BFS layer at which an uninfected node appears. Together, these two quantities characterize the separation between infected and uninfected regions around the candidate node. The third descriptor, \(\mathrm{FD}_v\), is the maximum depth of the BFS tree and serves as a reference scale for interpreting node locations across different propagation layers.  
These descriptors assume that the relevant node sets are nonempty and reachable from the candidate source. Fully infected snapshots and disconnected networks require an explicit convention for undefined or infinite distances.

PC evaluates each candidate-rooted BFS tree using the Rationality Observation Value (ROV), which combines observed node states with the layered structure of the constructed tree. The score reflects how well the infected region and its boundary agree with the propagation pattern implied by the candidate. Since the BFS tree does not necessarily represent the realized transmission history, ROV should be interpreted as a structural evaluation criterion rather than a source probability~\cite{yang2016locating}.
The frontier descriptors above summarize different parts of this geometry. Its role is to compare candidate-rooted explanations, with the BFS approximation subject to the path assumptions in Section~\ref{sec:problem_definition}.

The estimated source is the infected node whose BFS tree attains the largest ROV:
\begin{equation}
\hat v_{\mathrm{PC}}
=
\operatorname*{argmax}_{v\in V_I(T)}
\mathrm{ROV}\!\left(\mathcal{T}^{\mathrm{BFS}}_v\right).
\label{eq:pc_estimator}
\end{equation}

PC incorporates susceptible as well as infected nodes, whereas infected-subgraph centralities may discard the former. This additional constraint can alter rankings, especially near the infection boundary. The trade-off is the need to construct and score candidate-rooted propagation trees; its cost and accuracy depend on network size, candidate count, and the suitability of the BFS approximation.

Structure-based source detection has gradually evolved from global centrality measures to richer descriptions of local propagation patterns. Local neighborhood information is also useful for multiple-source detection. Wang et al.~\cite{wang2021localization} combined topological-potential-based overlapping community detection with likelihood estimation and contagion neighborhood bias to identify multiple sources. For a candidate node \(v\), the neighbor infection ratio is defined as
\begin{equation}
\rho_I(v)
=
\frac{|\mathcal{N}(v)\cap V_I(T)|}{|\mathcal{N}(v)|},
\label{eq:neighbor_infection_ratio}
\end{equation}
where \(\mathcal{N}(v)\) denotes the neighbor set of \(v\) in the original network \(G\). This metric quantifies the proportion of infected (or activated) neighbors around a candidate node and reflects the local propagation pattern surrounding it. As illustrated in Fig.~\ref{PC}, the method first uses the number of local maximum potential nodes to estimate the number of sources and then distinguishes candidate sources within each partition by combining source likelihood with contagion neighborhood bias. Although nodes 2 and 4 exhibit similar source likelihoods, node 2 has a more strongly infected neighborhood and is therefore selected as the detected source. By incorporating local neighborhood information in addition to propagation boundaries, this approach provides a finer structural characterization of candidate sources than methods relying solely on geometric distances.

Entropy measures have also been used to characterize the structural organization of observed infection patterns and identify plausible diffusion sources. Zhang et al.~\cite{zhang2021ite} proposed a structural entropy method in information-theoretic terms. By comparing the structural complexity of infection trees associated with different candidate sources, the method identifies the node with the minimum structural entropy as the detected source. Liu et al.~\cite{liu2022multi} subsequently introduced infection intensity entropy and infection adjacency entropy, and developed the multi-source detection method Source Localization Based on Neighborhood Entropy (SLBNE), which reduces interference among infected regions and improves detection performance in multi-source spreading scenarios.  
These approaches extend structure-based source detection by quantifying the organization of infection patterns at different structural scales. Structural entropy characterizes the arrangement of candidate infection trees, while neighborhood entropy captures variations in local infection configurations that may help distinguish overlapping spreading regions.

When the spreading model is unknown or its parameters are unavailable, Wang et al.~\cite{wang2017multiple} proposed the Label Propagation Source Identification (LPSI) method. Instead of relying on an explicit diffusion model, LPSI iteratively propagates infected and uninfected labels over the network and identifies source nodes from the converged label distribution. Hu et al.~\cite{HU2026131748} subsequently extended this framework to information--epidemic coupled multilayer networks by proposing the Motif-based Source Localization Method (MSLM). MSLM reconstructs weighted connections from motif structures in the information layer, initializes node labels using epidemic observations, and performs cross-layer label propagation to infer the diffusion source. This extension enables label propagation to handle coupled spreading processes in which information diffusion and epidemic transmission evolve simultaneously~\cite{mei2026coupled}.

\subsubsection{Probabilistic and Simulation-based Methods}

Probabilistic approaches evaluate candidate sources under an assumed diffusion model, accounting for unobserved transmission histories through exact calculations or suitable approximations. Simulation-based methods instead compare recorded spreading patterns with outcomes generated from candidate origins. Although both approaches use the forward diffusion process, they differ in how candidate sources are evaluated. In particular, marginal state probabilities, observation likelihoods, and simulation-based similarity scores have distinct interpretations.

\begin{figure}[htp]
    \centering
    \includegraphics[width=0.95\linewidth]{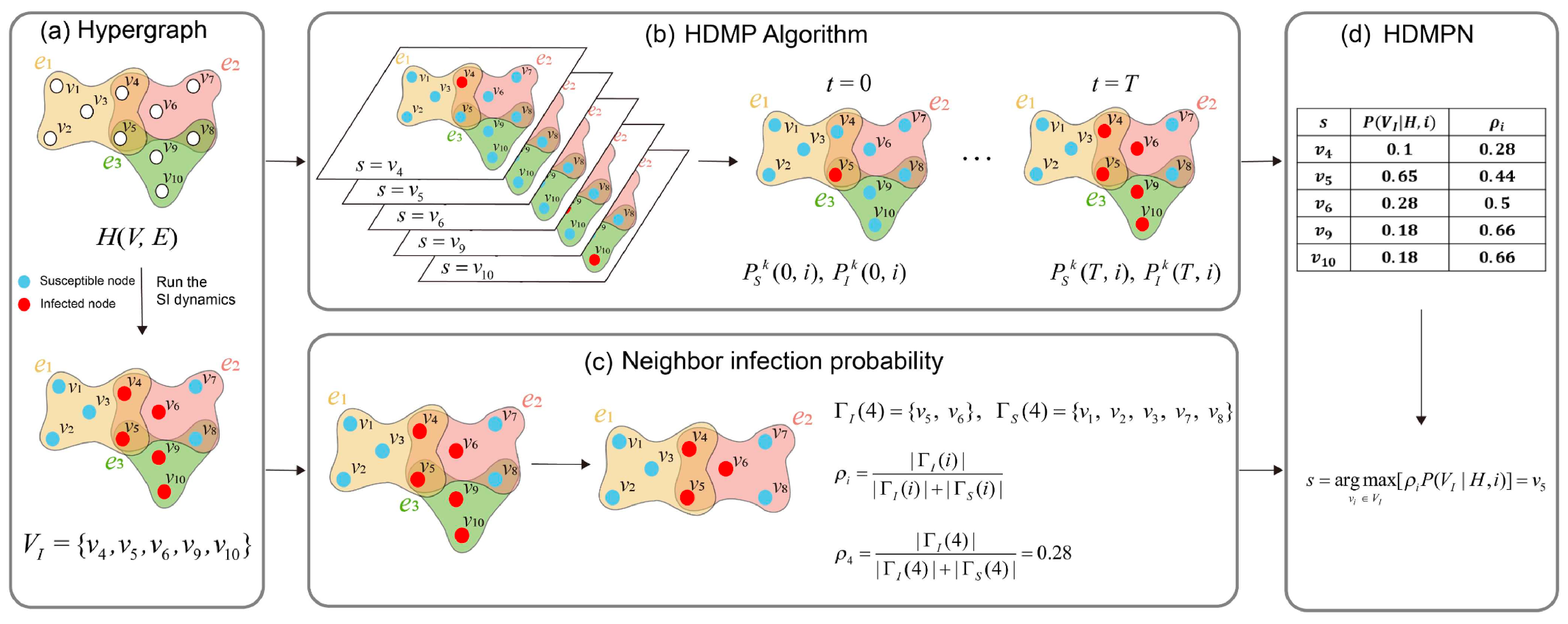}
    \caption{\rev{
Illustration of the HDMPN framework~\cite{ke2026source}. 
(a) Hypergraph diffusion model. 
(b) Dynamic message-passing procedure for estimating node infection probabilities. 
(c) Neighbor infection probability used to weight candidate-source scores.
(d) Source ranking based on the HDMP scores and neighbor infection probabilities.
}}
   
    \label{HDMPN}
\end{figure}

Among probabilistic and simulation-based methods, dynamic message passing (DMP) is one of the most representative approaches~\cite{lokhov2014inferring}. DMP approximates the diffusion process through recursive message updates along network edges. Given a source hypothesis and a specified spreading model, such as the SIR model, it iteratively estimates the marginal probability that each node is in the S, I, or R state at every time step:
\begin{equation}
p_i^{\sigma}(t;\mathcal{S})
=
P\!\left(X_i(t)=\sigma \mid \mathcal{S},G,\Theta\right),
\label{eq:dmp_marginal}
\end{equation}
where \(\sigma\in\{S,I,R\}\) denotes the state of node \(v_i\) at time \(t\). The quantity \(p_i^\sigma(t;\mathcal S)\) is the marginal probability defined by the diffusion model, while DMP evaluates it through recursive cavity messages. The accuracy of these estimates depends on the network structure and the assumptions underlying the message-passing equations. For each candidate source, the predicted state probabilities are then used to evaluate their agreement with the observed snapshot.

In DMP, \(\theta_{j\to i}(t)\) is a cavity message for the probability that infection has not passed from \(j\) to \(i\) by time \(t\). For the corresponding SIR assumptions,
\begin{equation}
p_i^S(t;\mathcal S)=p_i^S(0;\mathcal S)
\prod_{j\in\mathcal N(i)}\theta_{j\to i}(t).
\label{eq:dmp_susceptible}
\end{equation}
The messages are propagated for each candidate initialization. A commonly used source score is the product of observed-state marginals,
\begin{equation}
\mathcal L_{\mathrm{DMP}}(\mathcal S)=\prod_{i\in V}p_i^{X_i(T)}(T;\mathcal S).
\label{eq:dmp_likelihood}
\end{equation}
This factorized score approximates the joint snapshot likelihood. Even when tree-based DMP gives exact individual state marginals, the node states are not generally independent, so their product is not automatically an exact joint probability. Comparing this score across source initializations is a separate inference step from calculating the forward marginals~\cite{lokhov2014inferring}.

Most message-passing methods have been developed for pairwise diffusion on simple networks. In many real-world complex systems, however, spreading is driven by group interactions or higher-order contacts that cannot be adequately represented by pairwise edges~\cite{battiston2021physics,battiston2020beyond,iacopini2019simplicial}. To account for these higher-order transmission mechanisms, Ke et al.~\cite{ke2026source} extended dynamic message passing to hypergraph spreading models and proposed Hypergraph Dynamic Message Passing (HDMP), together with its neighbor-infection-probability-modulated variant, HDMPN (Hypergraph Dynamic Message Passing with Neighbor Infection Probability Modulation).

HDMP estimates node infection probabilities for each candidate source under the hypergraph diffusion model. HDMPN then weights the resulting likelihood-based source score by the candidate's neighbor infection probability, as illustrated in Fig.~\ref{HDMPN}.

%

For pairwise SI and SIR models on trees, cavity equations can yield exact node-state marginals under the specified transmission and initial-state assumptions. On networks with cycles, neglected dependencies may affect these estimates, while reinfection requires different message-passing equations. For HDMPN, accuracy also depends on how higher-order transmission is modeled and how the predicted probabilities are used to score candidate sources.

For source inference, the observation time can be treated as a fixed experimental input or as a quantity searched alongside the candidate. 
Forward message passing provides state marginals at successive times, allowing candidate observation times to be evaluated without assuming knowledge of the actual infection order. Negative states contribute as well as positive ones: a candidate that would have infected many recorded susceptible nodes can receive a low score. Under partial state observations, a factorized score includes only measured states, with any test-error model applied to their predicted probabilities. This adaptation changes the observation factor, not the underlying meaning of a forward message. It also illustrates the overlap between the complete-snapshot and incomplete-record chapters.
 
 
Belief propagation (BP) provides an observation-conditioned graphical-model approach~\cite{altarelli2014bayesian,Altarelli_2014}. Infection and recovery variables encode latent histories, while local transmission factors impose causal compatibility and observation factors encode recorded states. Messages summarize neighboring contributions to candidate infection times or initial states. After conditioning on the snapshot, the resulting marginals rank possible initial infections. 
Unlike DMP, which evaluates forward state probabilities under each candidate initialization, BP incorporates the observed states directly into a factor-graph representation of the diffusion process. Message updates then account for the constraints imposed by these observations when estimating latent transmission histories and possible source locations.
Tree-factorized settings admit exact sum-product calculations; loopy implementations are approximate and their convergence is not a general guarantee. A point estimate can be extracted from marginal beliefs, but those beliefs also describe alternatives supported by the fitted model. Computational demands depend on the time discretization, latent-state domains, network structure, and convergence behavior.

\begin{figure}[htp]
    \centering
    \includegraphics[width=0.8\linewidth]{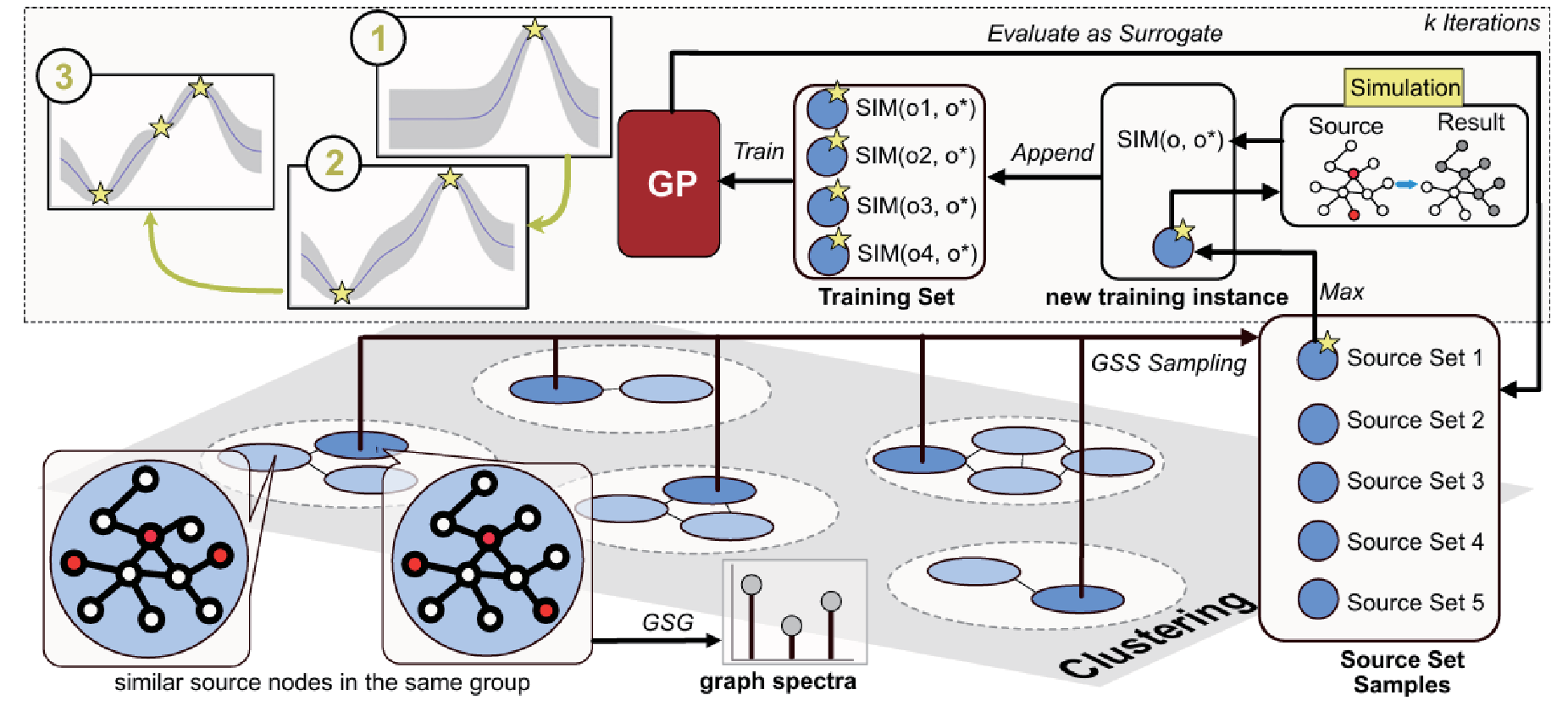}
    \caption{\rev{
Illustration of the BOSouL framework~\cite{zhang2024multiple}. Candidate source sets are represented in graph Fourier coordinates and clustered for stratified sampling. Simulation-based similarity evaluations update a Gaussian-process surrogate, whose expected-improvement criterion guides subsequent candidate evaluations.
}}
  
    \label{BoSoul}
\end{figure}

Monte Carlo methods initialize repeated forward simulations at each candidate and compare their outcomes with the observed snapshot. The MCSM framework of Antulov-Fantulin et al.~\cite{antulov2015identification} uses simulation-based comparison to estimate source plausibility. Exact matching of discrete snapshots would estimate a likelihood by relative frequency if sufficiently many simulations were available, but rare exact matches make that calculation impractical in large state spaces. Similarity or kernel-weighted matching relaxes equality and introduces a discrepancy measure and a tolerance or bandwidth. Normalized similarity scores should not automatically be called exact posterior probabilities. The inference depends on the propagation model, observation time, candidate prior, and simulation budget. Unlike an analytic likelihood calculation, this route can accommodate a simulator whose transition rules are available but whose trajectory sum is intractable. Its main computational burden is repeated outbreak generation; increasing the simulation count improves numerical approximation, not the information content of the observed outbreak.

The accuracy of simulation-based source inference depends on the assumed diffusion model and the number of simulated realizations. With limited sampling, candidate scores may fluctuate substantially, while errors in the transmission or recovery assumptions persist regardless of the simulation budget. The choice of discrepancy measure introduces a further distinction: comparing infected sets, complete node states, or event times may produce different source rankings because each retains different aspects of the observations. These factors need to be considered when interpreting simulation-based scores, particularly when some candidates produce no matches to the observed outbreak.

Repeated forward simulation becomes costly as the candidate-set space grows. Bayesian Optimization for Source Localization (BOSouL) addresses this search burden (Fig.~\ref{BoSoul})~\cite{zhang2024multiple}. Its formal target is a source-set posterior, reduced to likelihood maximization under a uniform source prior. The evaluated objective is instead a simulation similarity: for each candidate set, the method averages the best snapshot match attained within each simulated trajectory. Thus time is optimized within each realization before averaging; no observed diffusion duration is required. The Gaussian-process (GP) posterior describes uncertainty in this expensive response function, rather than posterior probabilities of source sets. With a supplied source count, BOSouL restricts candidates to fixed-size subsets of a pool selected by degree. It represents those subsets in graph Fourier coordinates and fits a GP with the proposed graph spectral Gaussian kernel. Graph stratified sampling draws from clusters of these representations, and expected improvement selects the next sampled candidate for simulation. Each new evaluation updates the GP; the final estimate maximizes its prediction over the restricted candidate sets. The computational trade-off is fewer expensive simulation evaluations in exchange for candidate construction, surrogate fitting, and prediction. The accuracy of source estimates depends on candidate-pool coverage, simulator accuracy, and the evaluation budget.

Random-walk approximations offer a different reduction. Jain et al.~\cite{jain2016fast} use hitting-time information to rank candidate origins. The input is the observed spreading pattern on a known network; the approximation summarizes propagation through walk-based accessibility instead of enumerating outbreak histories. Its usefulness depends on how well that accessibility represents the actual spreading process. A random-walk surrogate is not identical to an SI cascade simply because both operate on the same edges.

NETSLEUTH uses MDL to select a set of seeds explaining an infected snapshot~\cite{prakash2014efficiently}. It combines a graph-based candidate search with a coding objective that trades the cost of specifying sources against the cost of describing the induced infection pattern. This gives source-count selection an explicit role: adding a seed is justified only when the improved explanation offsets its coding cost. Its output is a source set, rather than a posterior distribution over all sets. It therefore remains a distinct route from both simulation matching and Bayesian optimization. The chosen encoding and propagation assumptions determine which explanation MDL selects. Source identifiability and the accuracy of the selected source set require separate assessment.

Prior-informed methods instead retain an explicit statistical decision rule. Chang et al.~\cite{chang2018maximum} incorporate prior information into candidate evaluation and approximate search in weighted networks. A prior can exclude impossible nodes or assign different initial plausibilities. It should be based on information available independently of the held-out source label. Comparisons with uniform-prior baselines must report that additional input, since an improvement can reflect useful prior knowledge as well as a better search procedure.

Source inference also depends on how unobserved transmission histories are handled. The likelihood of a candidate source may involve contributions from many possible propagation histories, whereas trajectory-based approaches may consider only the most probable one. A source associated with numerous moderately probable histories can therefore receive a higher likelihood than one supported by a single dominant history. This distinction matters when comparing message-passing methods, trajectory-based optimization, and shortest-path approximations. A related issue arises in multiple-source detection, where individual source probabilities do not necessarily determine the most probable source set. Selecting nodes independently may produce a set that violates a fixed source-count constraint. Choosing the \(k\) nodes with the highest marginal probabilities satisfies this constraint, but does not generally maximize the joint posterior probability of the source set. The selection procedure should therefore distinguish between ranking individual candidate sources and estimating their joint configuration.

\begin{figure}[htp]
    \centering
    \includegraphics[width=0.45\linewidth]{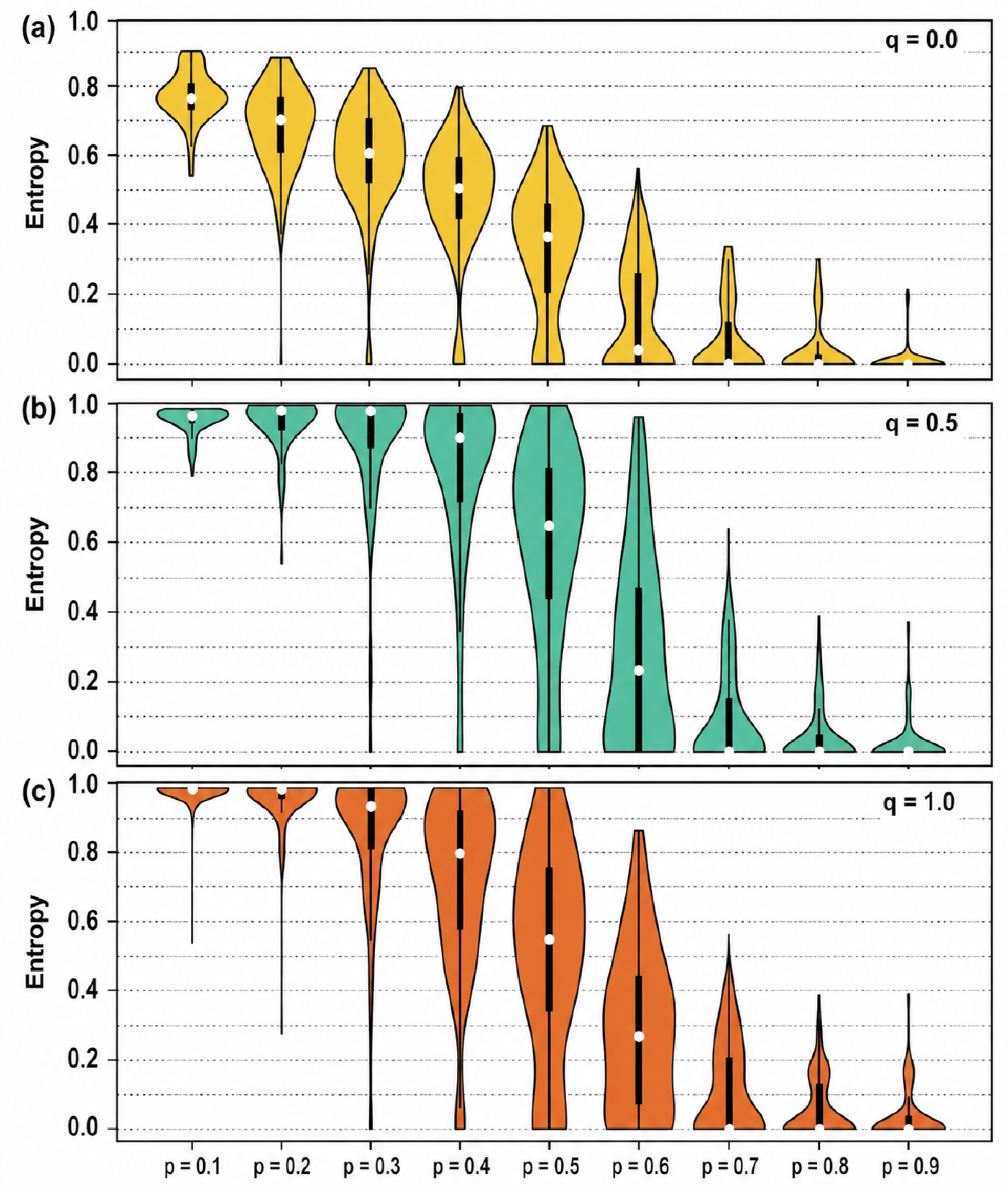}
    \caption{\rev{Posterior uncertainty in SIR source inference~\cite{antulov2015identification}. Panels (a)--(c) show distributions of normalized Shannon entropy \(H\) for infection probabilities \(p\), with recovery probabilities \(q=0\), \(0.5\), and \(1.0\), respectively. Lower entropy indicates greater concentration of the inferred source distribution; it does not establish that the concentration is around the true source.}}
   
    \label{fig:entropy_detectability_antulov}
\end{figure}

Antulov-Fantulin et al.~\cite{antulov2015identification} also use posterior entropy to examine source uncertainty across SIR parameter settings. Fig.~\ref{fig:entropy_detectability_antulov} shows how their numerical uncertainty measure varies with infection and recovery probabilities. Such a distribution is conditional on the simulation and observation protocol, including the source prior. It is not a general monotonic law relating transmission probability to source identifiability or detection accuracy. Entropy describes concentration under the inferred distribution; empirical accuracy against known simulated origins is a separate measure.

\FloatBarrier

\subsubsection{Learning-based Methods}

Learning-based methods fit a mapping from network observations to source labels or latent source states. Their inputs, training signals, and outputs differ across architectures. Source detection can be formulated as a supervised inverse learning problem, where the objective is to predict the true source set \(\mathcal{S}^{\ast}\), or the corresponding source scores, directly from the complete observation \(Q_T^{\mathrm{full}}\). In general, a learning-based model can be expressed as
\begin{equation}
    \hat{\mathcal{S}}
    =
    f_{\phi}\bigl(G,Q_T^{\mathrm{full}},\mathbf{Z}\bigr),
    \label{eq:learning_source_mapping}
\end{equation}
where \(f_{\phi}\) is a parameterized model with learnable parameters \(\phi\), and \(\mathbf{Z}\) denotes optional node attributes, edge attributes, or auxiliary diffusion features.

Existing learning-based methods have evolved along four main directions: network representation learning, self-supervised representation learning, deep generative probabilistic learning, and diffusion inverse learning. Network representation learning forms the foundation of modern learning-based source detection. Instead of relying on manually designed structural features, these methods learn node representations directly from the network topology and observed diffusion states using Graph Neural Networks (GNNs)~\cite{kipf2017semi,dong2019multiple,wu2021gnnsurvey}. The full observed snapshot, together with the underlying network structure and optional node or edge attributes, is used as the model input. Through iterative neighborhood aggregation, the learned representations encode both structural and diffusion information, allowing the model to estimate the probability or confidence score of each node being the diffusion source~\cite{hamilton2017inductive,xu2019powerful}.

Shu et al.~\cite{shu2021information} proposed the Multi-Channel Graph Neural Network (MCGNN), a representative network representation learning framework for source detection. MCGNN is motivated by the observation that source-related signals are encoded not only in node states but also in edge connectivity and local propagation patterns. To exploit these complementary sources of information, the model learns representations through separate node and edge channels before integrating them into a unified embedding. The node channel performs graph convolution on the original network \(G\), aggregating node states, structural positions, and neighborhood information to capture local diffusion characteristics. The edge channel operates on the corresponding line graph, where each edge in the original network is treated as a node, allowing the model to encode relationships among adjacent edges and potential transmission pathways. Let \(\mathbf{h}^{(V)}_i\) denote the representation learned from the node channel and \(\mathbf{h}^{(E)}_i\) the edge-channel representation mapped back to node \(v_i\). Their fusion is given by
\begin{equation}
    \mathbf{h}_i
    =
    \mathrm{Fuse}\!\left(\mathbf{h}^{(V)}_i,\mathbf{h}^{(E)}_i\right),
    \label{eq:mcgnn_fusion}
\end{equation}
where \(\mathrm{Fuse}(\cdot)\) denotes the feature fusion operation. The fused representation \(\mathbf{h}_i\) is then used to produce a source-classification score for node \(v_i\).

The separate channels encode complementary node and edge neighborhoods. Source labels supervise the final prediction; high scores need separate calibration before probabilistic interpretation~\cite{kull2019dirichlet}. Pairwise message passing also differs from a hypergraph representation in which group interactions are explicitly modeled~\cite{guo2025betweenness}, motivating comparison with the higher-order architecture below~\cite{zhang2026adaptive}.

To model higher-order diffusion directly, Cheng et al.~\cite{cheng2025hyperdet} proposed HyperDet (Source Detection in Hypergraphs via Interactive Relationship Construction and Feature-rich Attention Fusion), extending learning-based source detection from simple network structures to hypergraphs. HyperDet employs an Interactive Relationship Construction (IRC) module to jointly encode hypergraph topology, diffusion states, interactions between infected and uninfected nodes, and positional information. The resulting features are subsequently integrated through a Feature-rich Attention Fusion (FAF) module, which learns node representations while adaptively weighting their contributions to source detection. 
Unlike GNN architectures that operate solely on pairwise connections, HyperDet incorporates hyperedge-level relationships into the learned node representations. This allows group interactions to contribute directly to the identification of candidate sources.

\begin{figure}[htp]
    \centering
    \includegraphics[width=0.8\linewidth]{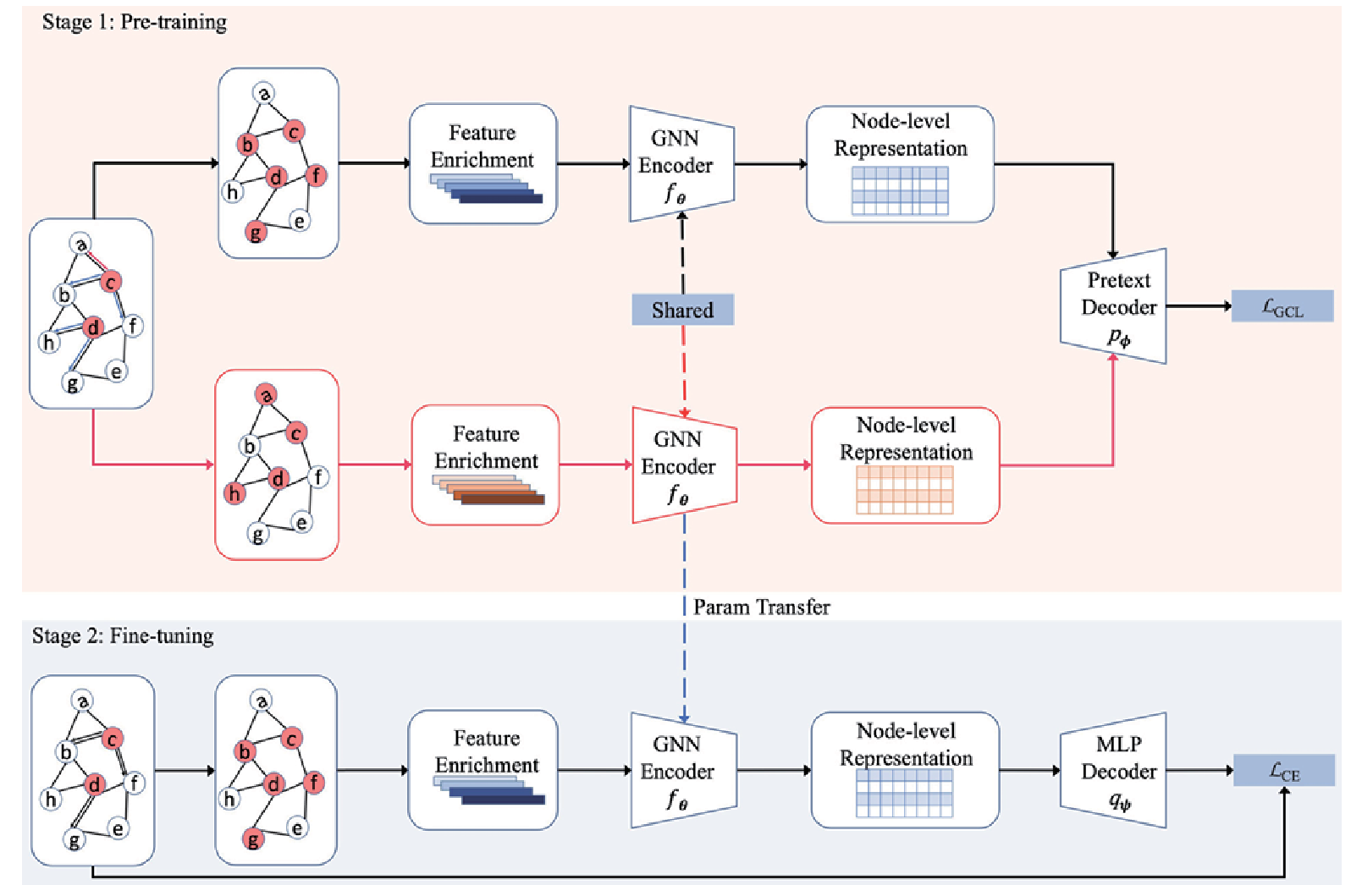}
    \caption{
Illustration of the SL-GCL framework~\cite{bao2024graph}. The model first performs self-supervised contrastive pretraining to learn node representations from augmented propagation views, followed by supervised fine-tuning that transfers the learned representations to source-node prediction.
}
    \label{SLGCL}
\end{figure}

{\emergencystretch=3em

Self-supervised learning offers another approach to extracting source-related information from diffusion observations, particularly when labeled training examples are limited. Graph contrastive learning uses different views of the data to learn representations that retain features shared across the chosen augmentations.} Source Localization with Graph Contrastive Learning (SL-GCL) is a representative method following this paradigm~\cite{bao2024graph}. As illustrated in Fig.~\ref{SLGCL}, the framework consists of a pretraining stage and a fine-tuning stage. 
During pretraining, different diffusion realizations generated from the same source set are used to construct related views for contrastive learning. Let \(Q_{T,a}^{\mathrm{full}}\) and \(Q_{T,b}^{\mathrm{full}}\) denote two views treated as a positive pair, and let \(h_\phi(G,Q)\) represent the embedding produced by the encoder. The contrastive objective can be written as

\begin{equation}
\mathcal{L}_{\mathrm{CL}}
=
-\log
\frac{
\exp\left(
\mathrm{sim}\left(
h_{\phi}(G,Q_{T,a}^{\mathrm{full}}),
h_{\phi}(G,Q_{T,b}^{\mathrm{full}})
\right)/\tau
\right)
}{
\sum_{Q'\in\mathcal{B}}
\exp\left(
\mathrm{sim}\left(
h_{\phi}(G,Q_{T,a}^{\mathrm{full}}),
h_{\phi}(G,Q')
\right)/\tau
\right)
},
\label{eq:slgcl_contrastive}
\end{equation}
where \(\operatorname{sim}(\cdot,\cdot)\) measures similarity between embeddings, \(\tau\) is the temperature parameter, and \(\mathcal B\) contains the positive view \(Q_{T,b}^{\mathrm{full}}\) together with the other comparison views in the training batch. The loss favors agreement between paired augmentations relative to other views. Its ability to preserve information useful for source detection depends on how the augmented observations are constructed and how the representations are subsequently fine-tuned.

Beyond contrastive pretraining, SL-GCL employs Laplacian-based feature enrichment and skip connections to expand the original node representations. This allows the encoder to incorporate structural information beyond the low-dimensional diffusion-state labels, providing additional features for distinguishing candidate sources.
During fine-tuning, the pretrained embeddings are fed into a decoder to estimate the probability of each node being the diffusion source. 
  The usefulness of the learned representations nevertheless depends on whether the augmentations retain information relevant to source detection and whether the training data adequately represent the diffusion conditions encountered during inference.

Deep generative probabilistic learning extends learning-based source detection by introducing latent-variable generative models. Rather than deriving an explicit diffusion likelihood or treating source detection as a deterministic node-classification task, these methods learn a probabilistic relationship between observed diffusion snapshots and their underlying source states~\cite{chen2025structure}. A representative example is the Source Localization Variational Autoencoder (SL-VAE)~\cite{ling2022source}. SL-VAE formulates source detection as an inverse problem on a network by introducing a variational autoencoder into the diffusion process. It learns latent variables from source--observation pairs and combines a generative model with a forward diffusion estimation module to approximate the conditional distribution of source states given an observed snapshot.
The resulting model provides an estimated distribution over possible source states, although the uncertainty it represents remains dependent on the learned generative model and the available training data.

In contrast to conventional probabilistic inference methods, SL-VAE does not require an analytically derived likelihood for each candidate source. It also differs from supervised GNN-based classifiers by modeling source detection through a probabilistic generative process instead of directly assigning deterministic scores to candidate nodes. Although SL-VAE combines ideas from probabilistic inference and representation learning, it is more naturally categorized as a learning-based approach in this review because source detection is achieved through data-driven training of latent variables and generative mappings rather than explicit posterior computation.

Invertible Validity-aware Graph Diffusion (IVGD) approaches source detection through a learned diffusion surrogate and an inverse module, with validity constraints refining candidate source states~\cite{wang2022invertible}. Its input is a diffusion observation on a network; training uses examples relating initial source indicators to resulting states. In schematic notation,
\begin{equation}
\widehat{\mathbf X}(T)=F_\phi(\mathbf X(0);G),\qquad
\widehat{\mathbf X}(0)=F_\phi^{-1}(\mathbf X(T);G).
\label{eq:ivgd_invertible}
\end{equation}
The inverse refers to the designed surrogate, not to a theorem that the underlying stochastic propagation realization is reversible. Validity-aware refinement constrains the inferred initial indicators and their consistency with the learned forward response. It can help avoid outputs inconsistent with the model's admissible source representation, but does not establish physical correctness under a mismatched simulator. Compared with direct node classification, this architecture explicitly couples a forward representation to inverse prediction. Both approaches still require validation on source labels not used during training, particularly under changed source counts or dynamics.

Zhao et al.~\cite{zhao2022improving} proposed Perturbed Graph Diffusion (PGD) to improve the adaptation of pretrained feature-construction models to source detection. Before fine-tuning, PGD applies matrix-wise uniform perturbations to the pretrained parameters, with their magnitudes determined by the scale and variance of the corresponding weight matrices. This strategy is intended to reduce dependence on the pretrained parameter configuration and improve adaptation to the downstream detection objective. PGD differs from IVGD in its treatment of the inverse problem: IVGD learns a forward diffusion representation with an inverse mapping, whereas PGD focuses on the adaptation of pretrained models for source prediction. The two approaches therefore address different aspects of learning-based source detection.

Training labels and output conventions are part of a learned source estimator. A single-source softmax imposes one mutually exclusive node label, whereas independent binary outputs can represent several initiating nodes but require a threshold or source-count decision. The two formulations cannot be compared solely through a single accuracy value. Class weighting changes the training objective; it does not by itself calibrate the output to a deployment prior. A model trained on uniformly sampled sources can become miscalibrated when actual introductions concentrate on a few accessible nodes. Likewise, separating training and test data by network tests a different capability from holding out cascades on the same network. These choices determine whether reported performance reflects memorization of a network's source frequencies, generalization to new outbreaks, or transfer to new structural and dynamical conditions.

These models exchange explicit likelihood calculations for training data, architectural assumptions, and optimization. Their computational comparison should include data generation and training as well as prediction time. Source imbalance, threshold selection, source-count assumptions, and score calibration matter in addition to the encoder architecture.

\subsection{Discussion and Outlook}

Complete-snapshot methods differ not only in their inference mechanisms but also in the assumptions used to interpret the observed spreading pattern. Structural estimators rely on the geometry of infected regions, whereas probabilistic approaches incorporate diffusion dynamics through message passing or likelihood-based evaluation. Simulation-based methods compare candidate sources through forward realizations, while learning-based methods draw on patterns acquired from training data. These approaches also differ in their outputs, ranging from individual source rankings to candidate sets and model-dependent probability estimates. Their relative performance must therefore be considered alongside the information and assumptions on which each estimate relies.

Computational demands vary substantially across these approaches. Structural methods may require repeated shortest-path or BFS calculations, while DMP and BP involve message updates whose cost depends on the network structure and temporal resolution. Simulation-based inference can become expensive when many candidate sources and diffusion realizations must be evaluated. Methods such as BOSouL reduce the number of direct simulations by introducing surrogate-based search, although constructing and updating the surrogate introduces additional costs. Learning-based methods shift much of the computation to data generation and model training, often allowing faster prediction once training is complete. Runtime comparisons should therefore distinguish the costs incurred at different stages and account for the candidate space, observation window, simulation budget, and computational environment. More efficient search can make source estimation computationally feasible without resolving ambiguities inherent in the observations.

The reliability of these methods also depends on how closely their assumptions reflect the actual spreading process. Structural criteria based on shortest paths may become less informative when transmission delays vary substantially across edges. Probabilistic estimators can favor incorrect candidates when transmission or recovery parameters are misspecified, while learned models may rely on structural or dynamical patterns that do not persist across networks. Complete knowledge of node states at the observation time does not eliminate these sources of error. Comparisons should therefore distinguish performance under correctly specified diffusion models from robustness to parameter variation and previously unseen network or diffusion conditions, using the evaluation measures introduced in Section~\ref{performance evaluation}.

\section{Source Detection with Incomplete or Noisy Observations}
\label{sec:partial-observability}

\subsection{Unified Formulation with Incomplete Observations}

Incomplete observations arise in different ways, including unrecorded node states, missing temporal information, uncertain network connections, and measurement errors. These forms of uncertainty have different implications for source inference. Missing node states restrict the evidence available about a diffusion realization, whereas uncertain connections affect the transmission routes considered by the model. Missing or inaccurate temporal records further limit what can be inferred about the sequence of spreading events. These distinctions need to be reflected in the observation model introduced in Eq.~~\eqref{eq:unified_observation}, with network uncertainty explicitly accounted for when the underlying topology is not fully known.

The statistical estimators introduced in Section~\ref{sec:problem_definition} remain applicable when the likelihood accounts for unobserved quantities and the recording process. Structural and learning-based methods may instead work directly with partial observations, although their treatment of missing data and measurement errors must be specified. Sensor-based inference often faces similar limitations, but sensor placement and data acquisition introduce additional design considerations, which are examined separately in Section~\ref{sec:sensor-based}.

\begin{figure}[htp]
    \centering
    \includegraphics[width=0.8\linewidth]{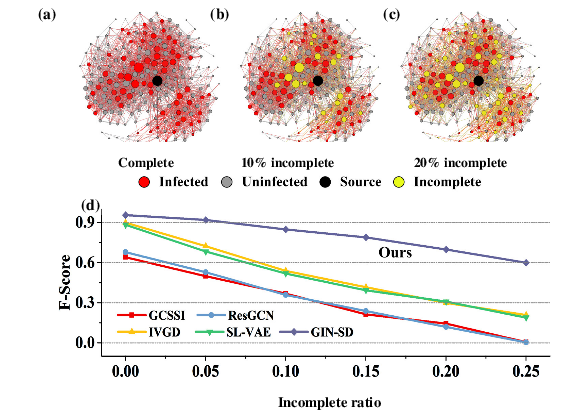}
    \caption{
Source detection under incomplete node observations~\cite{cheng2024gin}. (a)--(c) Diffusion snapshots with complete observations and with 10\% and 20\% missing node information, respectively. Red, gray, black, and yellow nodes denote infected, uninfected, source, and unobserved nodes. (d) Performance of GIN-SD as the proportion of missing node information increases.
}
    \label{fig:gin-sd-incomplete-nodes}
\end{figure}

Incomplete records can also change the apparent structure of an outbreak. Figures~\ref{fig:gin-sd-incomplete-nodes}(a)--(c) illustrate how missing node states alter the observed infection pattern, while Fig.~\ref{fig:gin-sd-incomplete-nodes}(d) shows the corresponding variation in source-detection performance. Uncertain network connections and missing temporal observations introduce different ambiguities by affecting the available transmission paths and the evidence about propagation order, respectively. The following sections examine these problems separately, considering missing node states, incomplete network structure, and partial temporal observations.

\subsection{Detection Methods with Incomplete Observations}

\label{subsec:partial-observation-characteristics}
\subsubsection{Methods for Missing Node States}
\label{subsubsec:missing-node-states}

We first consider source detection when the network topology is known but node states are recorded only for a subset \(V_{\mathrm{obs}}\subseteq V\). Missing states can obscure the extent of an outbreak and alter the apparent connectivity of infected nodes. Even when the underlying infection region is connected, its observed portion may consist of separate components, making distance-based and other structural evidence less reliable.

An unobserved node is not a susceptible node. Under SI, \(w\) missing binary states admit up to \(2^w\) configurations before diffusion constraints are imposed; this count illustrates a latent-state space, not a requirement to enumerate it. Likelihood methods may marginalize hidden states, geometric methods can use distances to observed infections, and learned models can encode missingness explicitly. A method can infer the source directly from the partial record or first reconstruct a complete snapshot.
 The observation mask can itself be informative when testing or reporting depends on infection, and should not silently be assumed random.

In a probabilistic treatment, missing states can be eliminated rather than filled with one guessed configuration. With a fixed observation mask and error-free measurements, the likelihood is obtained by summing over all full network states consistent with the recorded values. If the probability of observing a node depends on its state, the sampling mechanism must also be accounted for. With measurement errors, each possible full state is weighted by the probability of observing the recorded values given that state. Selecting a single imputed state replaces that sum with one contribution and may suppress alternative explanations. Direct geometric methods avoid such a sum by using constraints satisfied by the observed subset, while a classifier can include an explicit missingness category. These approaches make different approximations; none can be declared preferable from the missing fraction alone. In particular, a few reliable negative observations near a candidate can be more discriminating than many distant positive observations, depending on the propagation model.

The earliest methods for source detection with missing node states were based on the observation that, despite incomplete observations, the relative positions of infected nodes still preserve coarse geometric information about the underlying spreading process. Luo et al.~\cite{luo2014identify} generalized the Jordan infection center to partially observed SI diffusion and proposed the Reverse Greedy (RG) algorithm for practical source detection. Let $V_{\mathrm{obs}}^{I}\subseteq V_{\mathrm{obs}}$ denote the set of observed infected nodes, and let $d(v_i,v_j)$ be the shortest-path distance between nodes $v_i$ and $v_j$. For a candidate source $v_i$, its infection eccentricity with respect to the observed infected nodes is defined as

\begin{equation}
    \rho_I\!\left(v_i;V_{\mathrm{obs}}^{I}\right)
    = \max_{v_j\in V_{\mathrm{obs}}^{I}} d(v_i,v_j).
\end{equation}

Under the tree assumption, the diffusion source is estimated as the Jordan infection center, which minimizes the infection eccentricity defined over the observed infected nodes, i.e.,
\begin{equation}
    \hat{v}
    = \operatorname*{argmin}_{v_i\in V}
      \rho_I\!\left(v_i;V_{\mathrm{obs}}^{I}\right)
    = \operatorname*{argmin}_{v_i\in V}
      \max_{v_j\in V_{\mathrm{obs}}^{I}} d(v_i,v_j).
      \label{eq4.1.1.1}
\end{equation}

\begin{figure}[htp]
    \centering
    \includegraphics[width=0.9\linewidth]{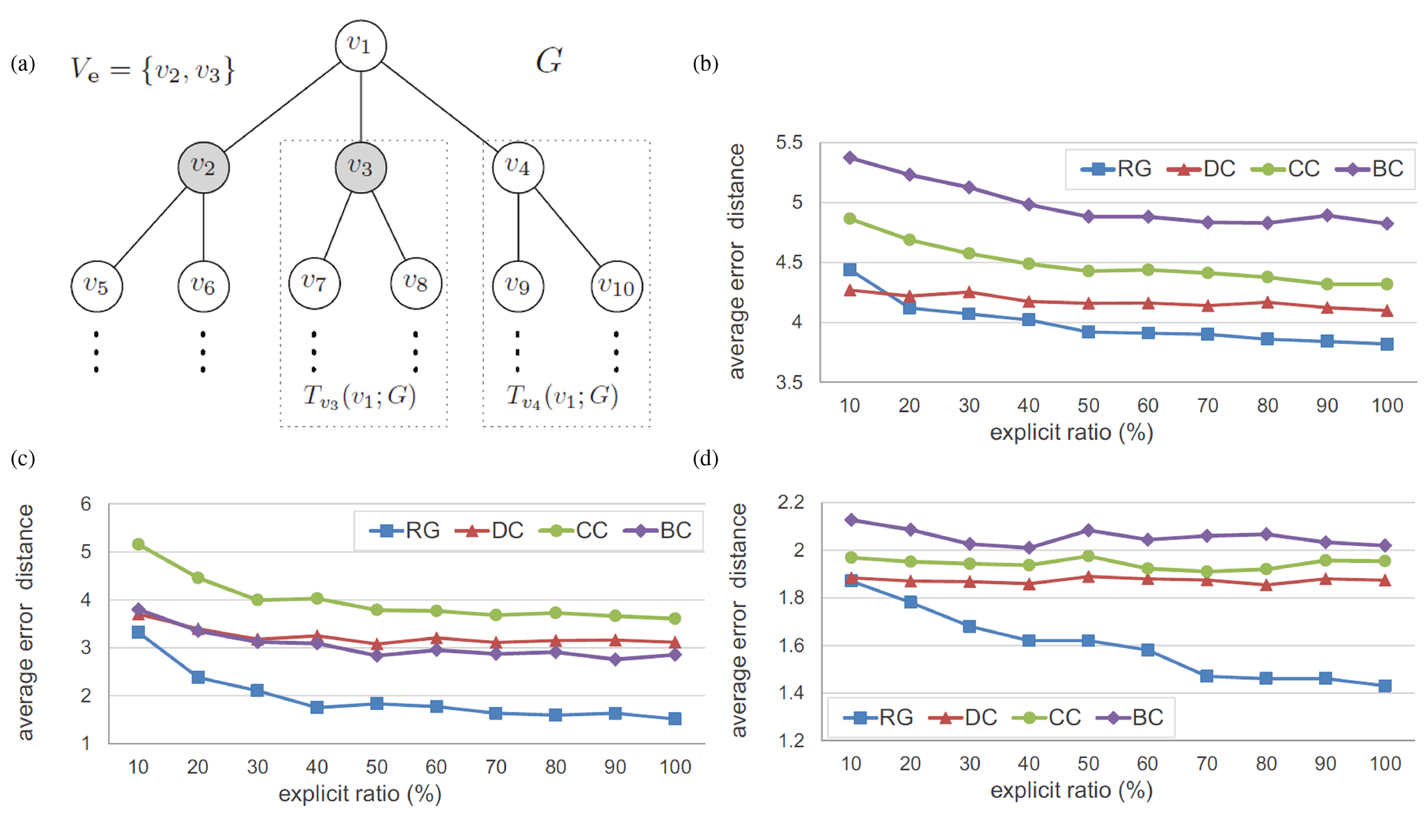}
    \caption{Illustration of source detection under limited observations~\cite{luo2014identify}. (a) Observation model in which only a subset of nodes is explicitly observed, while the remaining portions of the network are represented as hidden subtrees. (b)--(d) AED under different explicit observation ratios on three representative network topologies, comparing RG with degree centrality (DC), closeness centrality (CC), and betweenness centrality (BC).
}
   
    \label{fig:limited-observation-luo-rg}
\end{figure}

This criterion assumes at least one observed infected node and finite distances between the candidate and the observed infected nodes. The RG algorithm detects the source directly from the partial infection pattern without explicitly estimating the hidden spreading states (see Fig.~\ref{fig:limited-observation-luo-rg}(a)). This geometric formulation subsequently inspired a broad class of distance-based source-detection methods.
Figures~\ref{fig:limited-observation-luo-rg}(b)--(d) compare RG with several centrality-based estimators under different observation ratios. The AED generally decreases as more infected nodes become observable, although the improvement is not strictly monotonic. RG also achieves lower detection errors than the comparison methods in many of the tested settings. These results suggest that additional observed infections can strengthen the geometric constraints used by RG, with the extent of the improvement depending on network structure and the observation conditions.

A complementary result covers sparse positive reports under heterogeneous SIR dynamics. Zhu and Ying~\cite{zhu2014robust} prove that their most-probable-sample-path estimator is a Jordan infection center of the observed infected set on an infinite tree. Their constant-distance, high-probability error bound additionally requires branching and transmission lower bounds and an unbiased reporting rule that makes sufficiently large infected subsets likely to contain a report. This result supports geometric source estimation from sparse infection reports without reconstructing every missing state. Its sample-path objective selects a compatible history, whereas a marginal likelihood sums over compatible histories.

Spectral criteria offer another way to evaluate candidate sources. Fioriti et al.~\cite{fioriti2014predicting} proposed Dynamic Age (DA), which measures the change in the leading eigenvalue of an infected network when a candidate node is removed. Unlike the distance-based criterion used by RG, DA evaluates the contribution of a node to the spectral structure of the infection graph. Under incomplete observations, the same score can be computed on the subgraph induced by observed infected nodes or on a reconstructed infection region. Such an application, however, depends on how well the selected subgraph represents the underlying infection pattern. Let \(\widetilde A_I\) denote its adjacency matrix, with largest eigenvalue \(\lambda_{\max}(\widetilde A_I)\). For a candidate node \(v_i\), let \(\lambda_{\max}^{(-i)}\) denote the corresponding eigenvalue after removing that node. When \(\lambda_{\max}(\widetilde A_I)>0\), the dynamic age of node $v_i$ is defined as

\begin{equation}
DA_i
= \frac{\left|\lambda_{\max}(\widetilde{A}_I)-\lambda_{\max}^{(-i)}\right|}
{\lambda_{\max}(\widetilde{A}_I)}.
\end{equation}

A high DA score indicates that removing the candidate substantially changes the spectral radius of the infected subgraph. This provides a structural criterion for ranking potential sources, but spectral importance does not by itself establish that a node initiated the diffusion. When only part of the infection region is observed, the ranking may also depend on which infected nodes are missing. Its use under incomplete observations therefore requires care, particularly when missing states substantially alter the spectral structure.

RG and DA use structural information from observed infections in different ways. RG evaluates candidate sources through their shortest-path distances to observed infected nodes, whereas DA measures the change in the leading eigenvalue of the infected subgraph after removing a candidate node. Neither method explicitly reconstructs missing node states, and both may become less reliable when incomplete observations substantially alter the apparent structure of the infected region.

Recovery introduces an additional complication under SIR dynamics. In SI diffusion, infected nodes remain infected and can still indicate the affected region, whereas SIR diffusion allows infected nodes to recover. As a result, recovered nodes may no longer appear as infected in the observation, and the observed infection snapshot may provide only a fragmented record of the underlying spreading history. To account for this additional uncertainty, Zang et al.~\cite{zang2015locating} proposed a multisource detection framework that integrates reverse spreading with community partitioning. Let $V_I$ and $V_R$ denote the complete sets of infected and recovered nodes, respectively, and let $k$ be the number of sources. Given the network $G$ and the spreading parameters $\Theta$, the multisource maximum-likelihood estimator can be formulated as

\begin{equation}
    \hat{\mathcal{S}}
    = \operatorname*{argmax}_{\mathcal{S}\subseteq V_I\cup V_R}
      P\!\left(Q_T^{\mathrm{partial}}\mid \mathcal{S},G,\Theta\right),
    \qquad |\mathcal{S}|=k.
\end{equation}

Since the complete set $V_I\cup V_R$ is unavailable under partial observations, the method first applies a score-based reverse-spreading procedure to infer hidden infected and recovered nodes, yielding an approximate infected--recovered region, $V_I\cup V_R \approx \hat{V}_{IR},$ where $\hat{V}_{IR}$ denotes the reconstructed infected--recovered node set. Rather than searching exhaustively over all possible combinations of $k$ sources, the reconstructed region is further partitioned into multiple infected communities, allowing the multisource detection problem to be decomposed into a series of single-source detection tasks.

For a reconstructed community \(B_a\), a rumor-centrality-based step ranks candidates by \(R(v_i;B_a)\) and returns one representative per community, giving \(\widehat{\mathcal S}=\{\hat v_1,\ldots,\hat v_k\}\). This ranking is a structural component of the reconstruction procedure. Its equality to an exact likelihood does not follow from community partitioning: the regular-tree SI conditions discussed in Section~\ref{sec:full-observability} do not automatically hold for reconstructed SIR communities. Errors in reverse spreading or partitioning therefore propagate to the final source set.

\begin{figure}[htp]
    \centering
    \includegraphics[width=1\linewidth]{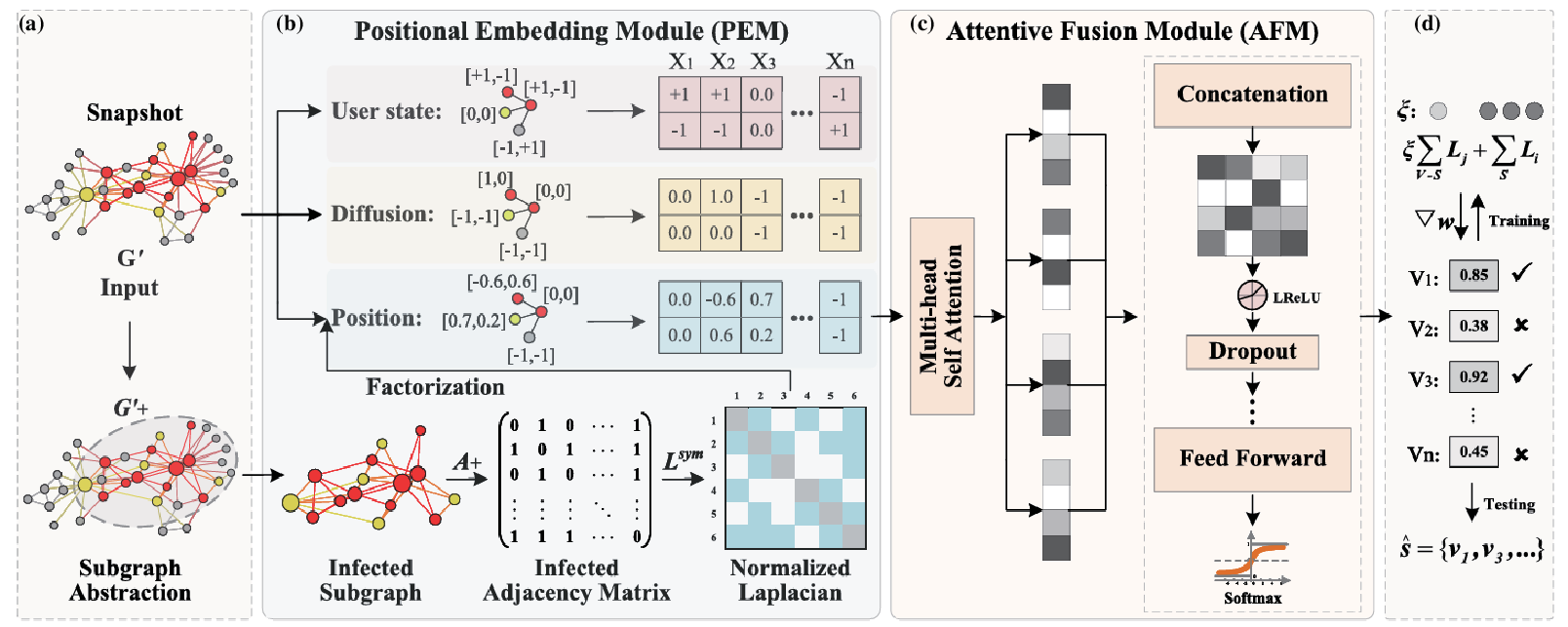}
\caption{
Illustration of the GIN-SD framework~\cite{cheng2024gin}. (a) An observed diffusion snapshot is transformed into an infected subgraph. (b) The Positional Embedding Module (PEM) encodes structural, diffusion, and node-state information into feature representations. (c) The Attentive Fusion Module (AFM) learns source-oriented node embeddings through multi-head self-attention. (d) A class-balanced training strategy is adopted for source prediction, yielding the estimated source set \(\hat{\mathcal{S}}\).
}
    
    \label{fig:gin-sd-architecture}
\end{figure}

Learning-based methods provide another approach to source inference when node information is incomplete. GIN-SD (Source Detection in Graphs with Incomplete Nodes via Positional Encoding and Attentive Fusion) addresses missing user information through a positional embedding module and an attentive fusion module~\cite{cheng2024gin}. As illustrated in Figure~\ref{fig:gin-sd-architecture}, the model combines observed node states, available propagation attributes, and structural position information to learn representations for source prediction. Its original formulation considers incomplete user and propagation records, rather than missing infection-state labels alone.

To explicitly represent missing node states, GIN-SD first assigns each node a state feature according to the observation snapshot. Let $V_{\mathrm{obs}}^{I}$ denote the set of observed infected nodes and $\Psi=V\setminus V_{\mathrm{obs}}$ the set of nodes with unknown states. The node-state encoding is defined as

\begin{equation}
    Y_i(T)=
    \begin{cases}
        +1, & v_i\in V_{\mathrm{obs}}^{I},\\
        -1, & v_i\in V_{\mathrm{obs}}\setminus V_{\mathrm{obs}}^{I},\\
        0,  & v_i\in\Psi.
    \end{cases}
\end{equation}

The three values represent infected, observed-uninfected, and unobserved nodes, respectively. Since node states alone provide limited information about the underlying diffusion pattern, GIN-SD further introduces positional encodings derived from the topology of the infected subgraph. Let $\widetilde{A}_I$ and $\widetilde{D}_I$ denote the corresponding adjacency and degree matrices. The symmetrically normalized graph Laplacian is

\begin{equation}
    \widetilde{L}_I
    = I-\widetilde{D}_I^{-1/2}\widetilde{A}_I\widetilde{D}_I^{-1/2},
\end{equation}
where $I$ is the identity matrix; the corresponding diagonal entry of $\widetilde{D}_I^{-1/2}$ is set to zero for an isolated node. Selected eigenvectors associated with the smallest nontrivial Laplacian eigenvalues provide positional information for nodes in the extracted infected subgraph. Nodes outside this subgraph receive a separate positional encoding. The resulting features are combined with node-state and available propagation information, while the attentive fusion module learns representations for source prediction. A class-balanced training objective further accounts for the imbalance between source and non-source nodes.

GIN-SD combines observed node states, structural information, and available propagation features through learned representations rather than predefined source-ranking criteria. This allows the model to use information from partially observed diffusion patterns without explicitly reconstructing every missing state. Its performance, however, depends on the training data and the extent to which the missing information affects the features used for source prediction.

These approaches differ primarily in how they use the information that remains observable. Distance-based estimators retain geometric constraints from reported infections, spectral methods evaluate structural properties of the available infection graph, and reconstruction-based methods estimate hidden states before locating candidate sources. Learning-based approaches instead use training data to extract source-related features from incomplete records. Their relative suitability depends on the missingness mechanism, the diffusion dynamics, and the reliability of the information used for inference.

\subsubsection{Methods for Incomplete Network Structure}

Source detection becomes more challenging when the network available for inference does not fully represent the underlying transmission structure. Let \(G=(V,E)\) denote the true network and \(\widetilde G=(V,\widetilde E)\) its observed counterpart, where missing or erroneous edges may distort the connectivity among nodes. Some studies assume that node states are fully observed, allowing the effects of structural uncertainty to be examined separately. In practice, incomplete network information may also coincide with missing node states or limited sensor records. The assumptions about these observations must be considered when assessing source-detection methods.

Incomplete network structure preserves the observed infection pattern while altering the topological substrate on which diffusion is interpreted. Missing edges may artificially disconnect genuine transmission routes, making infected regions that are close in the true network appear much farther apart. In contrast, spurious edges can create non-existent shortcuts that reduce the apparent separation between candidate sources and infected nodes. As a result, quantities derived from the observed topology, including shortest-path distances, infection boundaries, spreading trees, and node centralities, may deviate substantially from their true counterparts, reducing the structural discriminability of candidate sources.

Missing nodes add a further complication beyond missing edges. If an initiating node is absent from the represented node set, no estimator restricted to that set can return its correct label. Evaluation must then distinguish detection to a represented proxy, discovery of an unobserved origin, and inference over a known set of candidate nodes. With uncertain edges but known nodes, a joint probabilistic model would average source evidence over plausible networks, weighted by a topology-error model. A completion-first pipeline instead selects one network before source scoring. Its result can be sensitive to a small number of reconstructed shortcut edges, even when aggregate link-prediction accuracy is high. Source-oriented validation should therefore test how structural uncertainty changes candidate paths and rankings, rather than use network reconstruction quality as a substitute for source detection performance.

Existing approaches address incomplete network structure in two main ways. Some methods incorporate uncertainty about missing or erroneous connections into the source-detection procedure, using propagation evidence to assess plausible transmission routes. Others reconstruct the network before applying a source estimator. The distinction lies in whether structural uncertainty is considered during source inference or handled through a separate network-reconstruction step.

SDSI combines sensor deployment and Bayesian source inference to identify diffusion origins in structurally incomplete social networks~\cite{cheng2024sdsi}. Monte Carlo simulations are used to guide sensor placement, after which candidate sources are evaluated through maximum-a-posteriori estimation using the available network information and propagation observations. The method addresses source detection without requiring a complete reconstruction of the underlying network.

The use of propagation records allows SDSI to assess candidate origins under the available structural information. This differs from a reconstruction-first procedure, in which source detection is performed only after estimating the missing network connections. The reliability of the resulting source estimates depends on the sensor observations and the assumptions governing the incomplete network. In particular, successful detection does not imply that the missing transmission links have been uniquely recovered.

Sharing information supplies a different constraint. Under a message protocol in which a count of one certifies the first forwarding step, the sender of that record can directly identify an initiating node. Larger counts provide partial ancestry information and require further candidate evaluation. This interpretation depends on reliable counters and sender identities; a generic interaction count does not have the same meaning. The SDSI-S extension described with SDSI uses such behavioral records alongside temporal evidence~\cite{cheng2024sdsi}. It illustrates why structurally incomplete source detection can overlap with sensor-based inference without becoming simple network completion. Exact implementation details and calibration require the method's specified recording protocol.

An alternative is to reconstruct a network first and then apply a source estimator. This separates network learning from outbreak inference, but errors in the reconstructed topology must still be propagated or tested. The following network-learning studies are relevant background; their original tasks should not be reported as source-detection experiments.

Representative methods include hierarchical random graphs and generative network models. Link-prediction approaches use the observed network structure to score potential missing edges~\cite{zhan2020susceptible}. Clauset et al.~\cite{clauset2008hierarchical} employed hierarchical random graphs to characterize the multiscale organization of complex networks and infer missing links in partially observed networks.  Kim and Leskovec~\cite{kim2011network} proposed KronEM, which combines the Kronecker graph model with expectation--maximization to infer missing network structure and estimate the associated model parameters. Hierarchical random graphs and KronEM describe network formation through different generative assumptions, but both can provide structural estimates when the observed network is incomplete. Their application to source detection requires a subsequent inference procedure, whose results may be sensitive to uncertainty in the estimated topology.

Probabilistic graph learning provides another perspective on inference with uncertain network connections. Zhang et al.~\cite{zhang2019bayesian} developed Bayesian GCNN to account for uncertainty in graph structure during semi-supervised node classification. Wei et al.~\cite{wei2021towards} proposed EBGCN for rumor detection, incorporating Bayesian edge-reliability estimation and edge-level consistency learning to reduce the influence of unreliable propagation relationships. Although these methods were developed for different prediction tasks, their treatment of structural uncertainty offers possible directions for source detection. Applying them to source detection would require adapting the prediction objective and examining how errors in network structure affect the identification of diffusion origins.

Completion-first and uncertainty-aware approaches handle incomplete network information in different ways. A single completed network can make computation simpler while hiding uncertainty in candidate paths. Retaining several candidate network topologies or edge reliabilities can expose that uncertainty, at additional cost. Neither general link prediction nor rumor-veracity classification alone demonstrates more accurate inference of an unknown diffusion source. Source-specific validation needs outbreaks with known held-out origins and a controlled structure-error mechanism.

\subsubsection{Methods for Partial Temporal Observations}
\label{subsubsec:incomplete-temporal-observations}

In many applications, diffusion is observed at selected time points rather than continuously. Even when the states of all nodes are recorded at each sampling time, the transmission events and state changes between observations remain largely unknown. Different propagation histories may therefore be consistent with the same sequence of snapshots, making it difficult to distinguish among candidate sources. The information available for source inference depends on the diffusion dynamics, the timing of observations, and the variables recorded~\cite{kumar2017temporally}.

The timing of an observation can substantially affect source inference. Early snapshots may preserve the spatial concentration of an outbreak, but a small infected region may offer limited evidence for distinguishing candidate origins. Later observations can reveal a larger affected region, while recovery and widespread infection may obscure the initial spreading pattern. Shi et al.~\cite{shi2022effect} investigated this trade-off under discrete-time SIR dynamics using a snapshot of partially observed nodes. Their method jointly estimates the source and diffusion initiation time, with Monte Carlo simulations providing marginal node-state probabilities for a factorized snapshot score. The results show that the effect of observation time varies with network structure and the fraction of observed nodes. Although the study considers a single partial snapshot rather than a sequence, it illustrates why the timing of available observations matters for source inference.

Temporal information may be incomplete in several ways. A sequence of full snapshots leaves the intermediate transitions unrecorded, whereas asynchronous measurements describe different nodes at different times. Missing arrival-time records introduce a separate difficulty: when a functioning sensor has not detected the process by the end of monitoring, its observation provides censoring information rather than an unknown arrival time. An uncertain time origin introduces a common temporal offset, while node-specific clock errors can distort relative event times. These distinctions affect the information available for source inference and determine whether a method can use the observations retrospectively or must operate as data arrive.
 
Path inference with partial infection timestamps provides another distinct route. Zhu et al.~\cite{zhu2016locating} use observed infection times to constrain candidate propagation paths. A feasible path must respect both network connectivity and temporal order; timestamp constraints can reject a geometrically short route that cannot explain the observations. This input differs from an unordered complete snapshot, which is why the method belongs with temporal or sensor records. Topological shortest paths remain an approximation unless the delay model makes them appropriate.

FGSSI (Feature-representation-based Graph Sequential Source Identification) uses a sequence of observed network states to infer diffusion sources from temporal and structural features~\cite{hou2025fgssi}. By combining information from successive snapshots, the model can exploit changes in the spreading pattern that are unavailable from a single final observation. The intermediate propagation history nevertheless remains unobserved, even when node states are recorded throughout the network at each sampling time. If the snapshots also contain missing node states, their treatment introduces an additional observation assumption.

FGSSI combines graph-based feature extraction with recurrent processing and temporal attention. The recurrent module uses gated updates to retain information from earlier observations while incorporating features from the current snapshot. Attention then assigns different weights to the available temporal representations. 
Bidirectional processing uses information from both directions of the observed sequence to construct representations for source detection. This is suited to retrospective inference, while online prediction must rely only on snapshots available at the time of estimation.

FGSSI uses source labels for training and assigns different weights to source and non-source nodes to address class imbalance. Let \(x_i\in\{0,1\}\) indicate whether node \(i\) is a source, \(p_i\) denote its predicted source probability, and \(\rho=N^{-1}\sum_{i=1}^{N}x_i\) be the fraction of source nodes. The weighted classification loss is

\begin{equation}
\mathcal L_{\mathrm{cls}}=-\frac1N\sum_{i=1}^{N}
\left[(1-\rho)x_i\log p_i+\rho(1-x_i)\log(1-p_i)\right].
\label{eq:fgssi_classification}
\end{equation}

This expression gives the classification component rather than the complete training objective. The released implementation also includes a weighted squared-error term for an intermediate prediction. Both components contribute to model optimization and should be considered when reproducing the training procedure.

\begin{figure}[htp]
    \centering
    \includegraphics[width=1\linewidth]{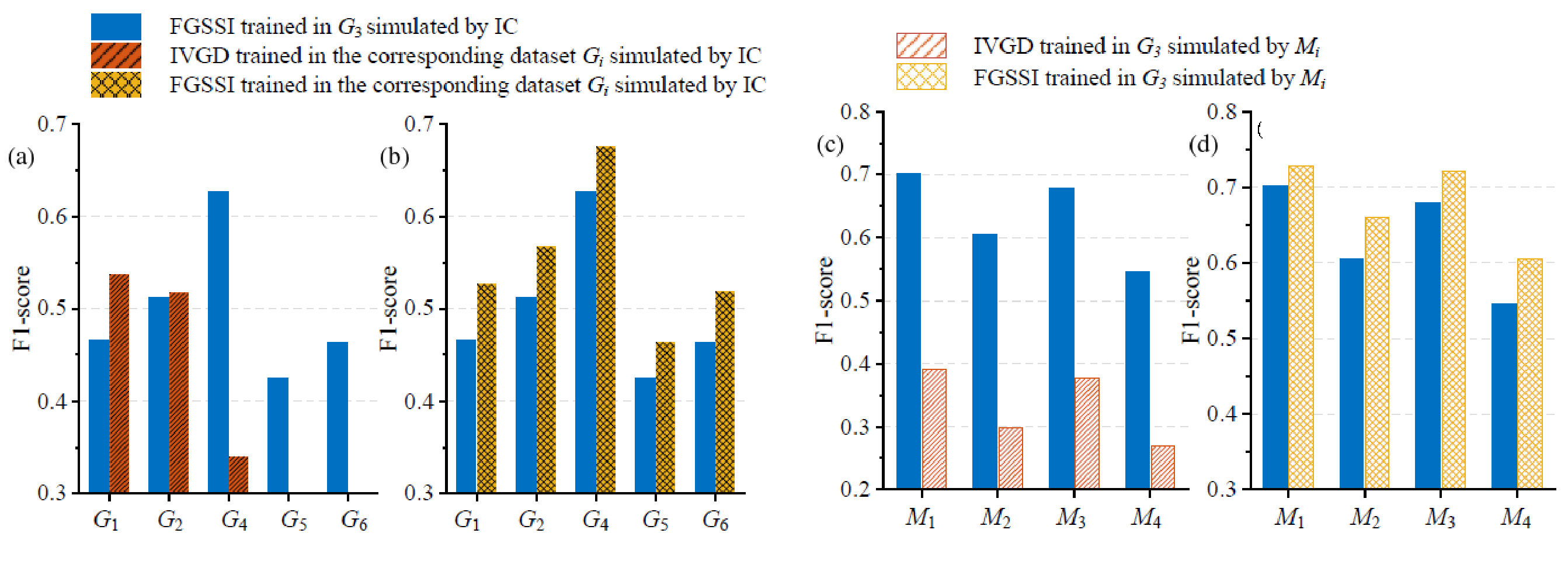}
        \caption{Transferability of FGSSI across networks and diffusion models~\cite{hou2025fgssi}. (a)--(b) Cross-network transfer. (c)--(d) Transfer across diffusion models. \(M_1\)--\(M_4\) denote homogeneous SI, homogeneous SIR, heterogeneous SI, and heterogeneous SIR, respectively.}
    \label{fig:fgssi_transferability_result}
\end{figure}

Figure~\ref{fig:fgssi_transferability_result} presents the transfer performance of FGSSI across networks and diffusion models~\cite{hou2025fgssi}. The reported \(F_1\) scores show how the method performs under the tested changes in network structure and spreading dynamics. These results provide evidence of transferability within the evaluated settings, while performance under other observation conditions remains to be established. Comparisons should also distinguish direct transfer from cases in which the target data are used for retraining or parameter adjustment.

\subsection{Discussion and Outlook}
\label{subsec:partial-observation-discussion}

The implications of incomplete observations for source detection depend on which information is missing and how the remaining evidence is used. Observed infections may still provide geometric constraints on candidate origins, while uncertain network connections affect the transmission routes considered during inference. Sparse temporal records introduce a different ambiguity by leaving intermediate spreading events unresolved. Existing methods address these limitations through structural scoring, state or network reconstruction, probabilistic inference, and learning from incomplete records~\cite{cheng2024gin,cheng2024sdsi,shi2022source,ru2023inferring}. Their performance cannot be compared solely by the fraction of missing observations, since the information retained by different recording schemes may vary substantially. Incomplete evidence can increase uncertainty about the source, but does not necessarily make distinct source hypotheses statistically indistinguishable.

Missing observations must also be distinguished from records that are inaccurate or delayed. An unobserved node state, an incorrectly reported infection, and an uncertain event time impose different constraints on possible diffusion histories. Measurements collected asynchronously are not inherently unreliable, provided that their sampling times are known and properly incorporated into the inference model. Errors in recorded states or timestamps, however, can alter the evidence used to evaluate candidate sources~\cite{Altarelli_2014,10.1145/357830.357849,10.1145/3335772.3335934}. Methods that perform well with missing data may therefore remain sensitive to measurement errors or reporting delays. A more systematic assessment of these effects, both separately and in combination, is needed to understand the robustness of source inference under imperfect observation conditions.

When several candidate origins remain compatible with the observations, reporting only the highest-ranked source may conceal substantial uncertainty. A more informative result may include alternative source candidates, estimated posterior probabilities, or uncertainty sets with clearly stated statistical interpretations~\cite{pmlr-v139-dawkins21a,Sadinle02012019,pmlr-v70-guo17a,stadler2021graph}. Posterior probabilities depend on the assumed prior and likelihood, whereas frequentist confidence sets require appropriate coverage guarantees. The reliability of these outputs must be assessed rather than inferred from the concentration of candidate scores alone. In applications where source estimates inform subsequent decisions, representing the remaining uncertainty can help determine whether additional observations are needed and which measurements would be most useful. This consideration connects inference under incomplete observations to the sensor deployment and observation design problems examined in Section~\ref{sec:sensor-based}.

\section{Sensor-based Source Detection and Observation Design}
\label{sec:sensor-based}

Sensor-based source detection uses local measurements collected at selected locations. Depending on the monitoring system, these records may include first-arrival times, node states, transmission directions, or forwarding metadata. Such measurements generally capture only part of the diffusion process, and their value for source inference depends on both sensor locations and the variables recorded. Source estimation can also be performed using fixed sensors, without adapting their locations during propagation.

Observation design and source estimation address different questions. Sensor locations and sampling rules may be specified in advance, optimized before diffusion begins, or adjusted as new observations become available. Source estimators then use the collected records and the corresponding observation model to evaluate candidate origins. Strategies designed for early detection, broad network coverage, or reconstruction of the system state do not necessarily yield the most accurate source estimates. The following subsections distinguish the objectives of sensor deployment from the inference methods applied to the resulting observations.

\subsection{Sensor Deployment Strategies}

\begin{figure}[htp]
    \centering
    \includegraphics[width=0.6\linewidth]{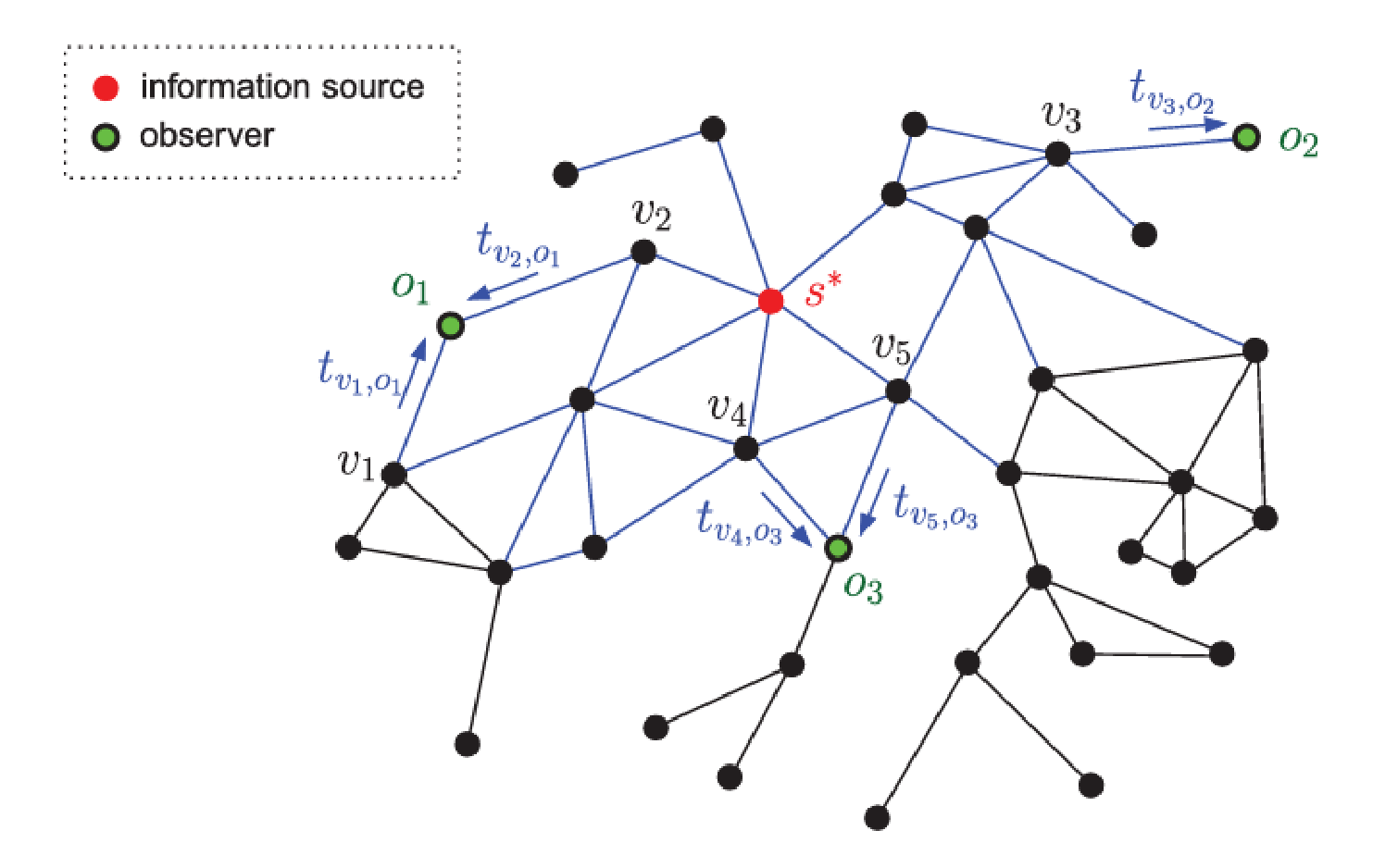}
    \caption{\rev{Illustration of sensor-based source detection on an 
    arbitrary network $G$~\cite{pinto2012locating}. At the unknown time $t = t^*$, 
    the information source $s^*$ initiates the diffusion. The blue 
    edges denote those over which information has already propagated. 
    In this example, there are three sensors, which record from 
    which neighbors and at what time they received the information. 
    The goal is to estimate, from these sensor observations, which 
    node in $\mathcal{G}$ is the information source. 
}}

    \label{fig:sensor}
\end{figure}

With a monitor budget, placement decides which parts of a propagation process can be recorded. Figure~\ref{fig:sensor} illustrates how local observations constrain a candidate origin. The information depends on both monitor locations and recorded variables; the same node set with timestamps or only final binary states constitutes two different experiments.

The sensor deployment problem can be formulated as the optimization of a sensor set under a sensing-budget constraint:
\begin{equation}
V_{\mathrm{sen}}^{*}
=
\operatorname*{argmax}_{V_{\mathrm{sen}}\subseteq V,\;|V_{\mathrm{sen}}|=K}
f(V_{\mathrm{sen}}),
\end{equation}
where \(K\) is the sensor budget and \(f(\cdot)\) denotes the deployment objective. Different choices of \(f(\cdot)\) correspond to different deployment principles, including maximizing network coverage, minimizing detection latency, increasing information gain, or enhancing the distinguishability of candidate sources.

Random placement is a reproducible baseline requiring little optimization. It can distribute monitoring across the network but does not explicitly target coverage, path separation, or source uncertainty. Comparisons should use repeated random placements under the same budget, because placement variability can be substantial~\cite{paluch2020optimizing}.

\subsubsection{Topology-aware Deployment}

Topology-based sensor placement often begins by ranking nodes using local structural measures. Degree centrality prioritizes nodes with many direct connections, whereas the clustering coefficient describes how closely connected their neighbors are~\cite{pinto2012locating,zhang2016identification}. Both measures can guide sensor selection at relatively low computational cost. However, selecting sensors according to individual node scores does not account for overlap in the information they collect. Highly ranked nodes may occupy similar neighborhoods, resulting in redundant coverage rather than a more informative set of observations.

Global distance measures provide another criterion for sensor placement. Betweenness centrality favors nodes traversed by many shortest paths, which may be useful for monitoring transmission between different network regions~\cite{fu2016multi}. Closeness centrality instead prioritizes nodes with small average distances to the rest of the network~\cite{xu2019identifying}. These measures describe different aspects of network accessibility, but neither directly quantifies the information collected about an unknown source. Their effectiveness depends on the extent to which shortest-path structure reflects the routes followed by diffusion.

Other structural measures characterize node importance through the organization of the network. Eigenvector centrality assigns greater weight to nodes connected to highly central neighbors, while K-shell decomposition identifies nodes according to their position in the network's core hierarchy~\cite{zhang2016identification,xu2019identifying}. Both can favor structurally well-connected regions, but their rankings need not coincide. A deployment concentrated in these regions may leave peripheral outbreaks poorly monitored, depending on the network structure and the location of the initiating source.

Collective betweenness centrality is a representative example~\cite{paluch2020optimizing}. Rather than evaluating the betweenness of individual nodes, it quantifies the fraction of topological shortest paths covered by the entire sensor set. The corresponding deployment objective is formulated as
\begin{equation}
V_{\mathrm{sen}}^*
=
\operatorname*{argmax}_{V_{\mathrm{sen}}\subseteq V,\ |V_{\mathrm{sen}}|=K}
\sum_{\substack{v_i,v_j\in V\\ v_i,v_j\notin V_{\mathrm{sen}}}}
\frac{\sigma_{ij}^{(V_{\mathrm{sen}})}}{\sigma_{ij}},
\end{equation}
where \(\sigma_{ij}^{(V_{\mathrm{sen}})}\) denotes the number of shortest paths between \(v_i\) and \(v_j\) intercepted by at least one sensor, and \(\sigma_{ij}\) is the total number of such paths. A path contributes only once to the collective score, regardless of how many selected sensors lie along it. The objective therefore accounts for overlap in path coverage, which is ignored when individual betweenness scores are simply added. Its computation, however, requires information about shortest paths across the network and can become expensive for large candidate sets.

Sensor spacing offers another way to address redundant placement. Seo et al.~\cite{seo2012identifying} introduced shortest-path distance constraints to prevent selected sensors from being concentrated in the same region. Their method combines these constraints with node in-degree, balancing structural importance against the separation between monitoring locations. The resulting placement is intended to provide a wider distribution of observations, although its value for source detection still depends on the diffusion process and the information recorded by the sensors.

Coverage-based deployment aims to maximize the portion of the network monitored by a limited number of sensors. Sensors are placed so that as many nodes as possible fall within their observation range. Although neighborhood coverage provides a useful criterion for sensor placement, greater coverage does not necessarily improve source identification. The information available for source inference also depends on the diffusion dynamics and the measurements collected.

The high-coverage strategy is a representative example, which seeks the sensor set covering the largest number of neighboring nodes. Let the sensor set be \(V_{\mathrm{sen}}=\{o_1,o_2,\ldots,o_K\}\)~\cite{zhang2016identification}. Its network coverage is defined as
\begin{equation}
C_s(V_{\mathrm{sen}})
=
\frac{
\left|
\displaystyle\bigcup_{i=1}^{K}\mathcal{N}(o_i)
\right|
}{N},
\end{equation}
where $\mathcal{N}(o_i)$ denotes the neighborhood of sensor node $o_i$. The corresponding deployment problem is formulated as
\begin{equation}
V_{\mathrm{sen}}^{*}
=
\operatorname*{argmax}_{V_{\mathrm{sen}}\subseteq V,\;|V_{\mathrm{sen}}|=K}
C_s(V_{\mathrm{sen}}).
\end{equation}

Neighborhood coverage measures access to nodes near monitors. It can miss the importance of intercommunity paths or source-separating timing patterns. Its relationship to detection accuracy therefore depends on the network and spreading process and should be evaluated rather than inferred from coverage alone.

The \(k\)-median strategy takes a distance-based approach to sensor placement~\cite{paluch2020optimizing}. It minimizes the total distance from network nodes to their nearest selected sensors, rather than maximizing the number of nodes covered within one hop. This objective favors spatial accessibility, but a small average distance to sensors does not necessarily imply that different sources can be distinguished from their observations. The optimization can also be computationally demanding on large networks.

\subsubsection{Optimization-driven Deployment}

Early-warning placement minimizes delay to the first recorded event. Such an objective can be useful operationally, but identifying an outbreak and identifying its initiating node are different stopping problems. A source-detection-oriented stopping rule must also specify when the accumulated evidence is sufficient for the desired source decision.

\begin{figure}[htp]
    \centering
    \includegraphics[width=0.7\linewidth]{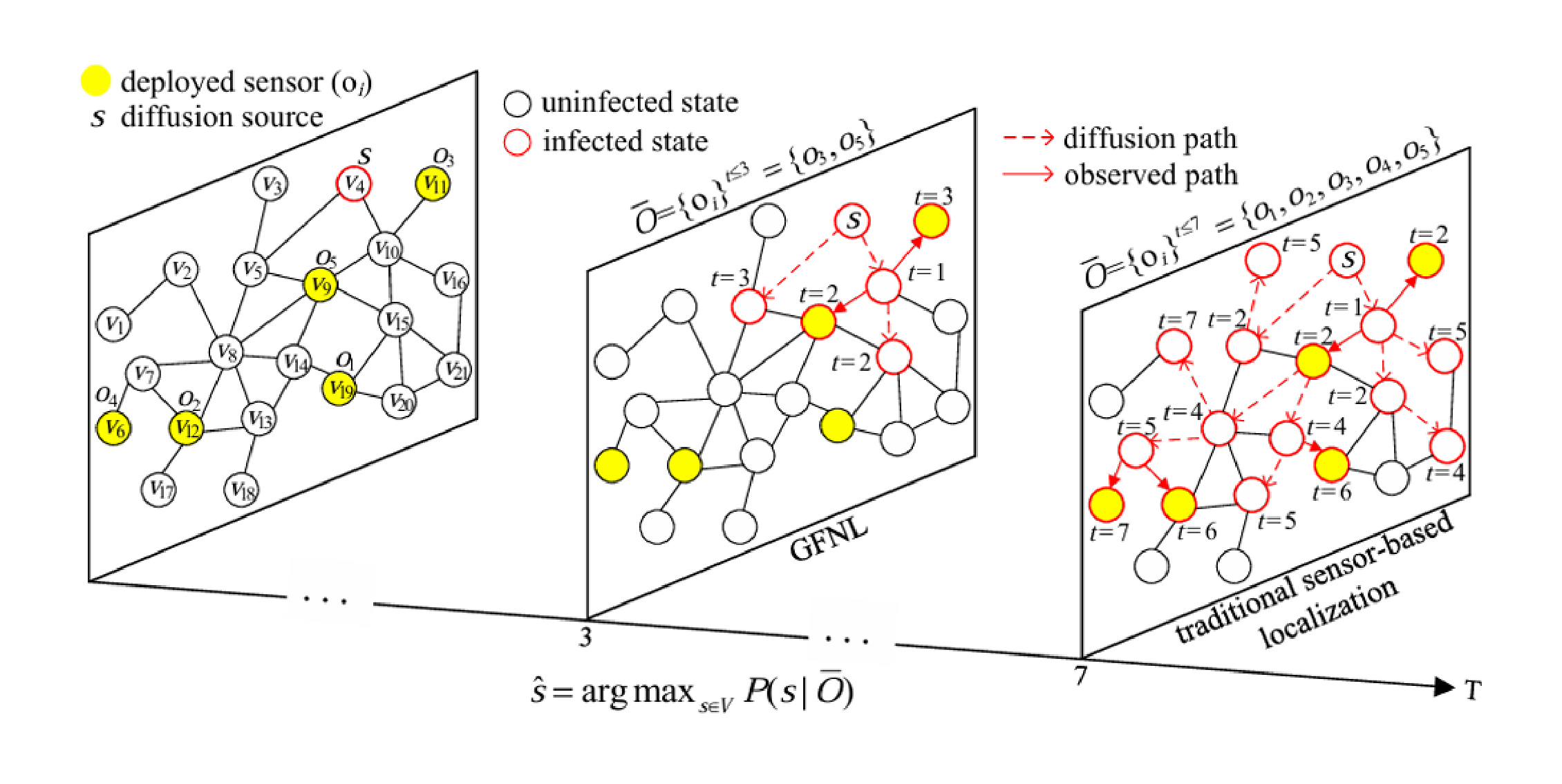}
\caption{\rev{
Comparison between GFNL and conventional sensor-based source detection~\cite{wang2022rapid}. 
The left panel shows the deployment of sensor nodes before diffusion begins. 
The middle panel illustrates the GFNL strategy, where source detection is initiated from sparse early sensor observations, allowing the candidate-source region to be narrowed before large-scale propagation occurs. 
The right panel depicts later inference from a larger accumulated set of sensor reports in this schematic comparison.
}}
   
    \label{fig:gfnl}
\end{figure}

Greedy full-order neighbor localization (GFNL) combines a greedy deployment strategy (DS) with an estimation strategy (ES)~\cite{wang2022rapid}. DS uses full-order neighborhood information to select monitors whose distance-based coverage complements the existing set; candidate gains are updated as monitors are added. This is a topology-based acquisition strategy. Figure~\ref{fig:gfnl} links the deployment stage with estimation from the monitors reached by the process. Its objective and the ES candidate score are distinct, and the method's early-stage experiments do not turn early alarm into a general guarantee of early and exact source detection.

Full-order neighborhood information has also been used to guide subsequent deployment strategies. FONC (Full-Order Neighbor Coverage) selects sensors greedily according to the coverage gained across different neighborhood orders~\cite{wang2020locating}. Hou et al.~\cite{hou2024random} considered a randomized strategy that combines hub nodes, leaf nodes, and sampled neighborhoods to diversify the selected monitoring locations. These approaches seek to improve the use of a limited sensor budget, although their deployment objectives and computational costs differ.

Diffusion-aware strategies incorporate spreading dynamics into sensor deployment. Cheng et al.~\cite{cheng2025efficient} proposed the Propagation-aware Sensor Deployment Module (SDM), which uses Monte Carlo simulations to evaluate candidate sensor locations and introduces a spreading-distance threshold to limit the influence of distant transmission events. Qiu et al.~\cite{qiu2022bpsl} adopted the IMM (Influence Maximization Martingale) algorithm to estimate propagation-related information through reverse reachable sets and guide sensor selection. Although both methods use diffusion information to inform deployment, their selection criteria rely on different approximations and assumptions about the spreading process.

Some studies further optimize how quickly spreading can be detected. Zhou et al.~\cite{ZHOU2018304} formulated sensor deployment as a combinatorial optimization problem that minimizes the expected detection time and reformulated it equivalently as the maximization of the expected remaining response time. The corresponding objective function is
\begin{equation}
f(V_{\mathrm{sen}}) = \sum_{v\in V} \pi(v) \, \mathbb{E}\left[ T_{\max} - \tau(v,V_{\mathrm{sen}}) \right],
\end{equation}
where $\pi(v)$ denotes the prior probability that node $v$ is the source, and $\tau(v,V_{\mathrm{sen}})$ is the random time at which the spreading process is first detected by the sensor set.  The finite monitoring horizon also requires a convention for outbreaks that do not reach any sensor before \(T_{\max}\). Their treatment determines the contribution of undetected realizations to the expected response time and must be consistent with the optimization model. By exploiting the submodularity of the objective function, the method admits an efficient greedy optimization algorithm, making the optimization scalable to larger networks~\cite{nemhauser1978submodular}.

Maximum Entropy Sampling (MES) is a representative information-theoretic deployment strategy~\cite{hu2024source}. It formulates sensor placement by maximizing the joint entropy of the selected observations, thereby preserving as much information about the network state as possible. 
Let $V = \{v_1, v_2, \dots, v_N\}$ denote the network node set. At a given observation time \(T\), let \(\boldsymbol{X}(T)=(X_1(T),X_2(T),\dots,X_N(T))\) denote the network-state vector, where \(X_i(T)\) is the state of node \(v_i\).

The uncertainty associated with the network-state distribution is quantified by the joint entropy

\begin{equation}
H(V) = -\sum_{\boldsymbol{X}(T)} p(\boldsymbol{X}(T)) \log_2 p(\boldsymbol{X}(T)),
\end{equation}
where \(p(\boldsymbol{X}(T))\) is the joint probability distribution of the network-state vector \(\boldsymbol{X}(T)\), and the summation extends over all possible realizations of \(\boldsymbol{X}(T)\). Let \(V_{\mathrm{sen}}\subseteq V\) denote the sensor set. The corresponding deployment objective under a fixed sensing budget is given by

\begin{equation}
V_{\mathrm{sen}}^* = \operatorname*{argmin}_{V_{\mathrm{sen}}\subseteq V,\ |V_{\mathrm{sen}}|=K} H(V\setminus V_{\mathrm{sen}} \mid V_{\mathrm{sen}}),
\end{equation}
By exploiting the chain rule of entropy, the above objective is equivalent to maximizing the entropy of the selected sensor set:
\begin{equation}
V_{\mathrm{sen}}^* = \operatorname*{argmax}_{V_{\mathrm{sen}}\subseteq V,\ |V_{\mathrm{sen}}|=K} H(V_{\mathrm{sen}}).
\end{equation}
 
Because maximizing the joint entropy is computationally intractable, MES constructs the sensor set using a greedy iterative strategy. At each iteration, the next sensor is selected by maximizing the conditional entropy with respect to the sensors that have already been deployed:

\begin{equation}
o_t = \operatorname*{argmax}_{v\in V\setminus V_{\mathrm{sen}}^{t-1}} H(v \mid V_{\mathrm{sen}}^{t-1}).
\end{equation}

To reduce the cost of evaluating conditional entropy, MES uses a tree-based approximation of diffusion dependencies, constructed through shortest-path calculations~\cite{hu2024source}. This simplifies the estimation of dependencies among node states during greedy sensor selection. The approximation is computationally useful, but its accuracy depends on how closely the selected tree represents the propagation relationships relevant to the observations.

MES selects informative measurements of the network state. In the entropy notation above, a node set denotes the random states recorded on that set; the two placement objectives are equivalent by the chain rule when the full-state entropy and sensor budget are fixed. This does not generally maximize information about the source. The latter is \(I(S;Q)=H(Q)-H(Q\mid S)\), and the second term can change with placement. State-reconstruction and source-detection objectives should therefore be assessed separately, even when the same greedy machinery is used.

Hu et al.~\cite{hu2019locating} proposed a minimum-sensor deployment strategy based on source distinguishability. The key idea is that different source nodes should produce distinguishable propagation-delay patterns at the deployed sensors. To quantify this property, the method first computes the propagation-delay matrix \(D^{\mathrm{delay}}\) and defines the pairwise propagation-delay difference with respect to a candidate source as
\begin{equation}
\Delta_{(o_a,o_b)}(v)
=
D^{\mathrm{delay}}_{o_a,v}
-
D^{\mathrm{delay}}_{o_b,v}.
\end{equation}
When the initiation time is unknown, candidate sources with identical delay-difference signatures cannot be distinguished from relative sensor arrival times alone. Based on this criterion, a propagation-delay difference matrix is constructed to identify indistinguishable source pairs, and sensors are progressively selected so as to maximize their separation. The deployment objective is formulated as
\begin{equation}
\min |V_{\mathrm{sen}}|,
\qquad
\text{s.t. }
V_{\mathrm{ud}}(V_{\mathrm{sen}})
=
\varnothing,
\end{equation}
 where \(V_{\mathrm{ud}}(V_{\mathrm{sen}})\) denotes the set of candidate sources whose delay-difference signatures coincide with those of at least one other candidate. The constraint requires every pair of distinct admissible sources to have different signatures under the assumed propagation-delay model. The authors use greedy search and a simulated-annealing-like procedure to construct sensor sets satisfying this criterion. Distinguishable signatures provide a structural condition for source separation, but do not by themselves guarantee accurate detection when transmission delays or recorded times are uncertain.

Spinelli et al.~\cite{spinelli2016observer} also consider placement criteria that depend on the assumed variability of propagation delays. At low variability, distinguishing source equivalence classes is a useful design objective; with greater timing uncertainty, constrained path coverage provides a more robust surrogate. Adaptation here concerns selection of the placement criterion for a noise regime. It should not be confused with online relocation or sequential acquisition in response to newly observed outbreak records.

Sensor placement must also account for network structures beyond static pairwise connections. Hypergraphs represent group interactions, multilayer networks distinguish different types of relationships, and temporal networks restrict when particular connections are available~\cite{ke2026source,cheng2025hyperdet,ke2026entropy,HU2026131748,kivela2014multilayer,7393814}. Uncertain networks introduce additional ambiguity about the routes that diffusion may follow~\cite{8299476}. These features can change the information collected at a given monitoring location, so deployment criteria developed for static simple networks may need to be adapted to the network and diffusion model under consideration.

\begin{figure}[htp]
    \centering
    \includegraphics[width=0.4\linewidth]{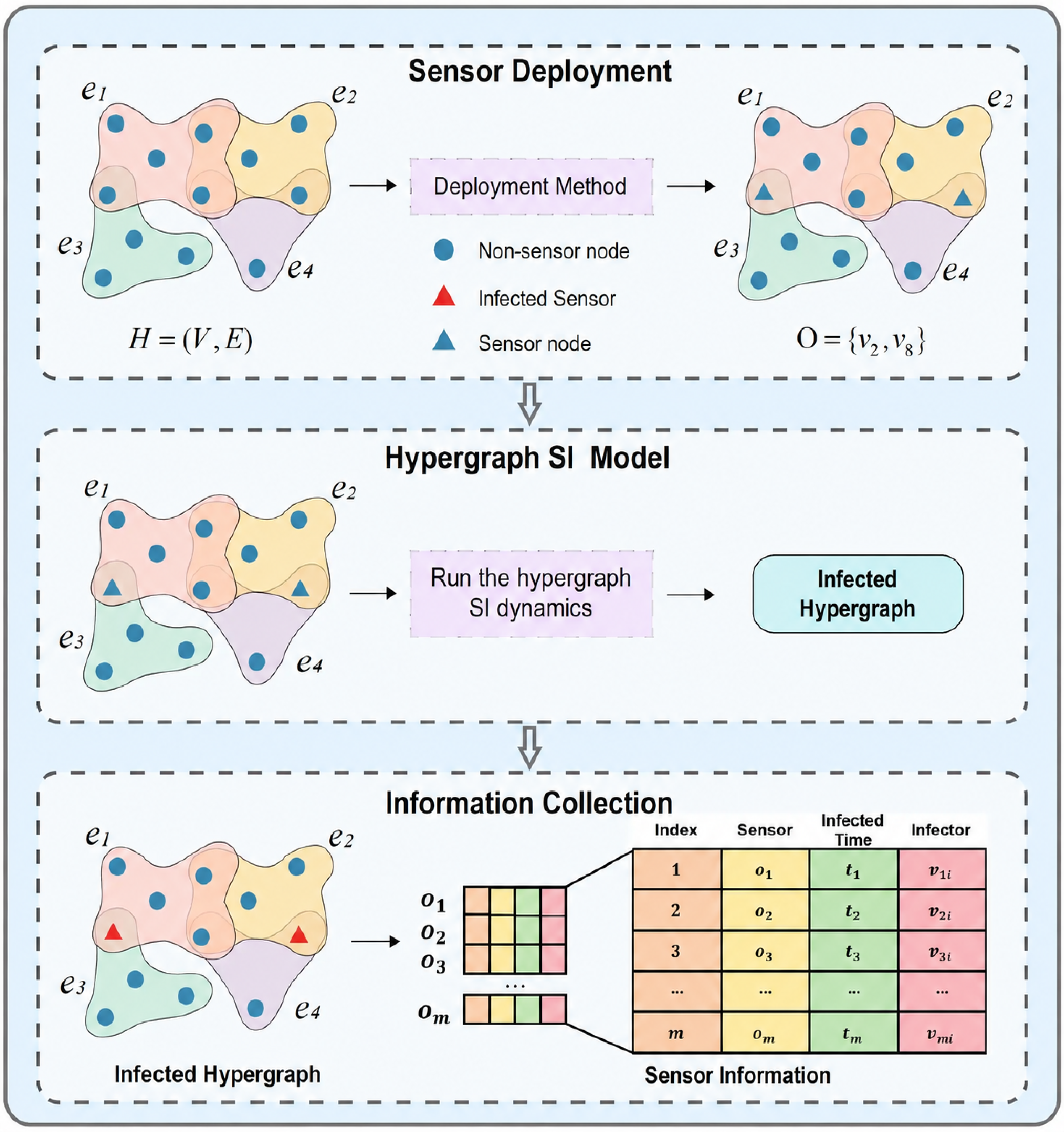}
    \caption{
    Illustration of sensor deployment and information collection on a hypergraph~\cite{ke2026entropy}. Sensors are first deployed on the hypergraph, after which diffusion is simulated and the observations recorded by the deployed sensors, including infection times and infecting neighbors, are collected for subsequent source detection.
    }

    \label{fig:SSDH}
\end{figure}

Taking higher-order networks as an example, Ke et al.~\cite{ke2026entropy} extended sensor deployment to hypergraph diffusion and proposed the entropy-driven framework SSDH (Sensor-based Source Detection in Hypergraphs). As illustrated in Fig.~\ref{fig:SSDH}, the deployment stage selects sensor nodes by maximizing a logarithmic coverage-based deployment score. To capture the higher-order interactions encoded in hyperedges, the deployment objective incorporates the diminishing marginal coverage gain associated with each additional sensor and defines its incremental contribution as
\begin{equation}
\Delta C(e)
=
\log(K_e+2)-\log(K_e+1),
\end{equation}
 where \(K_e\) is the number of sensors already selected in hyperedge \(e\). The logarithmic form assigns a smaller marginal gain to additional sensors in hyperedges that are already monitored. Sensors are selected greedily according to their contributions to the overall coverage score, encouraging deployment across different hyperedges while accounting for their overlap.
 
\begin{figure}[htp]
    \centering
    \includegraphics[width=1\linewidth]{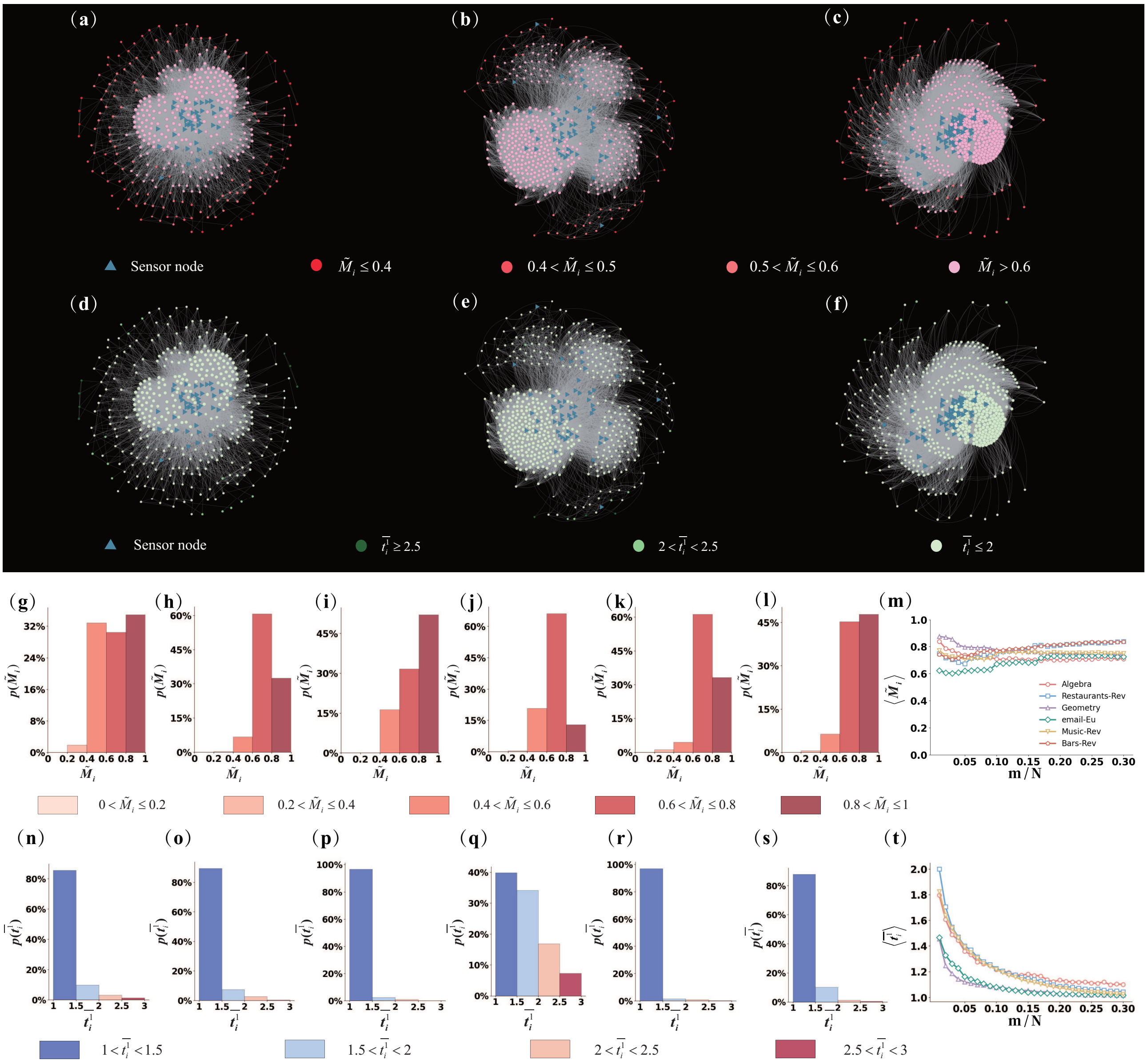}
\caption{
Sensor monitoring coverage on representative empirical hypergraphs~\cite{ke2026entropy}. 
(a)--(f) Spatial distributions of deployed sensors and their monitoring coverage on empirical hypergraphs (Algebra, Restaurants-Rev, and Geometry). 
(g)--(t) Statistical summaries of node monitoring intensity, first-sensor infection time, and their variation with sensor density, illustrating the observation quality achieved by the deployment strategy.
}
    
    \label{fig:SSDH_result}
\end{figure}

Figure~\ref{fig:SSDH_result} evaluates observation intensity through cumulative inverse-distance measures and the time to the first sensor record. The displayed coverage and arrival-time distributions describe acquisition performance at the tested sensing ratios. They do not directly establish the accuracy of an unknown-source estimator or equality between the deployment score and source mutual information.

Hu et al.~\cite{9807439} study observability-based placement for a specified linear multilayer diffusion system. PBH rank tests and spectral properties of its state matrix characterize the associated linear observation problem. The relevant conditions involve that matrix and the allowed measurement structure; they are not a universal sensor-count theorem for stochastic epidemics. Other network-specific designs address community, uncertain, signed, or temporal structure~\cite{8299476,zhao2025enhanced,Zhi-WeiMa:28902,HUANG2017434}. Their outputs should be compared against the objective each design actually optimizes.

\subsection{Source Detection Mechanisms}
\label{sec:sensor_mechanisms}

Once sensors have been deployed, source inference depends on the information they record. Some methods use infection arrival times to constrain candidate origins, while others combine temporal observations with transmission directions or node states. Source detection can also rely on repeated state measurements or forwarding records when infection timestamps are unavailable. These observations provide different constraints on the spreading process and require corresponding inference methods.

\subsubsection{Temporal-Information-based Methods}

Likelihood-based methods model the probability mass or density of sensor measurements conditional on a candidate source. Maximizing this likelihood ranks model compatibility; incorporating a source prior instead yields posterior-based inference. Li et al.~\cite{li2021sensor} provide a concrete sensor-based example on human mobility networks. Their model treats the vector of first-arrival times as a source-dependent multivariate Gaussian component, with means and covariances calibrated through stochastic epidemic simulations and smoothing. Combining these components with a source prior permits both a point estimate and a posterior region. Here the monitored nodes represent communities, and a first arrival denotes attainment of a specified incidence level. The Gaussian sensor likelihood is therefore tied to a disease and measurement model, illustrating how simulation-based calibration can support an explicit observation law.

GMLA (Gradient Maximum Likelihood Algorithm) reduces the cost of PTVA-LI (Pinto--Thiran--Vetterli Algorithm with Limited Information) by selecting observer records and locally searching candidate nodes~\cite{paluch2018fast}. Let \(K\) be the number of deployed sensors and \(K_{\mathrm{avail}}\) the number with available arrival-time records. GMLA retains the \(K_0\) earliest records, with \(2\leq K_0\leq K_{\mathrm{avail}}\leq K\) for the calculation below. This selection is made once, and the same retained sensors are used for every candidate comparison. Here ``nearest'' means earliest informed, rather than a known distance from the unknown source. The method evaluates a Gaussian timing approximation on candidate-rooted BFS trees using known edge-delay means and variances.

Choose any retained observer as the reference and label it \(o_{K_0}\), with the other retained observers labeled \(o_1,\ldots,o_{K_0-1}\). Denote the earliest retained observer by \(o_{\min}\). For \(i=1,\ldots,K_0-1\), define \(d_i=\tau_{o_i}-\tau_{o_{K_0}}\). Subtracting the reference arrival removes a common additive initiation time, but the resulting differences share that reference and are generally correlated. Thus \(\boldsymbol d\) and its candidate-dependent mean must have \(K_0-1\) components, with a covariance matrix for this same relative-time vector.

For this fixed record set, a general Gaussian timing likelihood approximation, up to a factor common to candidates, is

\begin{equation}
\mathcal{L}(v) = \frac{\exp\!\left(-\frac12 (\boldsymbol d - \boldsymbol\mu_v)^T \Lambda_v^{-1} (\boldsymbol d - \boldsymbol\mu_v)\right)}{\sqrt{\det(\Lambda_v)}},
\end{equation}
where \(\boldsymbol\mu_v\in\mathbb{R}^{K_0-1}\) is the candidate-source conditional mean of \(\boldsymbol d\), and \(\Lambda_v\in\mathbb{R}^{(K_0-1)\times(K_0-1)}\) is its covariance matrix. In Paluch et al.'s construction, the mean components are the edge-delay mean multiplied by differences of candidate-to-observer BFS path lengths; covariance entries use shared edges on reference-to-observer paths in that candidate's tree. The displayed form requires \(\Lambda_v\) to be positive definite; singular cases lie outside it. The determinant must be retained when covariance varies across candidates.

\begin{figure}[htp]
    \centering
    \includegraphics[width=0.55\linewidth]{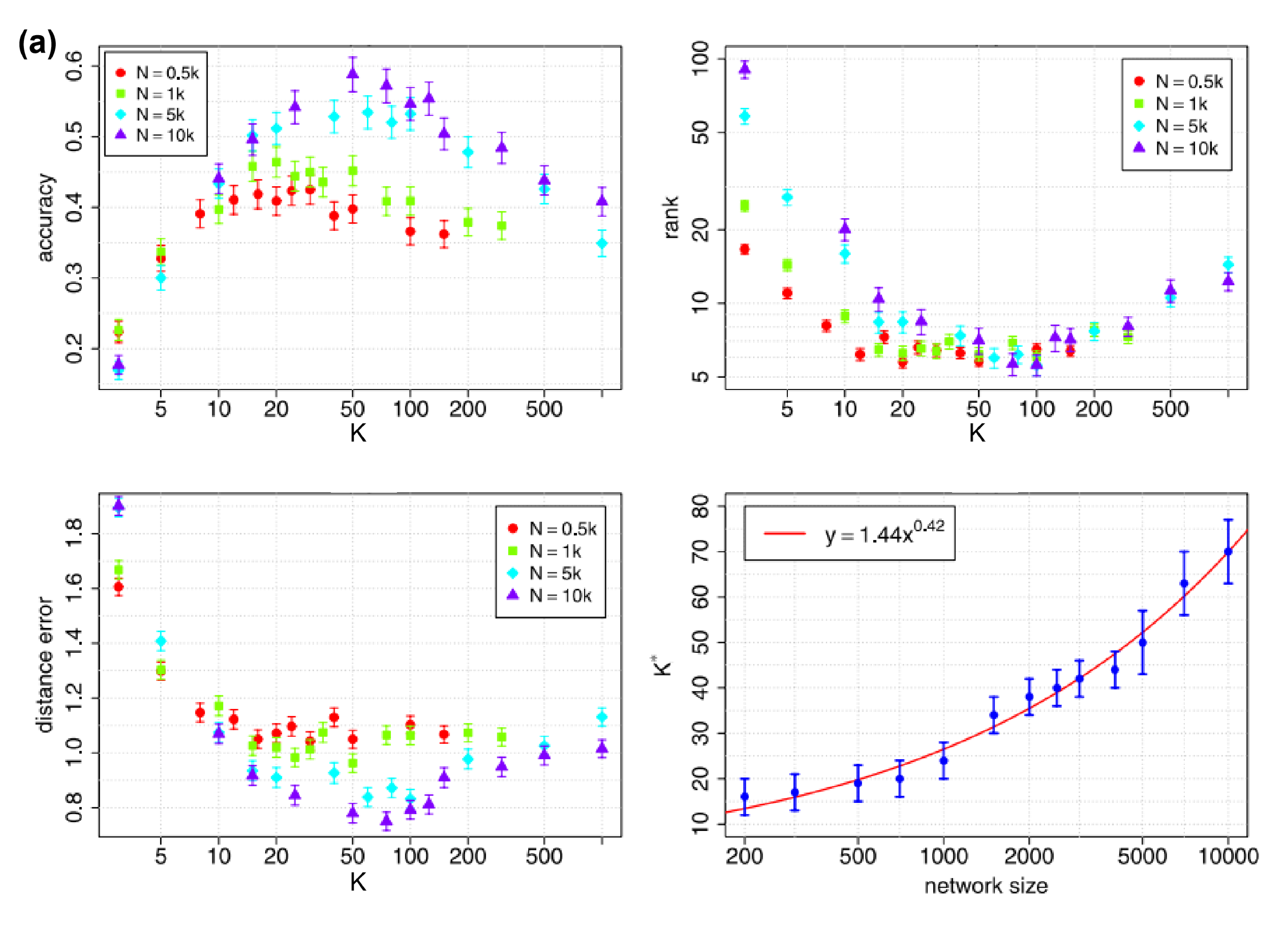}
    \includegraphics[width=0.55\linewidth]{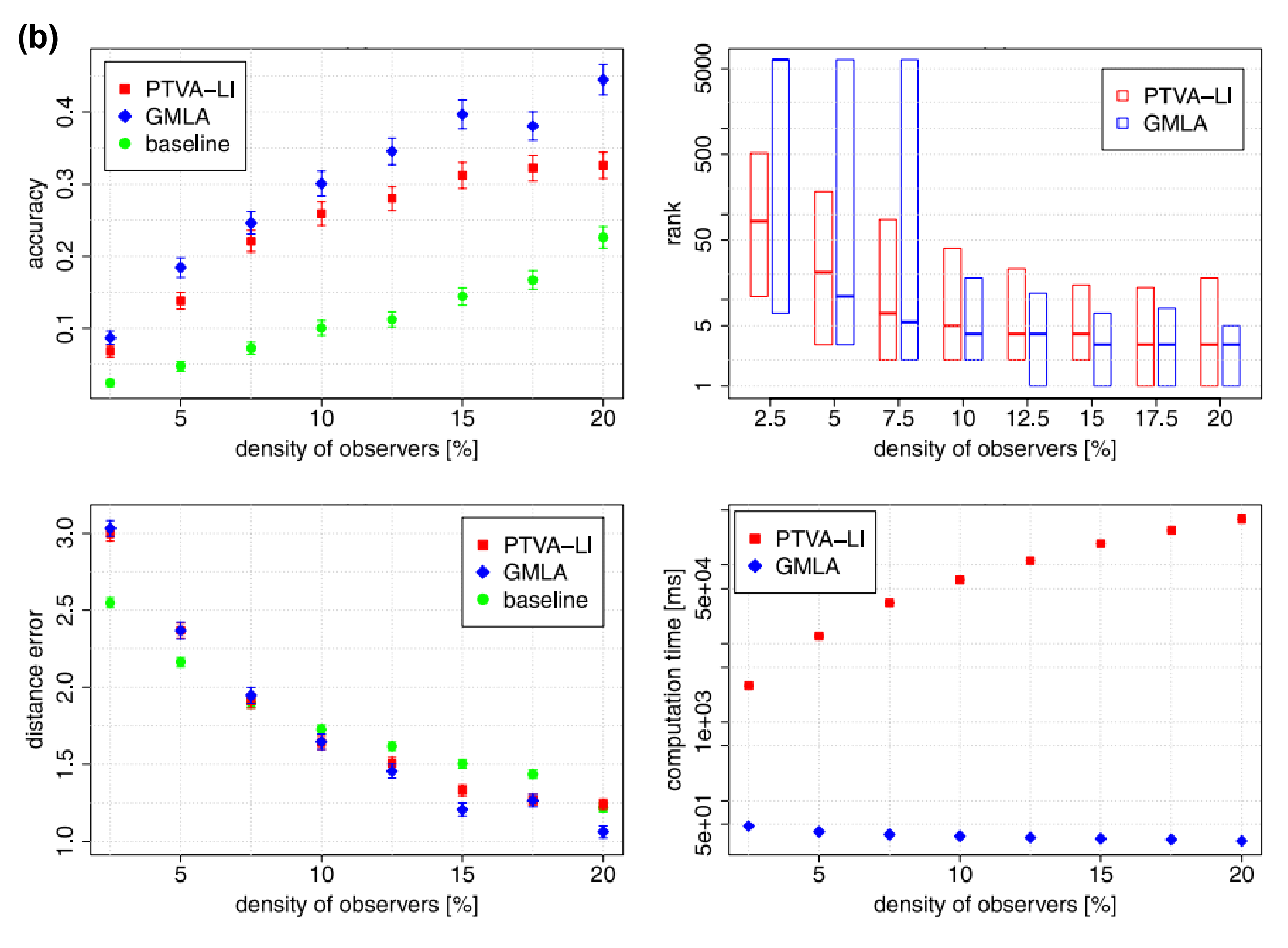}
    \caption{
\rev{Sensitivity of GMLA to observation conditions~\cite{paluch2018fast}. (a) Accuracy, true-source rank and distance error versus the retained earliest-observer count on Barab{\'a}si--Albert (BA) networks of varying sizes, together with the estimated optimal retained count versus network size. The reproduced axes \(K\) and \(K^*\) correspond to \(K_0\) and \(K_0^*\) in the original study and the present text; they denote selected records, rather than the total deployed budget. (b) GMLA and PTVA-LI on Gnutella (\(N=6299\)) as deployment density \(\rho=K/N\) varies, with GMLA's retained count fixed at \(K_0=30\). The panels report accuracy, true-source rank, distance error and computation time.}
}
    
    \label{fig:gmla}
\end{figure}

GMLA begins by scoring neighbors of the earliest observer \(o_{\min}\), then follows the highest-scoring neighboring candidates under its gradient-like search rule. Computed scores are stored to avoid repeated evaluation and to rank the tested nodes. This reduces the evaluated candidate set, but does not guarantee that it contains the true source or attains the global maximum over the network. Performance therefore depends on the retained count \(K_0\) and the observation conditions, as shown in Fig.~\ref{fig:gmla}.

The accuracy of timing-based likelihoods also depends on how propagation paths are represented. Gajewski et al.~\cite{GAJEWSKI201934} show why a single candidate-rooted BFS tree can overestimate first-arrival times on networks with competing routes: the first arrival is the minimum of several path travel times, and shared edges correlate these alternatives. Their equiprobable-path and equiprobable-link variants approximate the resulting means and covariances while accounting for multiple shortest paths. This provides an alternative to a single-tree approximation, with additional computational cost and assumptions about path overlap.

GLAD estimates nuisance timing quantities together with the source instead of assuming all delay parameters are known~\cite{app9183758}. Candidate nodes induce different expected path delays; fitting an onset time removes a common shift, while scale estimation adjusts the temporal residuals. Its modified form uses candidate-rooted BFS paths and a simplified covariance structure to reduce computation. These approximations exchange dependence information for tractability. The candidate search must state whether monitored nodes can themselves be sources; excluding them is an admissibility assumption, not a general consequence of observing their arrival times.

Further developments in likelihood-based source detection have addressed both propagation-delay modeling and candidate search. PESL (Pascal-based Epidemic Source Localization) uses a Pascal distribution to describe transmission delays, with particular attention to settings involving low transmission probabilities~\cite{li2019locating}. Spinelli et al.~\cite{spinelli2016observer} examine how the variability of propagation delays affects source inference, using source equivalence classes under low delay variability and truncated Gaussian approximations when delays are more uncertain. Another line of research seeks to reduce the cost of candidate evaluation. 
Louni et al.~\cite{8299476} employ a hierarchical search that first identifies a candidate source community and then locates the source within that community. This strategy reduces the candidate space for subsequent likelihood evaluation.

\begin{figure}[htp]
    \centering
    \includegraphics[width=0.8\linewidth]{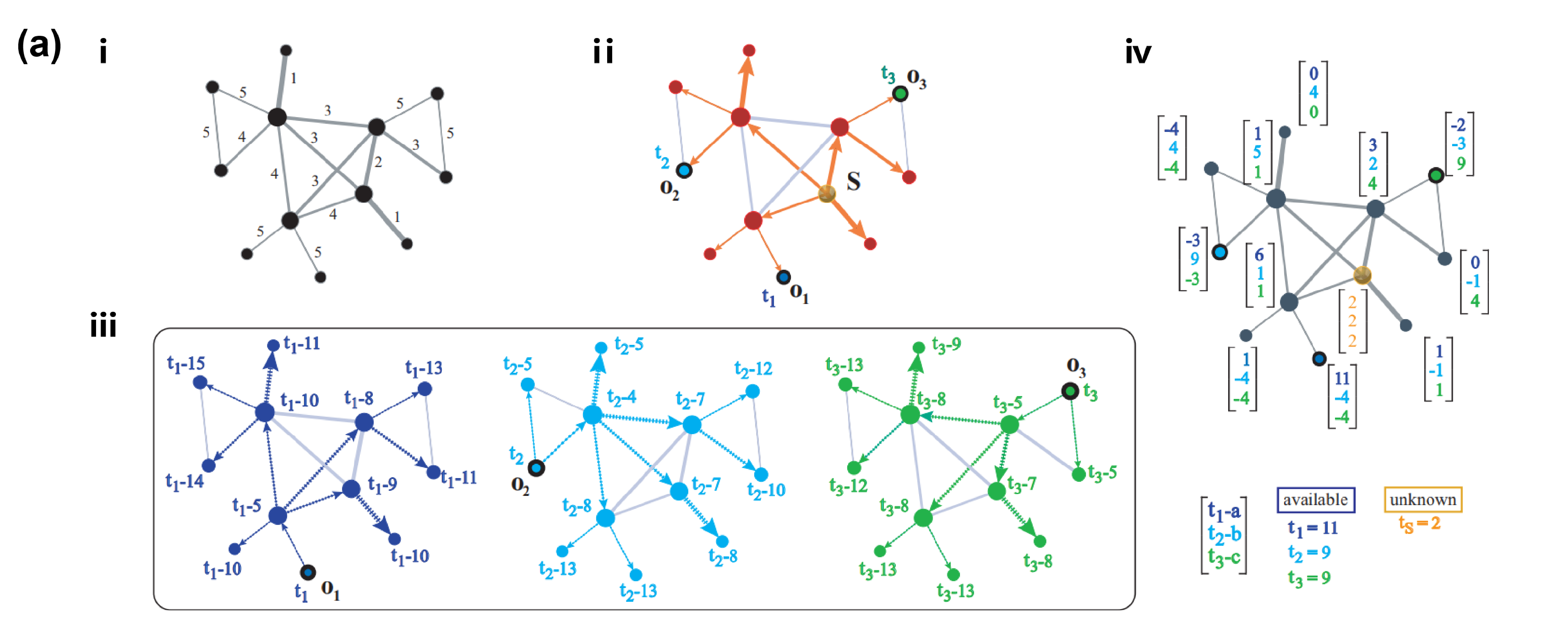}
    \includegraphics[width=0.8\linewidth]{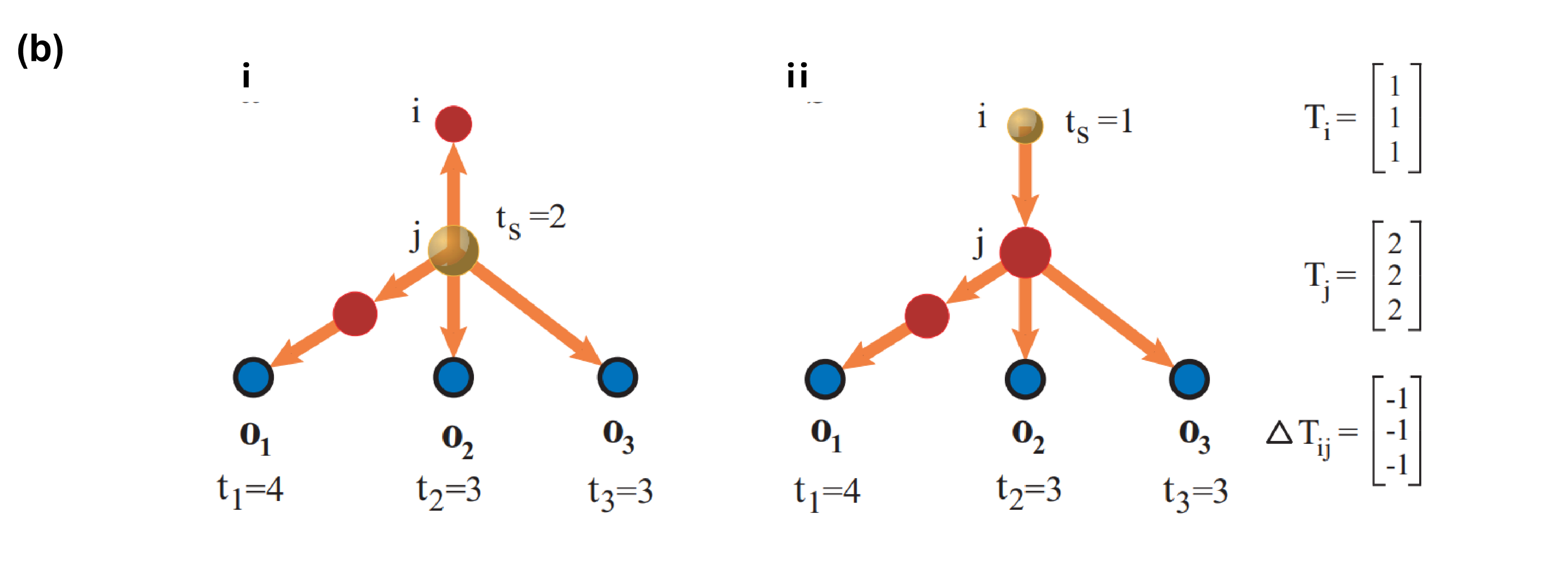}
    \caption{
Illustration of the TRBS method and its identifiability condition~\cite{shen2016locating}. 
(a) Source detection based on reverse propagation from sensor observations. Arrival times recorded by the sensors are projected back through weighted shortest paths to construct candidate triggering-time signatures for source evaluation.
(b) Identifiability limitation of sensor-based detection. Different candidate sources can produce indistinguishable sensor observations, resulting in triggering-time signatures that differ only by a constant shift and therefore cannot be uniquely separated.
}
   
    \label{fig:trbs}
\end{figure}

The Time-Reversal Backward Spreading (TRBS) method is a representative time-backtracking approach~\cite{shen2016locating}. For each candidate source, TRBS combines the arrival time recorded by every sensor with the corresponding shortest propagation delay to estimate the diffusion starting time implied by that sensor. These estimates are organized into an initial triggering time matrix \(T_{\mathrm{initial}}\), in which each row corresponds to a sensor and each column corresponds to a candidate node, i.e.,
\begin{equation}
\left(T_{\mathrm{initial}}\right)_{ij}
=
t_{o_i}
-
\tau_{v_j,o_i},
\label{eq:T_initial}
\end{equation}
where \(t_{o_i}\) is the arrival time observed at sensor \(o_i\), and \(\tau_{v_j,o_i}\) denotes the shortest propagation delay from candidate node \(v_j\) to sensor \(o_i\). For the true source, the backtracked starting times inferred from different sensors are expected to be nearly identical (Fig.~\ref{fig:trbs}(a)). TRBS therefore evaluates the temporal consistency of these estimates and identifies the node with the minimum variance as the detected source:
\begin{equation}
\hat v
=
\operatorname*{argmin}_{v_j \in V}
\operatorname{Var}_{o_i \in V_\mathrm{sen}}
\left(T_{\mathrm{initial}}\right)_{ij}.
\label{eq:TRBS_source}
\end{equation}

TRBS identifies candidate sources by comparing the consistency of the initiation times inferred from different sensors. When several candidates yield the same minimum variance, the available records may not distinguish a unique origin. As illustrated in Fig.~\ref{fig:trbs}(b), two candidates whose propagation-delay vectors differ by a common offset produce identical relative arrival times when the diffusion onset is unknown. This ambiguity arises because the observations constrain differences in travel times rather than their absolute values. If the initiation time is known, the backtracked estimates can instead be compared directly with that time. Under stochastic propagation delays, differences among the estimates also reflect variability in transmission times, which must be considered when interpreting the resulting source ranking.

MSPT (Multi Snowballing with Partial Timestamps) uses timestamps recorded at a subset of monitored nodes to identify multiple diffusion sources~\cite{devarapalli2024locating}. It determines a snowball radius from the available timestamps to narrow the search for potential sources. Beginning with the monitor that records the earliest arrival, MSPT searches the resulting candidate regions for likely source locations. This reduces the number of nodes considered during source detection.

Simulation-based approaches match observed temporal records to delay samples generated under a specified spreading model. Shi et al.~\cite{shi2022source} proposed the Gaussian Weighted Averaging Correlation Coefficient (GWACC) method for continuous-time diffusion with partially observed activation times. Monte Carlo simulations generate candidate-to-observer transmission delays, and Gaussian-weighted correlation measurements are used to estimate the initiating node and the diffusion start time. This combines a simulation model with statistical comparison of temporal patterns.

Let \(o_1,\ldots,o_m\) be the sensors with recorded activation times, where \(m\leq K\) and \(K\) remains the deployed sensor budget. Write \(\boldsymbol t=(t_{o_1},\ldots,t_{o_m})^T\) and \(\boldsymbol d_v^j=(d^j(v,o_1),\ldots,d^j(v,o_m))^T\) for the observed times and the simulated delay vector from admissible candidate node \(v\) in simulation \(j\). With \(\bar t=m^{-1}\sum_{i=1}^{m}t_{o_i}\) and \(\bar d_v^j=m^{-1}\sum_{i=1}^{m}d^j(v,o_i)\), their Pearson coefficient is
\begin{equation}
r(\boldsymbol t,\boldsymbol d_v^j)=
\frac{\sum_{i=1}^{m}(t_{o_i}-\bar t)(d^j(v,o_i)-\bar d_v^j)}
{\sqrt{\sum_{i=1}^{m}(t_{o_i}-\bar t)^2\sum_{i=1}^{m}(d^j(v,o_i)-\bar d_v^j)^2}}.
\end{equation}
The coefficient is undefined when fewer than two times are recorded or either vector has zero variation. GWACC aggregates correlations using Gaussian weighting and Monte Carlo delay samples; the simulations require a transmission-delay model, including the exponential and Rayleigh families considered in the study~\cite{shi2022source}. Its source estimate is a candidate node drawn from an admissible node domain. Pearson correlation is invariant to a common additive time shift, so the correlation definition alone does not specify the separate start-time estimate or a normalized observation likelihood. The aggregation kernel and admissibility convention must be taken from the original method specification.

Temporal correlations provide another way to evaluate candidate sources from recorded infection times. Wang~\cite{wang2019universal} proposed Spearman centrality, which compares the ordering of observed infection times with the ordering of shortest-path distances from each candidate source. The Sequential Neighbor Filtering (SNF) algorithm constructs local temporal vectors from candidate neighborhoods and evaluates their Pearson correlation with the recorded sensor timestamps~\cite{wang2020locating}. Candidate nodes are progressively filtered according to these temporal relationships. Both methods use correlations between network structure and observed infection times, avoiding repeated forward diffusion simulations during candidate evaluation.

GSSI uses the Gromov product to represent shared path structure in the covariance of partial infection timestamps~\cite{tang2018estimating}. It fits temporal quantities while accounting for candidate-dependent path relations rather than assuming an arbitrary fixed covariance. Candidate evaluation therefore combines the timing residual with the estimated dependence structure. The SCCE extension clusters temporal observations and estimates local candidates when the number of sources is unknown. Clustering and source-count decisions introduce uncertainty beyond the single-source fit. These methods illustrate a topology-informed statistical approach; comparison with independent-delay approximations should distinguish improved covariance modeling from changes in candidate search.

\subsubsection{Hybrid-Information-based Methods}

Pinto et al.~\cite{pinto2012locating} introduced the Pinto--Thiran--Vetterli Algorithm (PTVA), which combines infection arrival times and transmission directions to identify diffusion sources. On tree networks, directional observations are used to determine an active subtree \(T_a\), restricting the candidate sources to nodes within this region. Assuming independent Gaussian edge transmission delays with common parameters, PTVA constructs a likelihood from differences between sensor arrival times. When the covariance matrix is the same for all candidates, maximizing this likelihood is equivalent, up to candidate-independent terms, to

\begin{equation}
\hat v
=
\operatorname*{argmax}_{v\in\mathcal{T}_a}
\boldsymbol{\mu}_v^{\mathrm{T}}
\Lambda^{-1}
\left(
\boldsymbol{d}
-
\frac12
\boldsymbol{\mu}_v
\right),
\end{equation}
where \(\boldsymbol d\) denotes the observed arrival-time difference vector, \(\boldsymbol{\mu}_v\) is its expected value under candidate source \(v\), and \(\Lambda\) is the covariance matrix. For networks containing cycles, PTVA approximates propagation through candidate-rooted BFS trees and derives the corresponding timing statistics. When the covariance varies across candidates, the full Gaussian likelihood must be used, retaining both the quadratic residual and the covariance determinant.

Directional observations have also been used to refine the propagation structures employed in source estimation. DISGE (Direction-Induced Search based Gaussian Estimator) constructs a direction-induced search tree that incorporates recorded transmission directions into candidate path estimation~\cite{yang2020locating}. The resulting propagation structure is used to derive the timing statistics for Gaussian source scoring. Trust-GMLA (Trust-based Gaussian-based Maximum Likelihood Algorithm) further considers inter-node trust relationships when constructing propagation paths and estimating their covariance structure~\cite{ZHANG202228}. OSDIS (Observers-based Similarity Measures and Direction-Induced Search) combines temporal similarity and infection-order consistency, using Euclidean distance and Kendall rank correlation to evaluate candidates on direction-induced propagation trees~\cite{yang2023locating}. These methods incorporate different forms of relational and temporal evidence into the construction and evaluation of candidate propagation paths.

Another research direction relaxes the assumption of a single propagation tree. BPSL (Back Propagation Source Location) argues that the actual diffusion process generally involves multiple feasible propagation paths~\cite{qiu2022bpsl}. Instead of approximating the infected subgraph with a single propagation tree, BPSL allows multiple neighboring nodes to propagate temporal labels and path probabilities backward to each candidate node simultaneously. The algorithm then aggregates all propagation paths that satisfy the temporal constraints and identifies the source according to the cumulative probability of these feasible paths.

GFNL's ES stage scores non-sensor candidate nodes using distance, recorded reception times, and local sender/neighborhood evidence~\cite{wang2022rapid}. In the author-released implementation, a distance-based cost is combined with a time--distance discrepancy. The latter compares each reception time with the shortest-path distance to that sensor's recorded predecessor plus one hop; it therefore uses the sender record as well as the timestamp. Parameter-dependent powers adjust the contributions of distance and timing. An adjacent monitor that remains uninfected increases a candidate's cost, whereas a candidate recorded as the sender of a used monitor receives a discount. Candidates are ranked by increasing cost, and tied minima can produce a set of candidates. The implementation excludes monitored nodes from this search, which is an explicit admissibility assumption. This is a structural consistency score, not a calibrated posterior. Reproduction requires the deployment rule, activated-monitor selection, and implementation settings as well as the estimation stage; the reported early-stage performance depends on these choices.

Time--distance fusion has also been used to reduce the computation required for source detection and to make use of early-stage observations. GRSL (Greedy-coverage-based Rapid Source Localization) estimates propagation distances from predecessor information recorded by sensors, avoiding the reconstruction of complete propagation paths for individual candidates~\cite{10.1145/3543507.3583299}. RF-RSL (Random Full-Order-Coverage Based Rapid Source Localization) focuses on observations collected during the early stages of diffusion~\cite{hou2024random}. It first restricts the candidate region using a distance threshold and then ranks the remaining nodes through a distance-penalty score combining normalized arrival times and diffusion distances.

MASE (Multi-attribute Source Estimator) is a representative state--structure-fusion-based source detection method~\cite{10445791}. MASE incorporates the state information of uninfected sensors and formulates source detection as a multi-attribute decision-making problem. Candidate sources are first filtered according to the relative positions of infected and uninfected sensors:
\begin{equation}
\mathcal{C}_{\mathrm{fil}}
=
\left\{
v\in V\setminus V_{\mathrm{sen}}
\colon
\bar{d}(v,V_{\mathrm{sen}}^{+})
<
\bar{d}(v,V_{\mathrm{sen}}^{-})
\right\},
\end{equation}
where \(V_{\mathrm{sen}}^+\) and \(V_{\mathrm{sen}}^-\) denote the sets of infected and uninfected sensors, respectively, and \(\bar d(\cdot)\) represents the average distance from a candidate node to the corresponding sensor set. The filtering criterion requires both sets to be nonempty and the relevant distances to be finite. It retains candidates that are, on average, closer to infected sensors than to uninfected ones. The remaining nodes are evaluated using propagation delay, source purity, and infection-order violations. These attributes are combined through the Technique for Order Preference by Similarity to Ideal Solution (TOPSIS), which ranks candidates according to their relative closeness to the positive and negative ideal solutions:

\begin{equation}
C(v_i)
=
\frac{D_i^-}{D_i^+ + D_i^-},
\end{equation}
where \(D_i^{+}\) and \(D_i^{-}\) denote the distances from candidate node \(v_i\) to the positive and negative ideal solutions, respectively.  MASE therefore combines positive and negative sensor observations with temporal and structural attributes to rank candidate sources.

PrEF (Percolation-based Evolutionary Framework) incorporates percolation-related structural information into diffusion source detection~\cite{liu2023diffusion}. The method examines the evolution of infected clusters, including their boundaries, merging patterns, and the formation of a giant connected component when such a structure emerges. These features are considered together with infection states recorded by sensors and the consistency of candidate propagation sequences. PrEF thus evaluates possible origins using the large-scale organization of the spreading pattern, complementing methods based primarily on local propagation paths or arrival-time relationships.

\subsubsection{Direction, State, and Behavioral Observations}

Source inference can also draw on sensor records that do not contain exact infection arrival times. Transmission directions identify observed predecessors, node states recorded at known sampling times constrain the evolution of diffusion, and forwarding records provide information about possible propagation paths. These observations offer different ways to trace the origin of spreading.

PSLM (Path-based Source Localization on Multiplex Networks) combines observer-recorded transmission directions with node-state observations on a known multiplex network~\cite{cheng2022path}. During diffusion, each observer records the neighbor from which it receives information. In the original observation protocol, spreading is stopped once a prescribed number of observers have been infected, and the resulting node states are used to initialize positive and negative labels. The multiplex topology determines the connections available for label propagation, while the recorded transmission directions guide the flow of labels toward possible sources.

\begin{figure}[htp]
    \centering
    \includegraphics[width=0.8\linewidth]{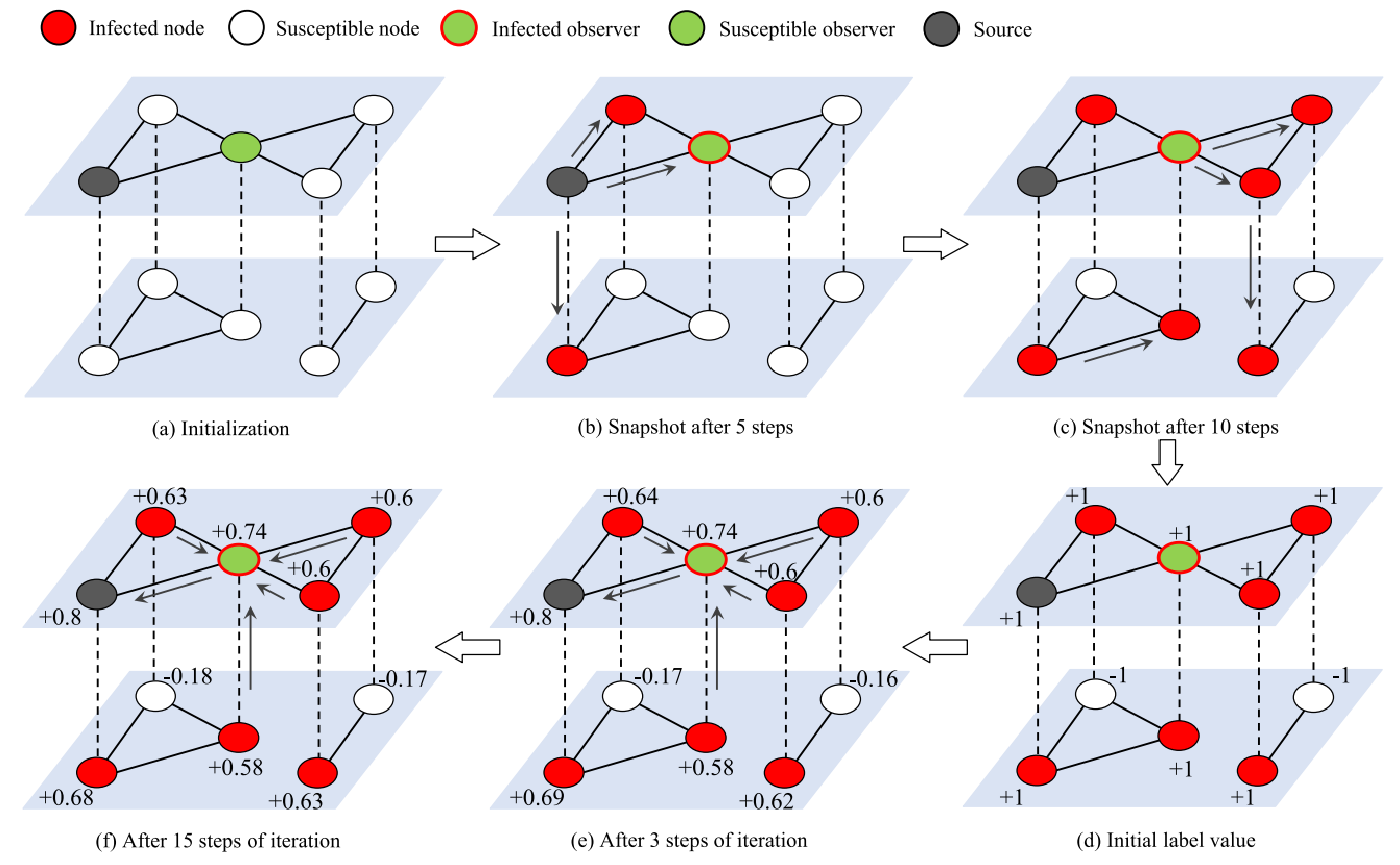}
    \caption{
\rev{PSLM on a two-layer multiplex network~\cite{cheng2022path}. (a) Initial source and observer placement. (b) and (c) Infection snapshots after five and ten spreading steps. (d) Labels initialized to \(+1\) and \(-1\) from the infected and susceptible states in (c). (e) and (f) Labels after three and fifteen iterations, respectively, with directional observations guiding label backflow. The panels follow the sequence (a)--(c), then (d)--(f), although the lower row is displayed from right to left.}
}
    
    \label{fig:pslm}
\end{figure}

To incorporate directional observations into label propagation, PSLM normalizes the supra-adjacency matrix \(\widetilde{\mathcal{A}}\) of the known multiplex topology:
\begin{equation}
W =
\widetilde{\mathcal{D}}^{-\frac{1}{2}}
\widetilde{\mathcal{A}}
\widetilde{\mathcal{D}}^{-\frac{1}{2}},
\end{equation}
where \(\widetilde{\mathcal{D}}\) is the supra-degree matrix. The indices \(i\) and \(j\) refer to node-layer pairs in this multiplex network representation. For an isolated node copy, the corresponding inverse-degree entry is set to zero. Let \(F_j^t\) be the label of neighboring node copy \(v_j\) at iteration \(t\). Writing its contribution to recipient \(v_i\) as \(M_{ij}^{t}\) makes both the receiving node and iteration explicit:
\begin{equation}
M_{ij}^{t}=
\begin{cases}
\beta W_{ij}F_j^t,&\text{Type 1},\\
\alpha W_{ij}F_j^t,&\text{Type 2},\\
0,&\text{Type 3},
\end{cases}
\end{equation}
where \(\alpha\in(0,1)\) and \(\beta\in(\alpha,1)\). In Type~1, observer \(v_j\) records receipt from \(v_i\): infection travels from \(v_i\) to \(v_j\), whereas the label contribution returns from \(v_j\) to \(v_i\) with weight \(\beta\). Type~2 retains the ordinary topology-based contribution with weight \(\alpha\) where the pair lacks a recorded directional constraint. Type~3 suppresses a contribution excluded by the directional rule. These are inference updates on the multiplex network representation, rather than a replay of the forward infection process.

At iteration \(t+1\), the label of node \(v_i\) is updated according to
\begin{equation}
F_i^{t+1}
=
\sum_{v_j\in\mathcal{N}(v_i)}
M_{ij}^{t}
+
(1-\alpha)F_i^0,
\end{equation}
where \(\mathcal{N}(v_i)\) is the neighbor set in the multiplex network representation and \(F_i^0\) is the state-derived initial label. The neighbor sum aggregates the directionally weighted contributions; the second term retains the initial infection-state evidence. Figures~\ref{fig:pslm}(d)--(f) show the initialization followed by three and fifteen label iterations. For single-source inference, the largest final label supplies the source estimate; for multiple sources, local label maxima relative to neighboring nodes supply candidate sources. Labels are updated repeatedly using the same recorded snapshot and directional constraints, rather than measured as new infections at each iteration.

PSLM uses recorded transmission directions to guide label propagation along possible reverse paths, while the multiplex topology allows information to pass through connections without directional records~\cite{cheng2022path}. Source candidates are consequently evaluated through the resulting label distribution, without explicitly estimating transmission-delay parameters. Other sensor-based methods use measurements collected at several times to reconstruct the initial network state.

Hu et al.~\cite{9807439} proposed a continuous-state-based source detection method for multilayer networks. The diffusion process is formulated as a linear dynamical system, where sensor states collected at multiple observation instants are assembled into the global measurement equation
\begin{equation}
\boldsymbol{U}
=
\Phi\boldsymbol{x}(0),
\end{equation}
 where \(U=[u(0)^\top,u(1)^\top,\ldots,u(\tau)^\top]^\top\) collects the sensor measurements at different observation times, and
\(
\Phi=
\begin{bmatrix}
C\\
CA_{\mathrm{dyn}}\\
\vdots\\
CA_{\mathrm{dyn}}^\tau
\end{bmatrix}
\)
is the corresponding observation matrix. The matrix \(A_{\mathrm{dyn}}\) describes the linear state transitions induced by intralayer diffusion and interlayer coupling, \(C\) maps network states to sensor outputs, and \(x(0)\) is the unknown initial state.

Suppose \(m\) scalar measurements are collected at each of \(\tau+1\) observation times. Let \(n_x\) denote the dimension of the initial state vector \(x(0)\). The stacked observation matrix \(\Phi\) then has \(m(\tau+1)\) rows and \(n_x\) columns. Thus, having fewer measurements than state variables at a single observation time does not necessarily prevent unique reconstruction. In the noiseless linear model, an arbitrary initial state can be uniquely recovered if and only if \(\operatorname{rank}\Phi=n_x\), consistent with the observability condition in Section~\ref{sec:linear_baseline}.

When \(\Phi\) is rank-deficient, additional constraints on the initial state may still permit source identification. If the initial state is assumed to be sparse, its estimation can be formulated as~\cite{donoho2006compressed,candes2006robust}
\begin{equation}
\min_z\|z\|_0
\quad\text{subject to}\quad
\boldsymbol U=\Phi z,
\label{eq:sparse_initial_recovery}
\end{equation}
where \(\|z\|_0\) counts the nonzero entries of \(z\). Uniqueness for all initial states with at most \(k\) nonzero entries requires that the null space of \(\Phi\) contain no nonzero vector supported on at most \(2k\) coordinates. This condition is weaker than full-state observability and concerns uniqueness within the restricted class of sparse states. Orthogonal Matching Pursuit (OMP) provides a computational approach to sparse reconstruction, but its accuracy and exact recovery guarantees depend on additional properties of the observation matrix and measurement noise~\cite{tropp2007omp}. Noisy observations also require an appropriate residual tolerance and an assessment of estimation stability. The nonzero entries of the estimated initial state indicate candidate source locations within the assumed linear model; these conditions do not directly establish source identifiability under stochastic SI, SIR, or SIS dynamics.

The null-space condition follows by subtracting two admissible sparse initial states: if both produce the same observations, their difference lies in the kernel and has at most twice the allowed support size. Conversely, a kernel vector with such a support can be split into two sparse states with identical measurements. Additional sign or amplitude constraints can further restrict this candidate class.

The observation schedule also enters the reconstruction operator on a time-varying network. Hu et al.~\cite{hu2018locating} construct stacked measurements using ordered products of transition matrices for linear diffusion, replacing powers of a fixed matrix. They combine structural observability with sparse reconstruction of nonzero initial states, making both measurement locations and times relevant to source inference. Their empirical tests also select monitors from an earlier segment of temporal contact data and apply them in a later segment. This illustrates one way historical interactions can guide deployment when future contacts are unavailable, while retaining a specified linear state model.

A two-state example illustrates the role of observations collected over time. Let \(C=(1,0)\) and suppose \(A_{\mathrm{dyn}}\) exchanges the two state coordinates. Measurements at \(t=0\) and \(t=1\) then give \(\Phi=I_2\), allowing both components of the initial state to be reconstructed from a single sensor. If \(A_{\mathrm{dyn}}=I_2\), the same sensor produces identical measurement rows at both times, leaving the second state unobserved. The distinction is therefore not simply the number of measurements, but whether observations collected over time provide independent information about the initial state.

A discrete-state-based label propagation framework was proposed in~\cite{hu2024source} for source detection using infection state observations. Because sensors are deployed on only a subset of network nodes, the states of many nodes remain unobserved. The framework therefore first infers the states of unobserved nodes from neighborhood observations and initializes node labels according to their infection states.  Specifically, nodes in the recovered (R) state are assigned the highest initial labels, followed by nodes in the infected (I) state, while nodes whose states remain unobserved receive zero initial labels. These assignments form the initial label vector \(F^0\), which is subsequently updated through label propagation until convergence:
\begin{equation}
\boldsymbol{F}^{*}
=
(1-\alpha)
\left(
I-\alpha\mathcal{M}
\right)^{-1}
\boldsymbol{F}^{0},
\end{equation}
where \(\mathcal M=D^{-1/2}AD^{-1/2}\) is the normalized label propagation matrix, \(A\) is the network adjacency matrix, \(D\) is the degree matrix, \(I\) is the identity matrix, and \(\alpha\in(0,1)\) is the propagation parameter. The node with the largest converged label is finally identified as the diffusion source. By combining state completion with label propagation, the framework propagates local observation information to previously unobserved regions of the network, mitigating the loss of information caused by limited sensor coverage.

The ROSE (Radial Observer Source Estimator) algorithm provides a representative example of spreading-behavior-based source detection~\cite{devarapalli2024estimating}. Designed for information diffusion in online social networks, ROSE uses the cumulative number of shares recorded by sensors as an indicator of diffusion progress, where the share count refers to the total number of forwarding events that have occurred before the message reaches a sensor, rather than the number of times the sensor receives the same message. Consequently, sensors with smaller share counts are expected to be located closer to the diffusion source. ROSE first identifies the two sensors with the smallest cumulative share counts from the sensor set \(V_{\mathrm{sen}}\):
\begin{equation}
o_b=
\operatorname*{argmin}_{o_i\in V_{\mathrm{sen}}}Q(o_i),
\qquad
o_r=
\operatorname*{argmin}_{o_i\in V_{\mathrm{sen}}\setminus\{o_b\}}Q(o_i),
\end{equation}
where \(Q(o_i)\) denotes the cumulative share count recorded by sensor \(o_i\), and \(o_b\) and \(o_r\) represent the best and second-best sensors, respectively. A candidate radial region is then determined from the shortest path connecting these two sensors. For each candidate node within this region, a Radial Infection Tree (RIT) is constructed, and the SERO (Shares Estimator at Radial Observers) model is employed to estimate the expected share counts at the two key sensors.

ROSE compares observed and candidate-predicted sharing counts at two selected sensors~\cite{devarapalli2024estimating}. The distance-normalized discrepancy is
\begin{equation}
\Gamma(v)=\frac{|Q(o_b)-\hat Q_v(o_b)|}{d(v,o_b)}+
\frac{|Q(o_r)-\hat Q_v(o_r)|}{d(v,o_r)}.
\end{equation}
This expression is defined only for candidates with positive finite distances to both sensors. Candidates coinciding with a selected sensor require a separate admissibility or scoring convention. Smaller values indicate smaller distance-normalized count discrepancies; the denominator does not imply that a closer candidate is always favored. The score is a compatibility measure whose interpretation depends on how sharing counts are recorded.

A count may describe the number of forwarding steps along a message path, the number of times an individual account shares a message, or an aggregate over several copies. These quantities contain different information about diffusion. Counts associated with a particular forwarding chain can provide evidence about propagation depth, whereas aggregated sharing activity does not identify a unique transmission path. Source estimation based on such records therefore requires a clear definition of how counts are updated and how message histories are preserved.

\subsection{Discussion and Outlook}

Sensor deployment and source detection address two closely related aspects of the same inference problem. Deployment determines where and how diffusion is observed, while source detection uses the resulting records to identify possible origins. A sensor configuration designed for early warning or broad network coverage may not provide sufficient evidence to distinguish candidate sources. Similarly, reconstructing an arbitrary initial state is a different task from identifying an initiating node within a restricted candidate set. These objectives require separate evaluation, with the recorded variables, sensor budget, and acquisition costs considered alongside detection accuracy.
 
The information available for source inference also reflects how measurements are collected. Missing arrival times, censored observations, clock offsets, and incorrect predecessor records affect the interpretation of propagation evidence in different ways. For example, a sensor that remains unreached during monitoring provides different information from one whose measurement is missing or incorrectly recorded. When sensors are selected or relocated during diffusion, their positions and sampling times may also depend on earlier observations. Accounting for this acquisition history is therefore necessary when evaluating the evidence used for source detection.

Table~\ref{tab:method_selection} compares representative source-detection methods from Sections~\ref{sec:full-observability}--\ref{sec:sensor-based} in terms of their required observations, model assumptions, inference procedures, and computational demands. The comparison highlights differences that are not captured by detection accuracy alone. Methods based on state snapshots do not have access to the arrival-time differences used by temporal estimators, while simulation-based approaches require a forward diffusion model to evaluate candidate origins. Learning-based methods instead use training data to construct an estimator that can be applied to subsequent observations. These differences in available information and computational requirements are essential when interpreting results obtained under different experimental settings.

An important direction is to connect sensor deployment more closely with the uncertainty remaining after source detection. When the available records support several possible origins, additional measurements can be selected to distinguish their predicted propagation patterns. This motivates sequential observation strategies in which source estimates guide subsequent sensor placement or sampling times, and newly acquired records refine the candidate ranking. Such a connection between observation design and inference extends beyond optimizing deployment and detection separately. Its evaluation should account for source detection accuracy, acquisition cost, decision time, and the uncertainty associated with the inferred origins. The experimental controls in Table~~\ref{tab:evaluation_observation_paradigms} provide a reference for comparing these strategies. These considerations also become relevant when source-detection methods are applied across different domains, as discussed in Section~\ref{sec:data-driven}.

\begingroup
\small
\ifrevisionmarked\color{red}\else\color{black}\fi
\setlength{\tabcolsep}{4pt}
\renewcommand{\arraystretch}{1.12}
\begin{longtable}{@{}>{\raggedright\arraybackslash}p{0.15\textwidth}>{\raggedright\arraybackslash}p{0.24\textwidth}>{\raggedright\arraybackslash}p{0.275\textwidth}>{\raggedright\arraybackslash}p{0.28\textwidth}@{}}
\caption{\rev{Method selection from observations and model knowledge. The entries describe representative constructions, not uniform guarantees or performance rankings for entire families.}}
\label{tab:method_selection}\\
\toprule
\textbf{Family and examples} & \textbf{Observations and extra knowledge} & \textbf{Candidate evaluation or inference} & \textbf{Output, conditions, and principal work}\\
\midrule
\endfirsthead
\caption[]{\rev{Method selection from observations and model knowledge (continued).}}\\
\toprule
\textbf{Family and examples} & \textbf{Observations and extra knowledge} & \textbf{Candidate evaluation or inference} & \textbf{Output, conditions, and principal work}\\
\midrule
\endhead
\midrule
\multicolumn{4}{r}{\emph{Continued on next page}}\\
\endfoot
\bottomrule
\endlastfoot
Structural scores~\cite{shah2011rumors,luo2014identify} & Infected snapshot or observed infected subset; known network. & RC counts infection orders; a Jordan center minimizes maximum distance to observed infections. & Node or tied candidates. Rumor-centrality likelihood equivalence assumes regular-tree homogeneous SI. Tree recursion or candidate distances dominate.\\
\addlinespace
DMP~\cite{lokhov2014inferring} & State snapshot; network, transition parameters, and specified or searched time. & Propagate cavity messages under each source hypothesis; score agreement of state probabilities with observations. & Candidate ranking or node. Forward marginals are exact on trees; their product approximates the joint likelihood. Candidate/time sweeps dominate.\\
\addlinespace
BP~\cite{altarelli2014bayesian,Altarelli_2014} & Possibly incomplete states; network, transition law, measurement model, and source prior. & Condition a factor graph of latent infection/recovery histories on observations and exchange messages. & Source marginals and node/set decisions. Exactness requires a tree factorization; loopy inference is approximate. Latent domains and iterations control cost.\\
\addlinespace
MDL: NETSLEUTH~\cite{prakash2014efficiently} & Infected snapshot; network and the coding model's propagation assumptions. & Balance seed-description length against the code length of the induced infection pattern. & Source set with count selection. Extra seeds must justify their coding cost. Candidate search and encoding evaluation dominate.\\
\addlinespace
Simulation matching~\cite{antulov2015identification} & Snapshot or supported record; network and a propagation/measurement simulator. & Simulate candidate outbreaks; aggregate exact matches or a specified discrepancy/kernel. & Ranking or node/set estimate; probabilistic meaning depends on the construction. Replications and candidate/time budgets dominate.\\
\addlinespace
Bayesian optimization: BOSouL~\cite{zhang2024multiple} & Single snapshot; network, simulator, supplied source count, and restricted candidate pool. & Average each simulated trajectory's best temporal match; fit a GP with a spectral kernel and acquire evaluations by expected improvement. & Fixed-size set selected by final GP prediction. GP uncertainty describes the similarity response. Simulation, surrogate fitting, and candidate sampling dominate.\\
\addlinespace
Learned inverse models~\cite{ling2022source,wang2022invertible,bao2024graph} & Model-specific snapshot/temporal features; network representation and representative training outbreaks. & Fit a source-pattern prior, inverse mapping, or learned representation and scoring function. & Node scores or source sets. Training amortizes repeated inference; transfer across networks, dynamics, and observation rules requires evaluation.\\
\addlinespace
Time/direction methods~\cite{pinto2012locating,zhu2016locating,cheng2022path} & Arrival times and/or recorded infectors; network, delay/path model, and method-specific state records. & Match delay differences, backtrack timed paths, or propagate state-initialized labels against recorded transmission. & Node or multiple-source candidates. Timing/direction errors and path approximations matter. Path, covariance, or label calculations dominate.\\
\addlinespace
Linear sparse recovery~\cite{9807439,astrom2008feedback} & Repeated sensor states; known transition and measurement matrices; sparsity when invoked. & Stack observations; solve a linear or sparsity-constrained inverse problem. & Initial state and source support. Full rank identifies arbitrary states; restricted kernel conditions address sparse states. Conditioning and solver assumptions govern recovery.\\
\end{longtable}
\endgroup

\section{Data-driven Source Detection and Cross-domain Perspective}
\label{sec:data-driven}
\subsection{From Model-driven to Data-driven Source Detection}
\label{sec:model_to_data_driven}
 
The methods reviewed in Sections~\ref{sec:full-observability}--\ref{sec:sensor-based} differ in how they relate diffusion observations to possible origins. Mechanistic estimators evaluate candidate sources using specified transmission and observation models, whereas data-driven approaches learn this relationship from examples. The distinction is not absolute. Simulated diffusion processes can provide training data for learned estimators, and structural or dynamical constraints can be incorporated into their design. In either case, the interpretation of the results rests on how observations and source labels are defined~\cite{ling2022source,wang2022invertible}.

The information used for learning also varies across applications. Infection timestamps constrain the timing of propagation, state snapshots describe the observed spreading pattern, and forwarding records may identify local predecessors. Additional attributes, such as textual content or spatial information, can help characterize an event without directly revealing its origin. These differences must be reflected in the training data and in the observations available when the model is applied.

Data-driven source detection typically learns a mapping from diffusion observations to candidate origins using labeled or simulated examples. This includes statistical learning, representation learning, generative models, and graph-based neural architectures. Although their training procedures differ, many of these methods produce candidate-source scores from which a source estimate is selected. A general score-based formulation is
\begin{equation}
    \mathbf{s}
    =
    f_{\phi}\bigl(G,Q,\mathbf{Z}_{\mathrm{aux}}\bigr),
    \qquad
    \hat{\mathcal{S}}
    =
    \operatorname{Select}_{\mathcal{C}}(\mathbf{s}),
    \label{eq:chapter5_data_driven}
\end{equation}
where \(Q\) denotes the diffusion observations defined in Section~\ref{sec:problem_definition}, and \(\mathbf{Z}_{\mathrm{aux}}\) represents additional information available for inference, such as node attributes, textual features, spatial information, genomic data, or transaction metadata. The parameters \(\phi\) are learned from training data, and \(\boldsymbol{s}\) contains the scores assigned to candidate sources. The selection operator \(\operatorname{Select}_{\mathcal C}\) converts these scores into a source estimate according to the admissible candidate space, source-count constraints, or a specified selection rule. The scores may be used for ranking without necessarily representing calibrated posterior probabilities.

Graph neural networks are widely used for this task because they can learn representations from network structure and observed node features. Recent studies have also explored dynamic graph neural networks, contrastive learning, generative models, invertible graph architectures, and state-space models~\cite{he2025tracing,bao2024graph,ling2022source,wang2022invertible,cheng2025sourcedetmamba}. These approaches differ in their training objectives and the information they encode. Some learn candidate-source scores directly, while others learn representations or generative relationships that support subsequent source inference.

Learned inverse models can encode nonlinear relations between initial and observed states, as in IVGD~\cite{wang2022invertible}. Their mathematical architecture may be invertible while the actual stochastic process is many-to-one at the observation level. Training and validity constraints determine which explanation is selected from that ambiguity. Predictive scores require calibration if they are to be reported as probabilities, and simulated supervision does not replace validation against independently established origins.

Sequential observations provide another source of information for learned estimators. FGSSI uses multiple network snapshots to capture changes in diffusion patterns and infer possible origins~\cite{hou2025fgssi}. Its evaluation concerns the recovery of source locations from observed spreading sequences, rather than the prediction of subsequent diffusion from known initiating nodes. These tasks may employ similar neural architectures but require different inputs and evaluation procedures.

The evaluation of multimodal models must reflect the information available at the time of inference. Features derived from the true source, future observations unavailable at the decision time, or transmission histories that were not recorded can introduce information leakage. Tests across different networks or diffusion models should preserve these observation constraints so that any reported transfer performance reflects the information the estimator could actually use.

Hybrid approaches can retain a simulator or mechanistic constraints while learning nuisance parameters, source representations, or candidate proposals. This division can improve computational tractability, but theoretical guarantees for individual components do not necessarily extend to the combined model.  
A learned proposal can reduce the cost of candidate search without changing the observation model, while a learned approximation to the likelihood requires validation before it is used for source inference.

\subsection{Source Detection in Epidemiological Systems}

In epidemiological studies, the source may refer to the first infected individual, the point of introduction into a region, or a node involved in disease transmission through animal movements. The target of inference must therefore be specified according to the network and transmission process under study. Reported cases do not necessarily reflect the order of infection, as diagnosis and reporting may occur well after transmission. When an outbreak involves multiple introductions, identifying a single source is insufficient, and the inference must account for multiple initiating events~\cite{pinto2012locating,pei2019detection}.

On individual contact networks, state-based and sensor-based estimators use infection patterns or timing to evaluate candidate origins under a transmission model~\cite{lokhov2014inferring,antulov2015identification,yu2022epidemic}. Their application requires more than a contact network: contacts must be temporally relevant, observations must reflect the modeled states, and candidates must include plausible unobserved infections. An intervention can change both the network and the probability that a case is detected, coupling model and measurement uncertainty.

Pei et al.~\cite{pei2019detection} examine avian influenza A(H7N9) through a live-poultry transportation network. The relevant candidate origins are transportation or market nodes associated with dissemination, not necessarily human index cases. Movement and disease records supply evidence about spread through the host-trade system. The example illustrates how the network's node semantics determine the source label and how human observations can be indirect measurements of an animal-mediated process. A regional origin estimate should therefore not be reported as identification of an individual first infection.

At the regional scale, epidemic spread is shaped by population movements between locations. Metapopulation models describe how local transmission interacts with mobility, including the effects of travel patterns on epidemic invasion and spatial synchronization~\cite{colizza2007invasion, balcan2009multiscale}. Brockmann and Helbing~\cite{brockmann2013hidden} use mobility flows to define an effective distance between locations. For a directed connection from \(n\) to \(m\), the link length is \(1-\log P_{mn}\), where \(P_{mn}\) is the fraction of travelers leaving \(n\) for \(m\). The effective distance between two locations is obtained by minimizing the sum of these lengths along directed paths.

\begin{figure}[htp]
    \centering
    \includegraphics[width=0.95\linewidth]{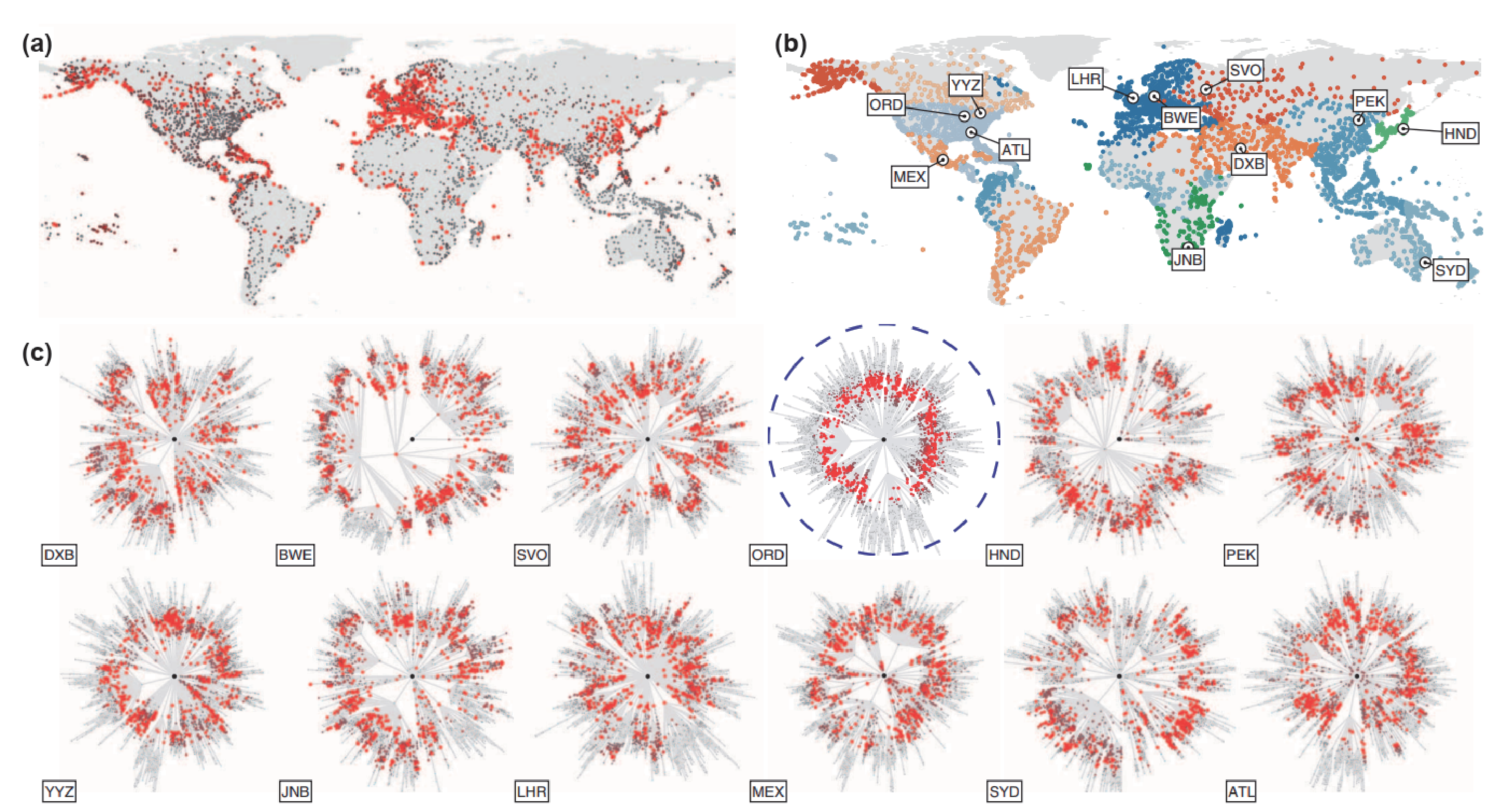}
    \caption{
    \rev{Outbreak-origin reconstruction in a simulated epidemic, adapted from Brockmann and Helbing~\cite{brockmann2013hidden} (their Fig.~3). (a) Simulated prevalence 81 days after an outbreak in Chicago (ORD). (b) Candidate outbreak airports from different geographic regions. (c) The same state on effective-distance shortest path trees rooted at those candidates. Chicago, circled in blue, yields the most concentric wavefront in this simulation.}
}
    
    \label{fig:epidemic_effective_distance}
\end{figure}

This representation allows candidate epidemic origins to be compared using the spatial progression of an outbreak. When arrival times are available, candidates can be assessed through their relationship with effective distances. For prevalence snapshots, the method examines the mean and variation of effective distances to affected locations. Figure~\ref{fig:epidemic_effective_distance} illustrates this approach using a simulated outbreak originating in Chicago, whose effective-distance representation produces a more concentric spreading pattern around the true origin. The study also examines historical H1N1, SARS, and EHEC outbreaks, illustrating both successful detection and cases where the inferred origin changes with the observation window.

Li et al.~\cite{li2021sensor} developed a probabilistic approach to regional source detection using infection arrival times recorded at a limited number of locations. Their method models the observed arrival times through a source-conditioned multivariate Gaussian distribution, with its mean and covariance estimated from stochastic cholera simulations on a human-mobility network. Combined with uniform or epidemiologically informed source priors, the model produces posterior estimates of possible outbreak origins and highest-posterior-density regions.

The method was applied retrospectively to the 2000--2002 cholera outbreak in KwaZulu-Natal, using records from nine infected communities with high weighted degree as observers for each epidemic wave. Since the actual introduction sites are not independently known for these historical outbreaks, the inferred origins cannot be validated directly against ground truth. Separate simulated outbreaks are used to assess detection accuracy and the coverage and size of the resulting credible regions.

\begin{figure}[htp]
    \centering
    \includegraphics[width=0.4\linewidth]{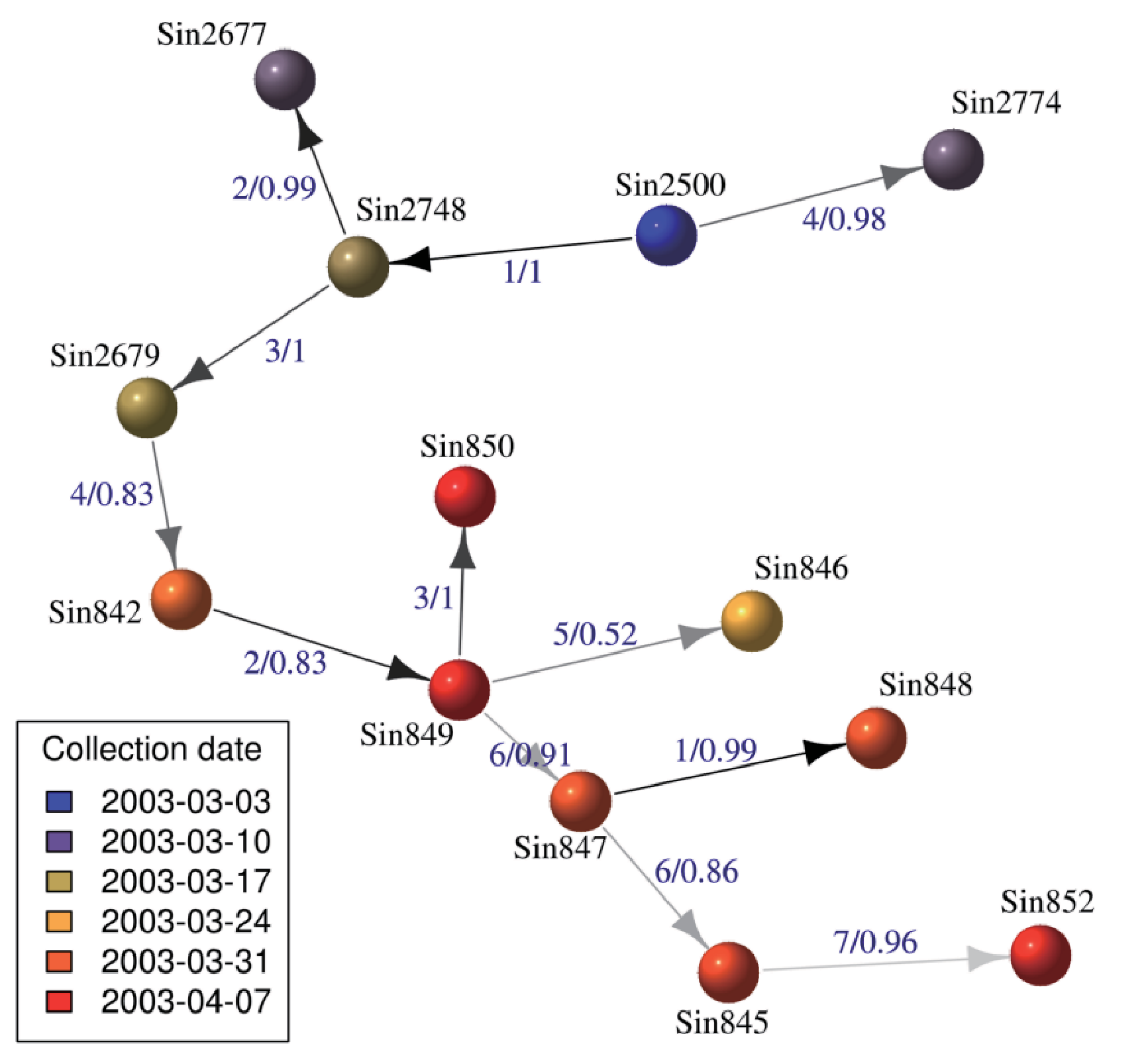}
    \caption{
Inference of SARS transmission pathways using epidemiological and genomic data~\cite{jombart2014bayesian}. Nodes represent sampled cases and are colored by collection date, while directed edges indicate inferred transmission relationships annotated with genetic differences and posterior support. The reconstructed transmission network combines temporal, genomic, and epidemiological evidence to infer hidden transmission pathways and identify plausible transmission origins in the absence of direct transmission records.
}
   
    \label{fig:epidemic_genomic_transmission_tree}
\end{figure}

Pathogen genomic data provide additional evidence for reconstructing epidemic origins and transmission histories. Genomic epidemiology and phylodynamic methods combine pathogen sequences with sampling times and geographic information to infer evolutionary relationships and patterns of dissemination~\cite{grenfell2004unifying,volz2013viral,grubaugh2019tracking,hadfield2018nextstrain}. Transmission-tree reconstruction further incorporates epidemiological evidence to estimate possible links between infected individuals and distinguish separate introduction events~\cite{jombart2014bayesian,didelot2017genomic,duault2022methods}. Figure~\ref{fig:epidemic_genomic_transmission_tree} illustrates the reconstruction of SARS transmission relationships from genetic differences, sampling dates, and posterior support for inferred transmission links.

The interpretation of genomic evidence requires a distinction between evolutionary ancestry and direct transmission. Closely related pathogen sequences do not establish direct transmission between sampled individuals, and the earliest sampled case need not be the first individual infected. Unsampled hosts and unobserved transmission events may separate cases that appear closely related in a reconstructed tree. Consequently, genomic analyses may identify ancestral lineages, possible introduction events, or transmission relationships without necessarily identifying the individual who initiated an outbreak. The target of inference and the evidence used to validate it should reflect these differences.

Digital surveillance can supplement conventional epidemiological records with proximity measurements, travel histories, and exposure notifications~\cite{ferretti2020quantifying}. These data help establish where and when contacts may have occurred, but a recorded contact does not necessarily indicate a transmission event. Moreover, systems developed for contact tracing or exposure notification are not necessarily designed to identify the first infected individual. Using their records for source detection therefore requires an inference model that relates observed contacts to possible transmission histories.

Epidemiological source inference must also account for the difference between infection events and their recorded consequences. Diagnosis and reporting times may lag behind transmission by intervals that vary across individuals, making the observed case order an imperfect record of epidemic spread~\cite{lauer2020incubation,he2020temporal}. Asymptomatic infections, unreported cases, and incomplete testing leave further gaps in the available transmission evidence~\cite{li2020substantial,moghadas2020implications}. Interventions, including isolation, vaccination, and changes in contact behavior, can alter both the spreading process and the likelihood that infections are observed. These features connect source inference to the surveillance process itself: reconstructing an epidemic origin requires consideration of how transmission occurred and how evidence of that transmission entered the available records.

\subsection{Source Detection in Information Systems}

Source detection in information systems involves two distinct tasks: tracing the origin of an online information cascade and reconstructing the initiating events of a cyberattack. Both rely on records retained by the platform or monitoring system. Social-media data capture selected posts, interactions, and reposts, while security logs record events according to the deployed audit policies. In each setting, the inferred source must be defined with respect to the event under investigation and the available observation window.

In online information spreading, the initiating account need not be the most influential participant in the subsequent cascade. Identifying this account is also distinct from determining whether the circulating information is true or false. Coordinated information operations may involve several accounts and related narratives~\cite{starbird2019disinformation}. Event matching determines which records belong to the same cascade, and incorrect grouping can change the candidate origins or create the appearance of multiple introductions.

Source inference on social platforms may use follower relationships, reposting records, timestamps, and content features. These observations describe different aspects of information spreading. Follower links indicate possible exposure, whereas recorded repost relationships provide evidence about the observed cascade. Even a reconstructed cascade covering all collected posts may omit earlier events or propagation across other platforms.

\begin{figure}[htp]
    \centering
    \includegraphics[width=0.8\linewidth]{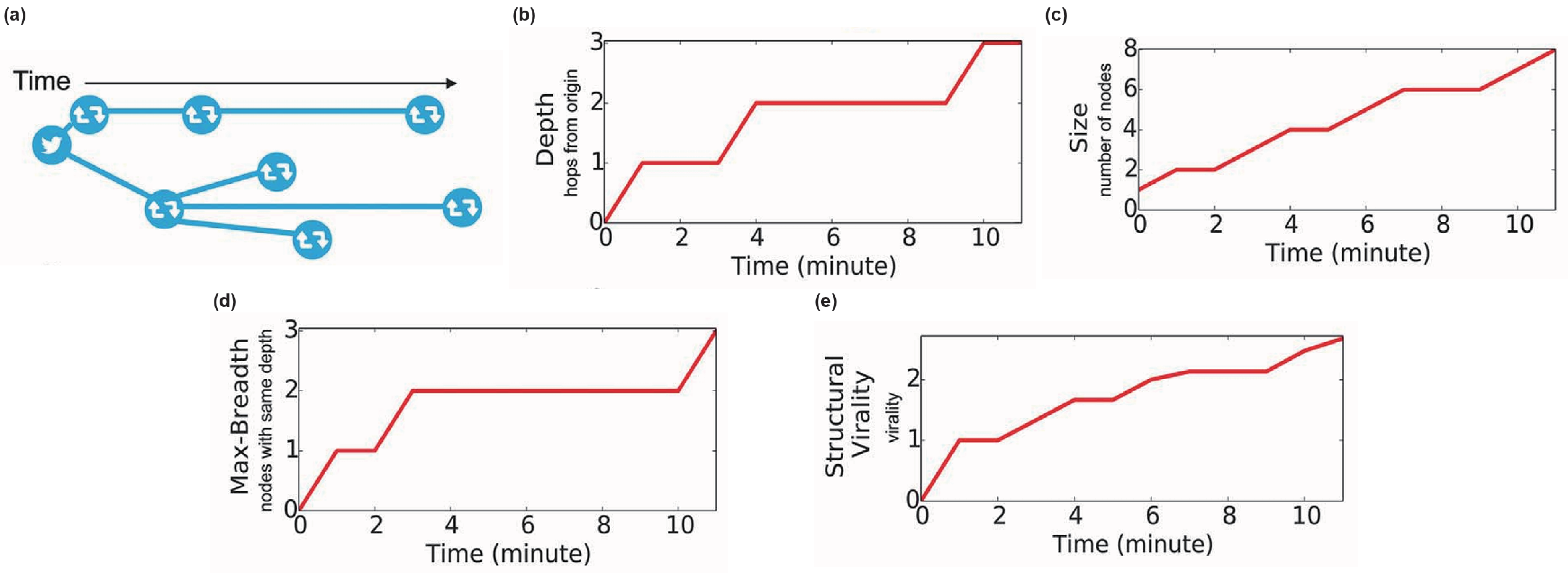}
    \caption{\rev{
Temporal evolution of structural characteristics in an online rumor cascade~\cite{vosoughi2018spread}. (a) A representative diffusion cascade with timestamped reposting events. (b)--(e) Evolution of cascade depth, cascade size, maximum breadth, and structural virality over time, respectively. These descriptors summarize cascade growth and shape. The figure summarizes forward propagation.
}}
   
    \label{fig:information_rumor_cascade}
\end{figure}

Vosoughi et al.~\cite{vosoughi2018spread} examined the spread of true and false news on social networks. Figure~\ref{fig:information_rumor_cascade} illustrates how cascade depth, size, breadth, and structural virality evolve over time. These characteristics describe the development of observed cascades and provide structural information relevant to source inference.

Source detection on real social platforms also reveals limitations of methods that favor structurally central candidates. Ali et al.~\cite{ali2025generalized} examined this problem using reconstructed rumor propagation networks, where the initiating account may lie away from the center of the observed cascade. They describe this discrepancy as source skewness and propose GEPA (Generalized Exoneration and Prominence-based Age) to evaluate candidates using local prominence and exoneration information. The study also introduces two real-world rumor-network datasets, Cov19-RN and Use20-RN, constructed from records associated with COVID-19 rumors and misinformation concerning the 2020 U.S. presidential election.

\begin{figure}[htp]
    \centering
    \includegraphics[width=0.5\linewidth]{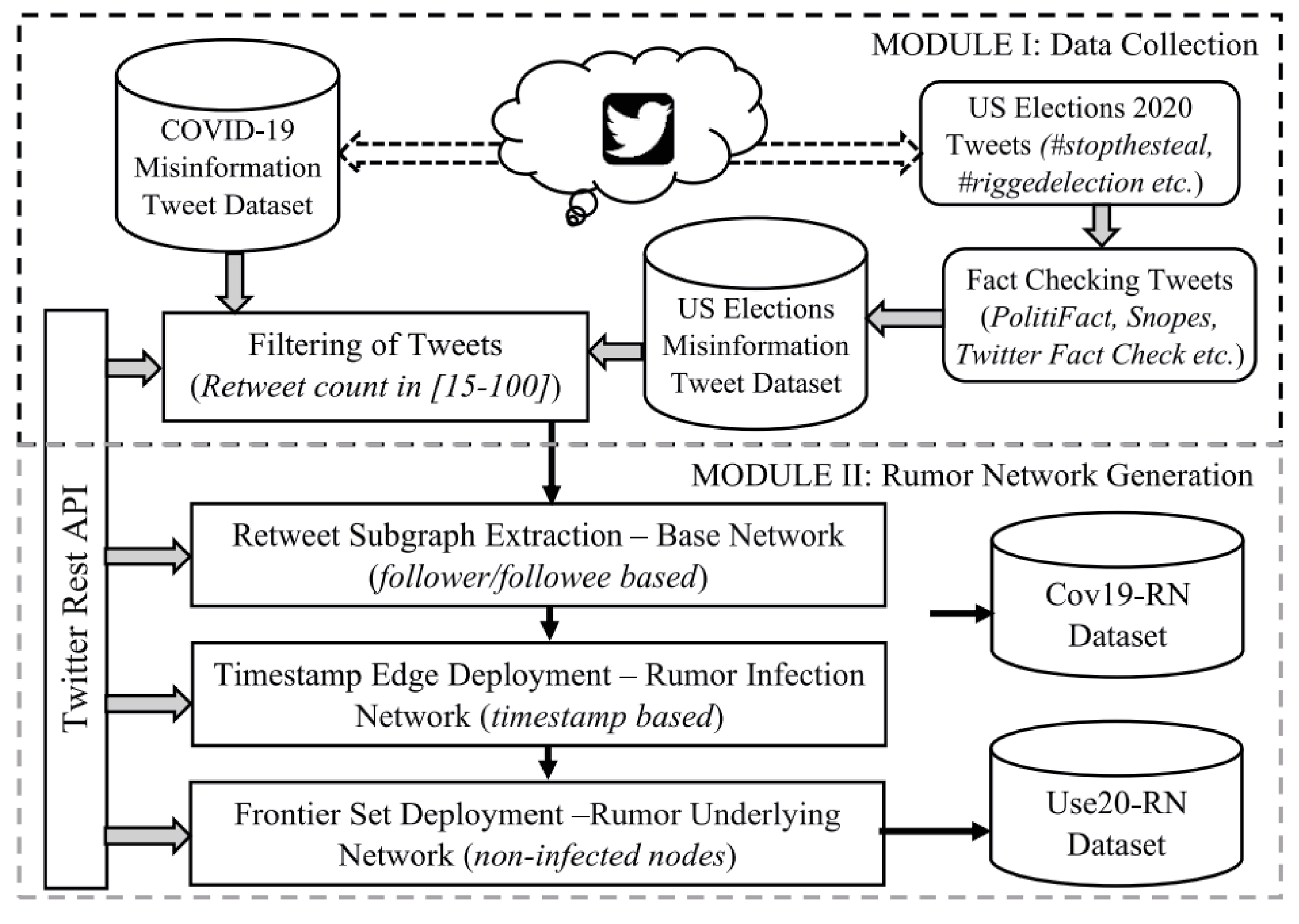}
    \caption{
Construction pipeline of real-world rumor-network datasets~\cite{ali2025generalized}. 
Module I identifies rumor events by integrating social-media data and fact-checking information. Module II reconstructs the observed diffusion network together with the underlying social network using reposting timestamps, follower--followee relationships, and frontier nodes, resulting in the Cov19-RN and Use20-RN datasets. This pipeline illustrates how source detection in real information systems relies on integrating heterogeneous observations rather than solely on simulated diffusion networks.
}
   
    \label{fig:real_world_rumor_network_generation}
\end{figure}

Figure~\ref{fig:real_world_rumor_network_generation} presents the construction of Cov19-RN and Use20-RN from social-media records. Rumor events are identified using tweets and fact-checking information, while reposting histories and follower relationships are used to reconstruct the observed cascade and its surrounding network. Frontier nodes are included to represent users connected to the cascade but without observed participation in its spread. The resulting datasets retain both the recorded propagation structure and additional network context that is absent from the observed cascade alone.

Let \(G_R=(V_R,E_R)\) denote the reconstructed rumor propagation network and \(G_U=(V_U,E_U)\) the corresponding underlying network, which includes additional social connections and frontier nodes. By construction, \(V_R\subseteq V_U\). The two networks provide different information for source detection: \(G_R\) represents the observed spreading pattern, while \(G_U\) supplies structural context for evaluating candidate origins. GEPA uses generalized local prominence to account for source skewness when ranking candidate origins.

In cybersecurity, the source of an attack may refer to the point of network entry, an initially compromised host, or an event that begins a sequence of malicious activities. These targets require different types of evidence. Network-level traceback reconstructs communication paths, while host-level analysis examines the relationships among processes, files, and other system entities recorded during an incident.

IP traceback methods, including those developed for distributed denial-of-service attacks, trace packet paths backward from the observed target toward possible traffic origins~\cite{savage2000practical,snoeren2001hash}. The reconstructed path identifies a network-level origin under the available traffic records, which need not reveal the identity of the attacker. Provenance-based approaches instead organize system audit records into dependency networks. Processes, files, and network connections form the entities of these networks, with recorded operations defining relationships among them. Tracing these dependencies helps reconstruct possible attack sequences and identify candidate initiating events within the monitored environment.

\begin{figure}[htp]
    \centering
    \includegraphics[width=0.96\linewidth]{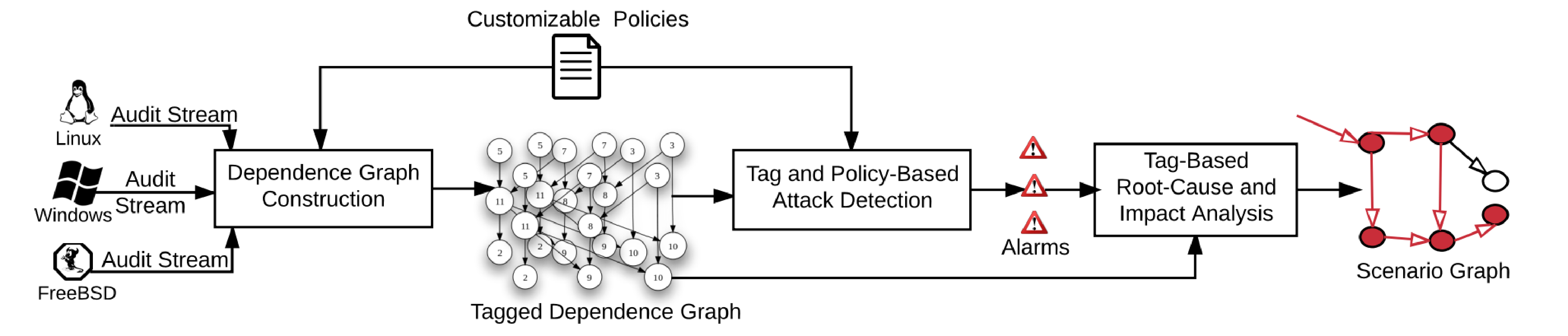}
    \caption{
    Provenance-based attack detection and root-cause reconstruction in the SLEUTH framework. Audit streams collected from different operating systems are transformed into a platform-independent dependency network. Tag- and policy-based analysis is then used for attack detection, root-cause identification, and impact analysis, ultimately producing a compact scenario network that summarizes the critical entities and causal relationships involved in the attack. Reproduced from~\cite{hossain2017sleuth}.
    }
    \label{fig:cyber_sleuth_overview}
\end{figure}

BackTracker traces system dependencies backward from evidence of an intrusion to identify events that may have contributed to the attack~\cite{king2003backtracking}. SLEUTH uses system audit records and policy-guided analysis to detect suspicious activities, reconstruct attack scenarios, and assess their consequences~\cite{hossain2017sleuth}. Figure~\ref{fig:cyber_sleuth_overview} illustrates how SLEUTH transforms audit streams into a dependency network and extracts a smaller scenario network containing relevant entities and relationships~\cite{hossain2017sleuth}. POIROT addresses a related but distinct task by matching attack behavior against provenance records for threat hunting~\cite{milajerdi2019poirot}. These systems use causal relationships inferred from system activity, but their detection targets and reconstruction procedures differ from those of source-ranking methods developed for epidemic diffusion.

The interpretation of reconstructed attack origins also depends on the scope of the available records. An initiating process identified within one host may have been triggered by an event on another machine, while an apparent first event may simply precede the beginning of log collection. Known attack traces can be used to assess whether the reconstruction retains relevant initiating events and their dependencies. A compact scenario network, however, may intentionally exclude unrelated or redundant activity. Its interpretability and size should therefore be evaluated separately from the completeness of the recovered attack history.

\subsection{Source Detection in Biological and Neural Systems}

Biological and neural network studies address several forms of inverse inference, including disease-gene prioritization, tumor lineage reconstruction, and the localization of neural activity. Although these problems all involve inferring unknown quantities from indirect observations, their inference targets differ from the initiating source of a diffusion process. In network medicine, molecular interaction networks and disease-association data are used to identify candidate disease-related genes~\cite{barabasi2011network}. Such approaches reveal how network structure can support inference from incomplete biological evidence, but gene prioritization does not necessarily involve reconstructing the origin of a spreading event.

Tumor evolution provides another example of reconstructing an unobserved history from present-day measurements. El-Kebir et al.~\cite{elkebir2016inferring} developed SPRUCE to infer tumor phylogenies from single-nucleotide variants and copy-number aberrations measured in mixed tumor samples. The method uses a multi-state perfect-phylogeny mixture model to reconstruct possible ancestral relationships among tumor cell populations. The inferred object is an evolutionary history rather than a diffusion origin on a fixed network.

EEG/MEG source imaging estimates the spatial distribution of neural electrical generators from measured electromagnetic fields using a forward head model~\cite{baillet2001electromagnetic,michel2019eegsource}. The lead-field matrix relates intracranial activity to sensor measurements, while regularization is used to stabilize the inverse solution. The reconstructed activity can be examined over time, but identifying the region that initiated a propagating event also requires information about the temporal recruitment of connected brain regions.

In epilepsy, regions showing strong activity during a seizure may have been recruited after its initiation. Bartolomei et al.~\cite{bartolomei2008epileptogenicity} introduced the Epileptogenicity Index to quantify regional involvement using intracerebral EEG recordings. The index combines spectral changes associated with fast oscillations and their timing relative to seizure onset, providing a measure of epileptogenic involvement across recorded brain structures.

\begin{figure}[htp]
    \centering
    \includegraphics[width=0.8\linewidth]{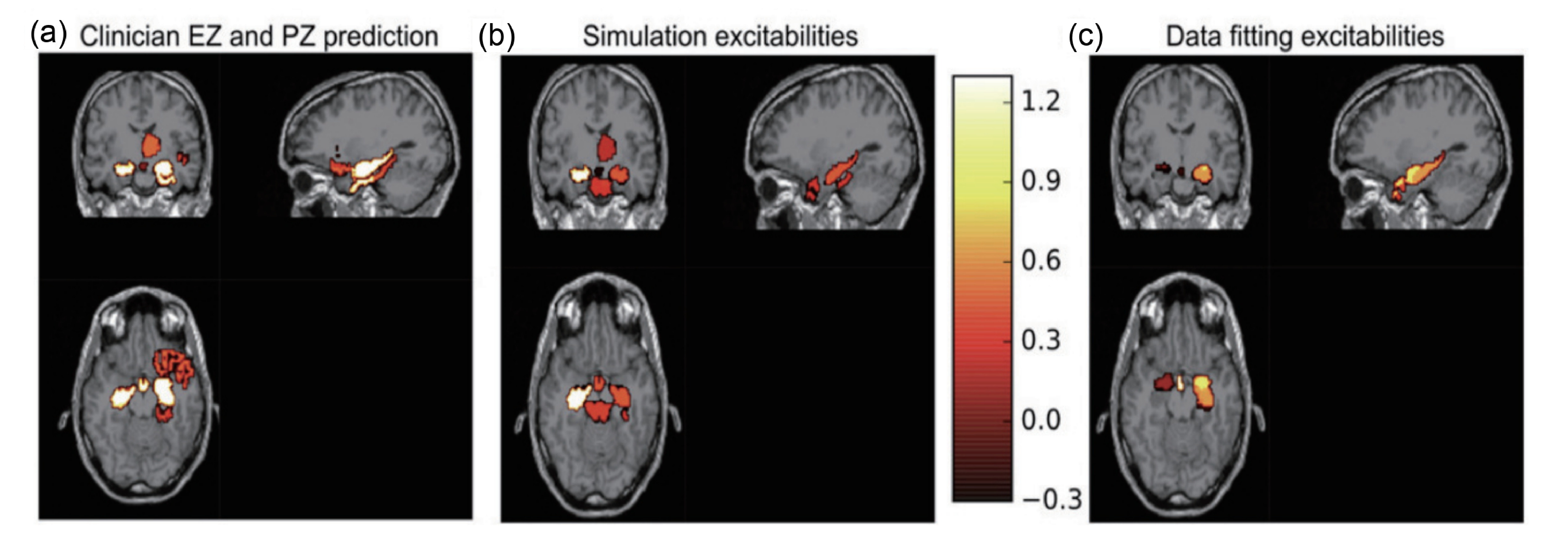}
    \caption{
 Clinical and model-based characterization of epileptogenic and propagation zones using the Virtual Epileptic Patient framework~\cite{jirsa2017virtual}. (a) Clinician-assessed epileptogenic zones (EZ) and propagation zones (PZ). (b) Regional excitability parameters used in patient-specific simulations. (c) Excitability parameters estimated by fitting the model to SEEG recordings.
}
    \label{fig:vep_ez_pz}
\end{figure}

The Virtual Epileptic Patient (VEP) framework combines patient-specific structural connectivity with dynamical models of seizure generation and propagation~\cite{jirsa2017virtual}. Regional excitability parameters are used to simulate seizure recruitment and can be estimated by fitting the model to stereoelectroencephalography (SEEG) recordings. Figure~\ref{fig:vep_ez_pz} compares the clinician's assessment of epileptogenic and propagation zones with the excitability distributions prescribed for simulation and obtained through data fitting. This approach relates observed seizure activity to a patient-specific model of neural propagation, allowing candidate epileptogenic regions to be examined in the context of their network connections.

\begin{figure}[htp]
    \centering
    \includegraphics[width=0.55\linewidth]{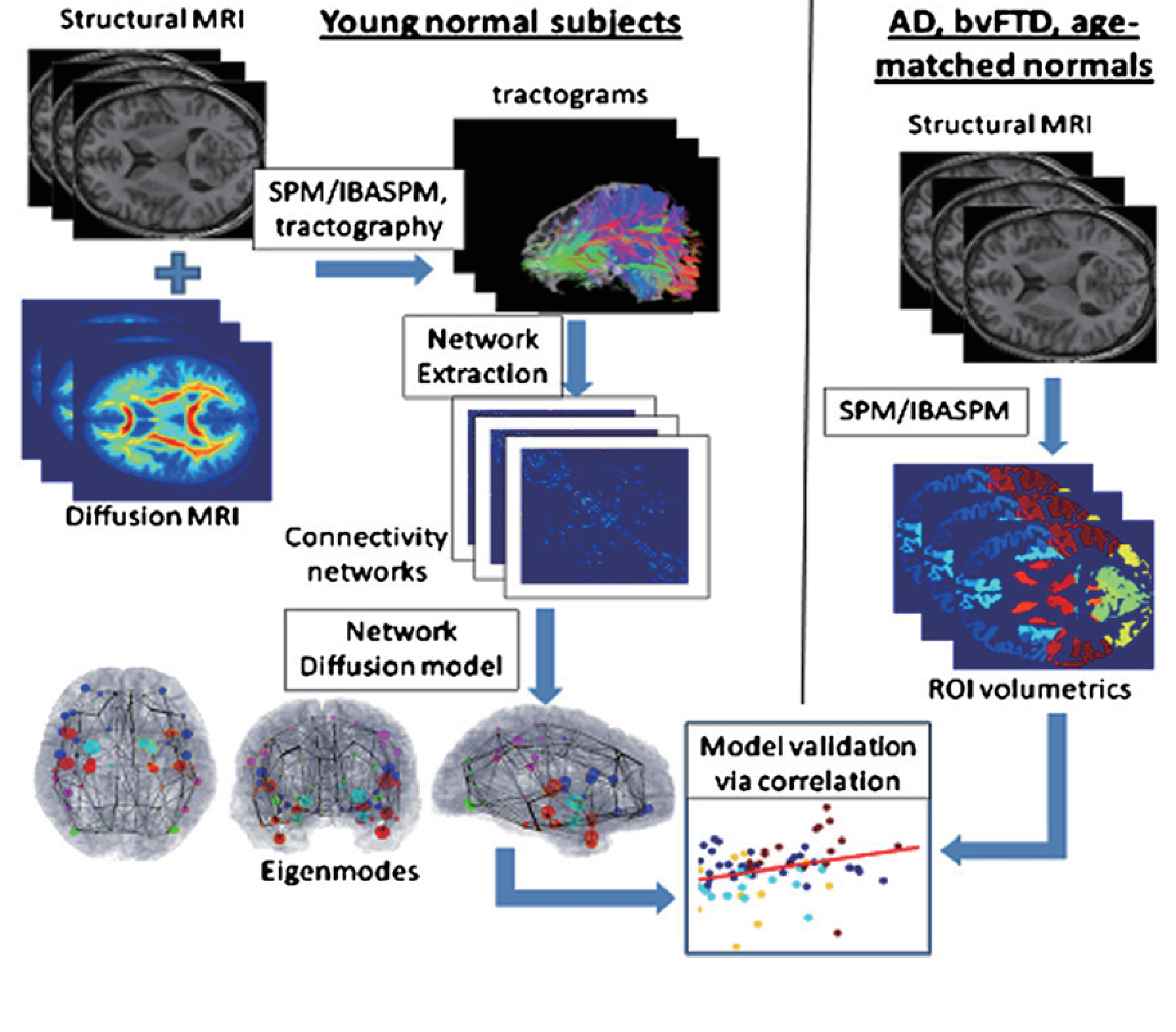}
    \caption{\rev{
Connectome-based diffusion modeling for neurodegenerative disease progression~\cite{raj2012network}. 
Structural and diffusion magnetic resonance imaging (MRI) data from healthy subjects are used to reconstruct the structural connectome, which serves as the substrate for modeling the spread of pathological burden. 
Predicted diffusion patterns are compared with regional volumetric measurements from patients to assess their agreement with observed patterns of atrophy.}}
    
    \label{fig:neuro_network_diffusion}
\end{figure}

Network diffusion models have also been used to study the spatial progression of neurodegenerative diseases. Raj et al.~\cite{raj2012network}  modeled the spread of pathological burden along anatomical brain connections and examined how the resulting diffusion patterns relate to regional atrophy observed in dementia. Figure~\ref{fig:neuro_network_diffusion} illustrates the construction of the structural connectome and the comparison between model predictions and measured atrophy patterns. This work provides a forward model of disease progression that could serve as the basis for a corresponding inverse problem, in which the initial distribution of pathological burden is inferred from later observations.

\FloatBarrier

\subsection{Source Detection in Supply Chain and Economic Systems}

Supply-chain records can help trace contamination through material and transaction flows. Economic-network studies also examine how shocks spread among firms connected through production and financial relationships~\cite{shi2026temporal}. These studies address different questions: where a disturbance originated, how it propagated, and which firms amplified its effects. Depending on the task, the source may be a contaminated batch, an initiating supplier, or a firm directly exposed to an external shock. Identifying the origin is distinct from assessing the extent of its consequences or the role of individual firms in amplifying them. Source labels used for evaluation should be established independently of the propagation outcomes used by the inference method.

Product traceability uses batch identifiers, production and transaction records, logistics histories, and inspection results to link downstream contamination with upstream candidates~\cite{golan2004traceability,aung2014traceability}. Directed flow and batch compatibility can exclude routes that an unweighted connectivity network would permit. Missing transactions, mixing, and shared suppliers can leave several plausible origins. The output may therefore be a traceback set requiring further inspection, rather than an automatically confirmed source.

Mass balance, direction of material flow, storage, and mixing constrain supply-chain propagation. A deterministic routing model can identify an upstream origin when its measurements and route constraints distinguish candidates.  Branching, mixing, and incomplete transaction histories may leave several upstream origins consistent with the observed contamination. Reliable traceback therefore requires information about material-flow directions, quantities, and transaction times, rather than supplier connectivity alone.

Manufacturing disruption studies analyze how shortages and delays spread through production dependencies~\cite{dolgui2018ripple}. Such ripple-effect models provide a propagation substrate for possible inverse analysis. Unless a study withholds and reconstructs the initiating disruption, its experiments evaluate consequences of a specified shock rather than detection of unknown sources. This distinction determines how its results can be used in a source-detection review.

Macroeconomic network models relate sectoral or firm-level outcomes to production linkages~\cite{acemoglu2012network}. Observed co-movement can reflect propagation, common shocks, or correlated exposures. Inferring an origin from it requires assumptions separating these explanations. Network centrality or estimated shock amplification alone does not identify which disturbance actually initiated the observed event.

\begin{figure}[htp]
    \centering
    \includegraphics[width=1\linewidth]{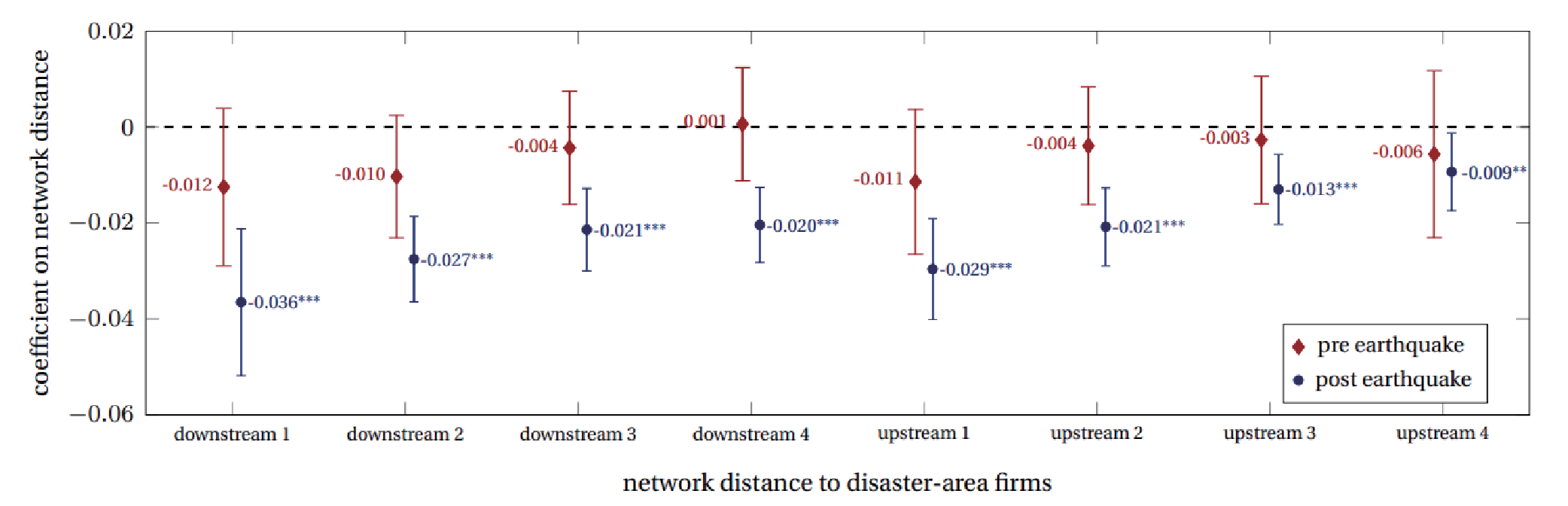}
    \caption{
Propagation of supply-chain disruptions following the Great East Japan Earthquake~\cite{carvalho2021supply}. Red diamonds and blue circles represent estimated coefficients before and after the earthquake, respectively. The horizontal axis shows upstream and downstream network distances from directly affected firms. The post-earthquake estimates indicate disruption effects across several tiers of the supply chain, with their magnitudes varying across directions and distances. The directly affected firms are known in this analysis.
}
    \label{fig:supply_chain_shock_propagation}
\end{figure}

Carvalho et al.~\cite{carvalho2021supply} examined how disruptions caused by the 2011 Great East Japan Earthquake spread through firm-level supply chains. Using directly affected firms as the observed exposure group, they estimated the effects on upstream suppliers and downstream customers. Figure~\ref{fig:supply_chain_shock_propagation} shows the estimated effects at different network distances before and after the earthquake. The results indicate that disruptions can extend across several tiers of supply-chain relationships, illustrating how an external shock propagates beyond directly affected firms.

Network traceability concerns how precisely an unknown origin can be identified from downstream observations. Lu et al.~\cite{lu2018universal} introduced Network Traceability Entropy (NTE) to characterize this property in food supply-chain networks. The measure is based on the posterior probabilities of feasible contamination sources and summarizes the uncertainty remaining after a specified set of downstream reports. Averaging this uncertainty across possible reporting configurations provides a network-level assessment of traceability under the assumed flow and observation model.

Let \(S\in V\) denote the random single contamination source and \(\mathcal{O}\) a set of reported contaminated retailers. The posterior probability that node \(v_i\) is the contamination source is defined as
\(
p_i(\mathcal{O})
=
P\!\left(
S=v_i
\mid
\mathcal{O},G
\right),
\)
the uncertainty associated with observation \(\mathcal{O}\), together with its average over all observation sets containing \(\lambda\) affected nodes, is given by
\begin{equation}
\begin{aligned}
H_{\mathrm{src}}(\mathcal{O};G)
&=
-\sum_{v_i\in\mathcal{C}(\mathcal{O})}
p_i(\mathcal{O})\log_{2}p_i(\mathcal{O}),\\
E^{\lambda}(G)
&=
\frac{1}{|\mathfrak{O}_{\lambda}|}
\sum_{\mathcal{O}\in\mathfrak{O}_{\lambda}}
H_{\mathrm{src}}(\mathcal{O};G),
\end{aligned}
\label{eq:network_traceability_entropy}
\end{equation}
where $\mathcal{C}(\mathcal{O})$ denotes the admissible candidate contamination-source nodes associated with the reports $\mathcal{O}$ under the assumed flow model and \(\mathfrak{O}_{\lambda}\) contains the retailer-report combinations of size \(\lambda\). The equal-weight average in this network index differs from the probability-weighted conditional entropy in Section~\ref{sec:information_identifiability}, unless those combinations are equiprobable. A smaller value describes more concentrated source uncertainty under the assumed flow model and candidate prior; it does not by itself certify that an inferred origin is correct.
\begin{figure}[!htp]
    \centering
    \includegraphics[width=0.8\linewidth]{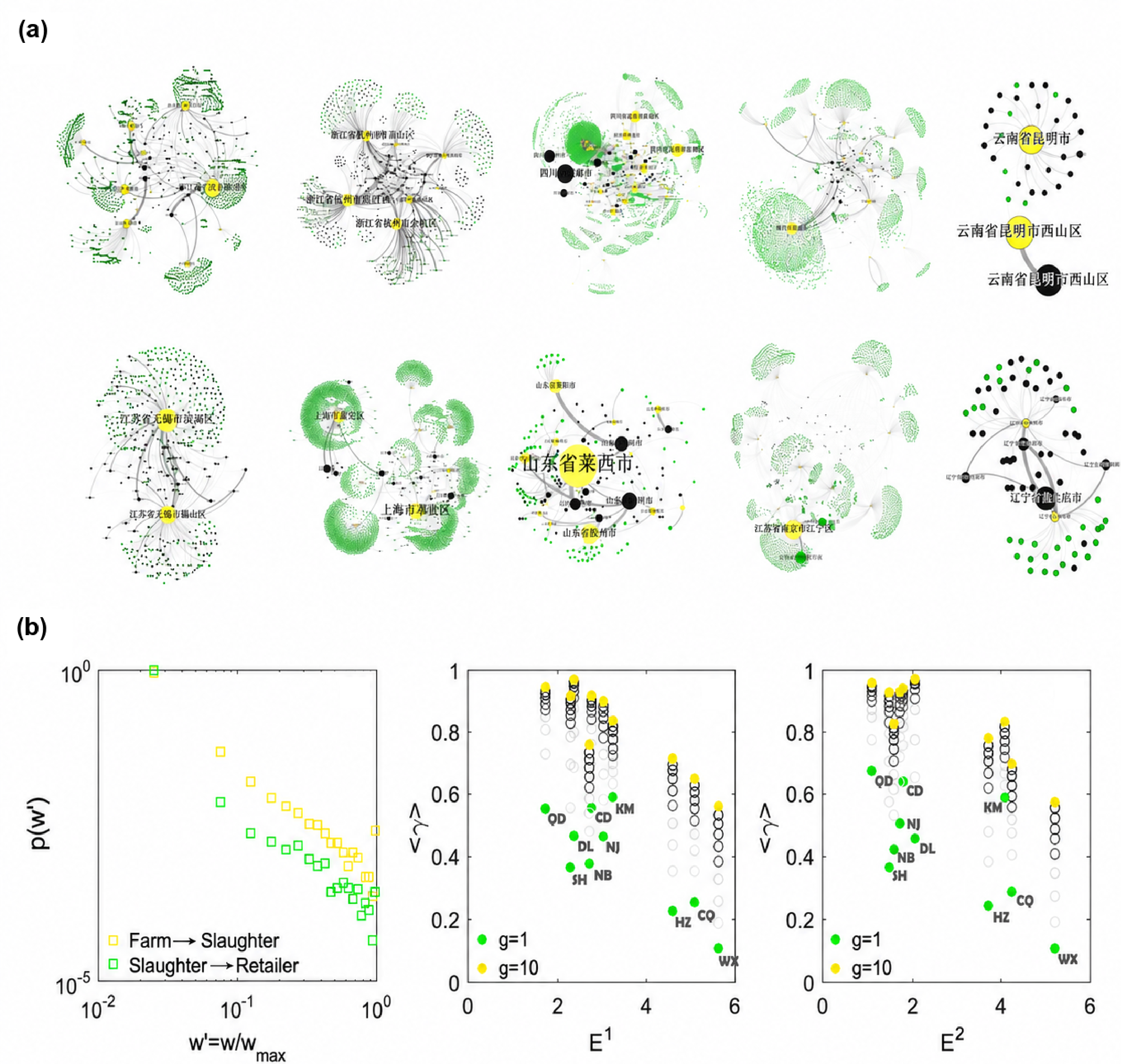}
    \caption{\rev{
Traceability analysis using empirical pork supply-chain networks~\cite{lu2018universal}. 
(a) Supply-chain networks reconstructed from transaction records for ten Chinese cities, where farms, slaughterhouses, and retailers are represented by black, yellow, and green nodes, respectively. 
(b) Network flow characteristics and source-tracing performance. The panels show the distribution of normalized transaction volumes together with the relationship between source-detection accuracy and the traceability measures \(E^{1}\) and \(E^{2}\) in simulated contamination scenarios with one and two reports, respectively. Network structure and transaction flows come from empirical records; the source-detection assessment uses the modeled contamination process.
}}
    
    \label{fig:nte_real_supply_chain}
\end{figure}
Figure~\ref{fig:nte_real_supply_chain}(a) shows the empirical networks used by Lu et al.~\cite{lu2018universal}; panel (b) relates their traceability indices to detection accuracy in the tested contamination scenarios. The comparison links network flow structure to source uncertainty within that model. A retrospective evaluation of source inference for actual contamination incidents would require independently confirmed origins.

Financial disturbances can spread through creditor--debtor relationships, common asset holdings, payment obligations, and price-mediated feedback. Defaults, liquidity shortages, and asset sales may transmit losses across institutions, while shared exposures can also produce correlated losses without direct contagion. Financial network models examine how these mechanisms contribute to the propagation and amplification of shocks~\cite{allen2000financial,eisenberg2001systemic,gai2010contagion}. For source inference, the central question is which event initiated the observed disturbance and how its effects reached other institutions. An institution that amplifies financial losses or occupies a systemically important position is not necessarily the origin of a particular episode. Distinguishing an initiating shock from subsequent contagion and common external exposures is therefore essential when interpreting financial source estimates.

\FloatBarrier

\section{Unifying Perspectives and Future Directions}
\label{sec:unifying-future}
\subsection{Toward an Information-centric View of Source Detection}\label{7.1}

The reviewed methods differ in what they observe, how they represent propagation, and how they search for an origin. These differences matter because methods that seek the same source may rely on different observations. A complete state snapshot describes the spatial distribution of node states without revealing the full transmission history, while sensor records may contain arrival times or transmission directions~\cite{pinto2012locating,cheng2022path}. Learning-based estimators draw on patterns represented in their training data. Such differences in the available information must be accounted for when comparing source-detection methods.

Across these settings, source inference targets the actual initiating node or set of nodes. Table~\ref{tab:method_selection} summarizes how different methods use observations, diffusion assumptions, and computational procedures to evaluate candidate sources. Structural approaches examine the geometry of observed spreading patterns, whereas message-passing and simulation-based methods incorporate specified diffusion dynamics~\cite{shah2011rumors,zhu2014information}. Learning-based methods use training examples to estimate source-related patterns and may reduce the cost of repeated inference. These approaches address the same underlying task, but their results must be interpreted in relation to the information available under each observation setting.

Eq.~\eqref{eq:unified_observation} distinguishes the latent diffusion trajectory from the observations generated by a specified measurement process. Section~\ref{sec:information_identifiability} further separates source identifiability under the observation model from ambiguity in a single realization and numerical instability. These distinctions point to different ways of improving source inference. Additional measurements may distinguish candidates that cannot be separated using the original records, while a more accurate diffusion model can improve candidate evaluation~\cite{hu2019locating}. More efficient search procedures can reduce computational costs without changing the information contained in the observations.

Observation design should be assessed against its intended decision. State reconstruction, early alarm, and source discrimination can favor different monitor locations~\cite{hu2024source,leskovec2007outbreak}. Likewise, more records can be redundant, and a restricted but well-chosen record can distinguish a small candidate set. These relationships are conditional on the source prior, dynamics, and measurement model, rather than consequences of node coverage alone.

\subsection{Open Challenges}

\textbf{(1) Limits of source inference.} Theoretical limits of source inference need to be stated for a specified candidate space, observation model, treatment of unknown parameters, and error criterion.
Distributional identifiability alone does not quantify finite-outbreak error, while posterior concentration can depend strongly on the prior~\cite{cover2006elements}. Relating these objects to explicit network and sampling conditions would make theoretical results more comparable without assuming a universal threshold across SI, SIR, SIS, and other dynamics.

\textbf{(2) Propagation model mismatch and mechanism uncertainty.} Most source detection methods rely on predefined diffusion mechanisms, whereas real propagation processes are often time-varying, heterogeneous, and only partially observed. As a result, reliable inference may become difficult when the assumed spreading model deviates from the actual diffusion process~\cite{GAJEWSKI201934,spinelli2016observer}. The relationship between mechanism uncertainty and source identifiability therefore remains insufficiently understood, limiting the applicability of existing methods in realistic environments.

\textbf{(3) Observation uncertainty and incomplete information.} Incomplete observations affect source detection far beyond reducing the amount of available data. Missing node states, hidden transmission paths, and structured sampling bias may fundamentally alter the evidence available for detection and increase ambiguity among candidate sources~\cite{luo2014identify,zhu2017catch}. How observation mechanisms reshape source identifiability, especially when coupled with diffusion uncertainty, remains an open theoretical problem.

\textbf{(4) Multiple sources.} When the number of sources and their initiation times are unknown, overlapping spreading regions make it difficult to attribute observations to individual origins. Methods evaluated with a fixed source count may behave differently when this count must also be estimated. Evaluation should distinguish missed sources, additional sources, and errors in their estimated locations. Approximate search procedures should also specify which source configurations are excluded from consideration~\cite{zhang2024multiple}.

\textbf{(5) Source detection on temporal, multilayer, and higher-order networks.} Temporal interactions constrain the order of possible transmission, while interlayer connections introduce additional propagation routes~\cite{zhan2025measuring,scholtes2014causality,zhang2025multilayer,boccaletti2014multilayer}. Higher-order interactions can also produce spreading patterns that are not captured by pairwise network representations. These structural features may change which candidate sources are consistent with the available observations. An important question is when static or pairwise approximations preserve source distinguishability and when temporal, layer-specific, or higher-order information is needed for reliable inference.

\textbf{(6) Scalability and stopping.} Large networks increase candidate search and simulation costs, while early decisions use shorter records~\cite{paluch2018fast,antulov2015identification}. The time to detect propagation differs from the time to identify its origin. A principled stopping policy should specify an error or uncertainty target and account for acquisition delay, computation, and changing candidate sets. Faster optimization of a fixed score does not automatically improve its statistical information.

\textbf{(7) Trustworthy data-driven source detection.} Learning-based source estimators are often trained on simulated diffusion data generated under a limited range of network structures, spreading parameters, and observation settings. Performance under these conditions does not establish how well an estimator will generalize to different networks or recording protocols. Evaluation should examine both detection accuracy and the reliability of uncertainty estimates when these conditions change. Predictive uncertainty under distribution shift is a broader concern in machine learning, but its implications for diffusion source inference require further investigation~\cite{ovadia2019uncertainty}.

\textbf{(8) Unified benchmarks and application-oriented evaluation.} Source-detection methods are often evaluated under different diffusion models, candidate spaces, observation budgets, and experimental protocols, making their reported accuracy difficult to compare directly~\cite{jiang2016identifying,gajewski2022comparison}. A common benchmark should distinguish improvements arising from additional observations from those obtained through more effective inference. Beyond detection accuracy, evaluation should examine uncertainty, robustness to model mismatch, computational cost, and the practical consequences of source-estimation errors.

A reusable benchmark should preserve the realized observation process as well as source labels. For each outbreak, it should record which nodes could have been sampled, which were sampled, the sampling times, the state or event definitions, and any censoring. For methods using training data, it should also record the simulator and source prior that produced those data. This allows a fair distinction between additional information and improved use of the same input. Reporting only a final accuracy table conceals these differences and makes apparently contradictory results difficult to interpret.

\subsection{Future Research Directions}

Future studies should distinguish improvements due to inference algorithms from those arising from additional observations. One approach is to compare estimators using the same diffusion realizations and observation records. Another is to vary the observation protocol while keeping the underlying diffusion process and evaluation conditions fixed. Such experiments can reveal whether performance differences arise from the available information or the inference procedure. Independent diffusion realizations with known or independently validated sources can also be used to evaluate uncertainty estimates. Reported candidate sets should be assessed by both their coverage of the true source and their size, since high coverage alone does not imply accurate source detection~\cite{li2021sensor}.

Mechanistic models and learning-based components can be combined at different stages of source inference, including parameter estimation, candidate selection, approximation of forward dynamics, and observation design. These applications require different forms of validation. Candidate-selection models can be evaluated by how reliably they retain the true source, while learned approximations of diffusion dynamics should be tested against the corresponding propagation model~\cite{wang2022invertible}. Adaptive observation policies require evaluation under the measurements collected through their own sampling decisions. Examining these components separately can help determine where improvements in source inference originate.

Applications also require a disciplined definition of initiation. A first infected individual, the location of an epidemic introduction, a contaminated batch, and a malicious process can all be meaningful sources, but they are not interchangeable labels. Studies of known-shock propagation, disease-gene prioritization, neural source imaging, and systemic importance can inform inverse modeling, even when their primary objectives differ from identifying the origin of a diffusion process~\cite{carvalho2021supply}. Validating source inference requires evaluating estimates of an unknown initiating source. Future comparisons should state where the label comes from and which alternatives the observations can exclude.

Adaptive monitoring is promising when later measurements can resolve specific remaining candidate ambiguities. The policy should account for monitor mobility, acquisition cost, censoring, and how quickly propagation changes relative to sampling. A useful comparison holds the total acquisition budget fixed and contrasts an adaptive policy with optimized fixed placement, reporting both source detection error and decision time~\cite{paluch2020optimizing}.

Source estimates should be reported together with the evidence and uncertainty associated with them. A point estimate identifies one candidate, while a ranking can guide further investigation. Posterior credible sets and frequentist confidence sets provide different forms of uncertainty assessment and should be evaluated according to their respective definitions. Relevant measures include detection error, set size and coverage, probability calibration, and sensitivity to uncertain model parameters. When several candidate sources remain plausible, identifying observations that could distinguish them may also help guide subsequent investigation.

Progress in source detection requires advances in both inference methods and the collection of informative observations. Distinguishing the true initiating source, its identifiability under a specified diffusion and observation model, and the estimator used to recover it is essential for interpreting theoretical results, experimental comparisons, and practical applications.

\section*{CRediT authorship contribution statement}
\textbf{Xiu-Xiu Zhan:} Conceptualization, Methodology, Writing -- original draft, Writing -- review \& editing, Project administration, Resources, Funding acquisition, Validation.
\textbf{Qiao Ke:} Conceptualization, Methodology, Investigation, Formal analysis, Visualization, Writing -- original draft, Writing -- review \& editing.
\textbf{Chengjun Zhang:} Investigation, Validation, Writing -- review \& editing.
\textbf{Miao He:} Investigation, Validation, Writing -- review \& editing.
\textbf{Chenwei Xie:} Investigation, Validation, Writing -- review \& editing.
\textbf{Seba Contreras:} Investigation, Validation, Writing -- review \& editing.
\textbf{Zhongliang Zhao:} Investigation, Methodology, Validation, Writing -- review \& editing.
\textbf{Yixiu Kong:} Investigation, Visualization, Validation, Writing -- review \& editing.
\textbf{Cong Li:} Investigation, Validation, Writing -- review \& editing.
\textbf{Chuang Liu:} Conceptualization, Supervision, Project administration, Resources, Funding acquisition, Writing -- review \& editing, Validation.

\section*{Declaration of competing interest}
The authors declare that they have no known competing financial interests or personal relationships that could have appeared to influence the work reported in this paper.

{\emergencystretch=3em
\section*{Acknowledgments}
This work was supported by Key Scientific \& Technological Project of XPCC (Grant No. 2025AA018), the National Natural Science Foundation of China (Grant No. 62573133, U23A20331, 62473123), the Scientific Research Fund of Zhejiang Provincial Education Department (Grant No. Y202558115), Beijing Social Science
Fund (Grant No. 24JCB040), Research Funds for the Central
Universities from the Beijing University of Posts and Telecommunications (Grant No. 505022019), Shanghai Institute for Mathematics and Interdisciplinary Sciences (SIMIS) (Grant No. SIMIS-ID-2025-NW), and Beihang University (Grant No. 606QNQR2024102002, 501RCPT2025102001).\par}

\bibliographystyle{elsarticle-num}
\bibliography{ref}

\end{document}